\documentclass[11pt]{article}
\usepackage{amsfonts,amssymb,amsmath,amscd,dsfont}
\usepackage{latexsym}
\usepackage[onehalfspacing]{setspace}
\usepackage{theorem}
\usepackage{natbib}
\usepackage[colorlinks=true,urlcolor=black,linkcolor=black,pdfpagemode="None",pdfstartview=FitH]{hyperref}
\usepackage{graphicx}
\usepackage{caption}
\usepackage{subfig}
\providecommand{\U}[1]{\protect\rule{.1in}{.1in}}
\newtheorem{theorem}{Theorem}

\newtheorem{Assumption}{Assumption}
\newcommand{\I}{\mathds{1}}
\hypersetup{pdfborder = {0 0 0},colorlinks=true,linkcolor=blue,citecolor=blue}
\renewcommand{\cite}{\citet*}
\newcommand{\diff}{\mathrm{d}}
\begin{document}

\title{Local Regression Distribution Estimators\thanks{Prepared for ``Celebrating Whitney Newey's Contributions to Econometrics'' Conference at MIT, May 17-18, 2019. We thank the conference participants for comments, and Guido Imbens and Yingjie Feng for very useful discussions. We are also thankful to the handling co-Editor, Xiaohong Chen, an Associate Editor and two reviewers for their input. Cattaneo gratefully acknowledges financial support from the National Science Foundation through grant SES-1947805, and Jansson gratefully acknowledges financial support from the National Science Foundation through grant SES-1947662 and the research support of CREATES.}\bigskip}
\author{Matias D. Cattaneo\thanks{Department of Operations Research and Financial Engineering, Princeton University.}
\and Michael Jansson\thanks{Department of Economics, UC Berkeley and CREATES.}
\and Xinwei Ma\thanks{Department of Economics, UC San Diego.}}
\maketitle

\begin{abstract}
This paper investigates the large sample properties of local regression distribution estimators, which include a class of boundary adaptive density estimators as a prime example. First, we establish a pointwise Gaussian large sample distributional approximation in a unified way, allowing for both boundary and interior evaluation points simultaneously. Using this result, we study the asymptotic efficiency of the estimators, and show that a carefully crafted minimum distance implementation based on ``redundant'' regressors can lead to efficiency gains. Second, we establish uniform linearizations and strong approximations for the estimators, and employ these results to construct valid confidence bands. Third, we develop extensions to weighted distributions with estimated weights and to local $L^{2}$ least squares estimation. Finally, we illustrate our methods with two applications in program evaluation: counterfactual density testing, and IV specification and heterogeneity density analysis. Companion software packages in \texttt{Stata} and \texttt{R} are available.

\end{abstract}

\bigskip

\bigskip
\thispagestyle{empty}
\setcounter{page}{0}
\newpage\doublespacing

\section{Introduction}\label{section:Introduction}

Kernel-based nonparametric estimation of distribution and density functions, as well as higher-order derivatives thereof, play an important role in econometrics. These nonparametric estimators often feature both as the main object of interest and as preliminary ingredients in multi-step semiparametric procedures \citep{Newey-McFadden_1994_Handbook,Ichimura-Todd_2007_HandbookCh}. Whitney Newey's path-breaking contributions to non/semiparametric econometrics employing kernel smoothing are numerous.\footnote{See, for example, \cite{Newey-Stoker_1993_ECMA}, \cite{Newey_1994_ECMA}, \cite{Newey_1994_ET}, \cite{Hausman-Newey_1995_ECMA}, \cite{Robins-Hsieh-Newey_1995_JRSSB}, \cite{Newey-Hsieh-Robins_2004_ECMA}, \cite{Newey-Ruud_2005_BookCh}, \cite{Ichimura-Newey_2020_wp}, and \cite{Chernozhukov-Escanciano-Ichimura-Newey-Robins_2020_LocallyRobust}.} This paper hopes to honor his influential work in this area by studying the main large sample properties of a new class of \emph{local regression distribution estimators}, which can be used for non/semiparametric estimation and inference.

The class of local regression distribution estimators is constructed using a local least squares approximation to the empirical distribution function of a random variable $x\in\mathcal{X}\subseteq\mathbb{R}$, where the localization at the evaluation point $\mathsf{x}\in\mathcal{X}$ is implemented via a kernel function and a bandwidth parameter. The local functional form approximation is done using a finite-dimension basis function. When the basis function contains polynomials up to order $p\in\mathbb{N}$, the associated least squares coefficients give estimators of the distribution function, density function, and higher-order derivatives (up to order $p-1$), all evaluated at $\mathsf{x}\in\mathcal{X}$. If only a polynomial basis is used, then the estimator reduces to the one recently proposed in \cite{Cattaneo-Jansson-Ma_2020_JASA}.

We present two main large sample distributional results for the local regression distribution estimators. First, in Section \ref{section:Pointwise Distribution Theory}, we establish a pointwise (in $\mathsf{x}\in\mathcal{X}$) Gaussian distributional approximation with consistent standard errors. Because these estimators have a U-statistic structure with an $n$-varying kernel, where $n$ denotes the sample size, we construct a fully automatic Studentization given a choice of basis, kernel, and bandwidth. Furthermore, we show that when the basis function includes polynomials, the associated density and its higher-order derivatives estimators are boundary adaptive without further modifications. This result generalizes \cite{Cattaneo-Jansson-Ma_2020_JASA} by allowing for arbitrary local basis functions, which is particularly useful for efficiency considerations.

To be more precise, for the special case of local polynomial density estimation, \cite{Cattaneo-Jansson-Ma_2020_JASA} showed that the asymptotic variance of the estimator is of the ``sandwich'' form, which does not reduce to a single matrix (up to a proportional factor) by a choice of kernel function. This finding indicates that more efficient estimators can be constructed via a minimum distance approach based on ``redundant'' regressors, following well-known results in econometrics \citep{Newey-McFadden_1994_Handbook}. In Section \ref{subsection:Efficiency}, we present a novel minimum distance construction for estimation of the density and its derivatives, and obtain an efficiency bound for the new minimum distance density estimator. Furthermore, we show that the efficiency bound coincides with the well-known asymptotic variance lower bound for kernel-based density estimation \citep*{Granovsky-Muller_1991_ISR,Cheng-Fan-Marron_1997_AoS}. We also show that this efficiency bound is tight: we construct a feasible minimum distance procedure exploiting carefully chosen redundant regressors, which leads to an estimator with asymptotic variance arbitrarily close to the theoretical efficiency bound. These results offer not only a novel theoretical perspective on efficiency of classical nonparametric kernel-based density estimation, but also a new class of more efficient boundary adaptive density estimators for practice. We also discuss how these results generalize to other local regression distribution estimators in the supplemental appendix.

Our second main large sample distributional result, in Section \ref{section:Uniform Distribution Theory}, concerns uniform estimation and inference over a region $\mathcal{I}\subseteq \mathcal{X}$, based on either the basic local regression distribution estimators or the associated more efficient estimators obtained via our proposed minimum distance procedure. More precisely, we establish a strong approximation to the boundary adaptive Studentized statistic, uniformly over $\mathsf{x}\in\mathcal{I}$, relying on a ``coupling'' result in \cite{Gine-Koltchinskii-Sakhanenko_2004_PTRF}; see also \cite{Rio_1994_PTRF} and
\cite{Gine-Nickl_2010_AoS} for closely related results, and \cite{Zaitsev_2013_RMSurveys} for a review on strong approximation methods. This approach allows us to deduce a distributional approximation for many functionals of the Studentized statistic, including its supremum, following ideas in \cite{Chernozhukov-Chetverikov-Kato_2014a_AoS}. For further discussion and references on strong approximations and their applications to non/semiparametric econometrics see
\cite{Chernozhukov-Chetverikov-Kato_2014b_AoS}, \cite{Belloni-Chernozhukov-Chetverikov-Kato_2015_JoE}, \cite{Belloni-Chernozhukov-Chetverikov-FernandezVal_2019_JoE}, and \cite{Cattaneo-Farrell-Feng_2020_AoS}, \cite{Cattaneo-Crump-Farrell-Feng_2021_Binscatter}, and references therein.

We employ our strong approximation results for local regression distribution estimators to construct asymptotically valid confidence bands for the density function and derivatives thereof. Other applications of our results, not discussed here to conserve space, include  specification and shape restriction testing. As a by-product, we also establish a linear approximation to the boundary adaptive Studentized statistic, uniformly over $\mathsf{x}\in\mathcal{I}$, which gives uniform convergence rates and can be used for further theoretical developments. See the supplemental appendix for more details.

In addition to our main large sample results for local regression distribution and related estimators, we briefly discuss several extensions in Section \ref{section:Extensions}. First, we allow for a weighted empirical distribution function entering our estimators, where the weights themselves may be estimated. Our results continue to hold in this more general case, which is practically relevant as illustrated in our empirical applications. Second, we present and study an alternative class of estimators that employ a non-random $L^{2}$ loss function, instead of the more standard least squares approximation underlying our local regression distribution estimators. These alternative estimators enjoy certain theoretical advantages, but require ex-ante knowledge of the boundary location of $\mathcal{X}$. In particular, we show in the supplemental appendix how these alternative estimators can be implemented to achieve maximum asymptotic efficiency in estimating the density function and its derivatives. Third, we also discuss incorporating shape restrictions using the general local basis function entering the local regression distribution estimators.

Finally, in Section \ref{section:Applications}, we illustrate our methods with two applications in program evaluation \citep[for a review see][]{Abadie-Cattaneo_2018_ARE}. First, we discuss counterfactual density analysis following \cite{DiNardo-Fortin-Lemieux_1996_ECMA}; see also \cite{Chernozhukov-FernandezVal-Melly_2013_ECMA} for related discussion based on distribution functions. Second, we discuss specification testing and heterogeneity analysis in the context of instrumental variables following \cite{Kitagawa_2015_ECMA} and \cite{Abadie_2003_JOE}, respectively; see also \cite{Imbens-Rubin_2015_Book} for background and other applications of nonparametric density estimation to causal inference and program evaluation. In all these applications, we develop formal estimation and inference methods based on nonparametric density estimation using local regression distribution estimators implemented with weighted distribution functions. We showcase our new methods using a subsample of the data in \cite{Abadie-Angrist-Imbens_2002_ECMA}, corresponding to the Job Training Partnership Act (JTPA).

From both methodological and technical perspectives, our proposed class of local regression distribution estimators is different from, and exhibits demonstrable advantages over, other related estimators available in the literature. For the special case of density estimation (i.e., when the basis function is taken to be polynomial), our resulting kernel-based density estimator enjoys boundary carpentry over the possibly unknown boundary of $\mathcal{X}$, does not require preliminary smoothing of the data and hence avoids preliminary tuning parameter choices, and is easy to implement and interpret. \citet*{Cattaneo-Jansson-Ma_2020_JASA} gave a detailed discussion of that density estimator and related approaches in the literature, which include the influential local polynomial estimator of \citet*{Cheng-Fan-Marron_1997_AoS} and related estimators \citep[][and referneces therein]{Zhang-Karunamuni_1998_JSPI,Karunamuni-Zhang_2008_SPL}. The class of estimators we consider here can be more efficient by employing minimum distance estimation ideas (Section \ref{section:Pointwise Distribution Theory}), easily delivers intuitive estimators of density-weighted averages (Section \ref{subsection:Reweighted Distribution}), and allows for incorporating shape and other restrictions (Section \ref{subsection:otherbasis}), among other features that we discuss below. Last but not least, some of the technical results presented herein for the general class of estimators, such as asymptotic efficiency (Section \ref{subsection:Efficiency}) and uniform inference (Section \ref{section:Uniform Distribution Theory}) are new, even for the special case of density estimation in \citet*{Cattaneo-Jansson-Ma_2020_JASA}.

The rest of the paper proceeds as follows. Section \ref{section:Setup} introduces the class of local regression distribution estimators. Section \ref{section:Pointwise Distribution Theory} establishes a pointwise distributional approximation, along with a consistent standard error estimator, and discusses efficiency focusing in particular on the leading special case of density estimation. Section \ref{section:Uniform Distribution Theory} establishes uniform results, including valid linearizations and strong approximations, which are then used to construct confidence bands. Section \ref{section:Extensions} discusses extensions of our methodology, while Section \ref{section:Applications} illustrates our new methods with two distinct program evaluation applications. Section \ref{section:Conclusion} concludes. The supplemental appendix (SA) includes all proofs of our theoretical results as well as other technical, methodological and numerical results that may be of independent interest. Software packages for \texttt{Stata} and \texttt{R} implementing the main results in this paper are discussed in \cite{Cattaneo-Jansson-Ma_2021_JSS}.

\section{Setup}\label{section:Setup}

Suppose $x_{1},x_{2},\dots,x_{n}$ is a random sample from a univariate random variable $x$ with absolute continuous cumulative distribution function $F(\cdot)$, and associated Lebesgue density $f(\cdot)$, over its support $\mathcal{X}\subseteq\mathbb{R}$, which may be compact and not necessarily known. We propose, and study the large sample properties of a new class of nonparametric estimators of $F(\cdot)$, $f(\cdot)$, and derivatives thereof, both pointwise at $\mathsf{x}\in\mathcal{X}$ and uniformly over some region $\mathcal{I}\subseteq\mathcal{X}$.

Our proposed estimators are applicable whenever $F(\cdot)$ is suitably smooth near $\mathsf{x}$ and admits a sufficiently accurate linear-in-parameters local approximation of the form:
\begin{equation}\label{cdf approximation, generic}
\varrho(h,\mathsf{x}) = \sup_{|x-\mathsf{x}|\leq h} \Big|F(x)- R(x-\mathsf{x})'\theta(\mathsf{x})\Big|\qquad \text{is small for $h$ small},
\end{equation}
where $R(\cdot)$ is a known local basis function and $\theta(\mathsf{x})$ is a parameter vector to be estimated. As an estimator of $\theta(\mathsf{x})$ in \eqref{cdf approximation, generic}, we consider the local regression estimator
\begin{equation}\label{Local regression estimator}
\hat{\theta}(\mathsf{x})=\operatorname*{argmin}_{\theta}\sum_{i=1}^n W_{i}\left(\hat{F}_{i}-R_{i}'\theta\right)^{2},
\end{equation}
where $W_{i}=K((x_{i}-\mathsf{x})/h)/h$ for some kernel $K(\cdot)$ and some bandwidth $h,$ $R_{i}=R(x_{i}-\mathsf{x})$, and 
\begin{equation}\label{cdf estimator}
\hat{F}_{i}=\frac{1}{n}\sum_{j=1}^n \I(x_{j}\leq x_{i})
\end{equation}
is the empirical distribution function evaluated at $x_i$, with $\I(\cdot)$ denoting the indicator function.

The generic formulation \eqref{cdf approximation, generic} is motivated in part by the important special case where $F(\cdot)$ is sufficiently smooth, in which case
\begin{equation}\label{cdf approximation, interior point}
F(x)\approx F(\mathsf{x})+f(\mathsf{x})(x-\mathsf{x})+...+f^{(p-1)}(\mathsf{x})\frac{1}{p!}(x-\mathsf{x})^{p}\qquad\text{for }x\approx
\mathsf{x},
\end{equation}
and $f^{(s)}(\mathsf{x})=\left.  \diff^{s}f(x)/\diff x^{s}\right\vert _{x=\mathsf{x}}$ are higher-order density derivatives. Of course, the approximation (\ref{cdf approximation, interior point}) is of the form \eqref{cdf approximation, generic} with $R(u)=(1,u,\cdots,u^{p}/p!)'$, and hence $\theta(\mathsf{x})=(F(\mathsf{x}),f(\mathsf{x}),\cdots,f^{(p-1)}(\mathsf{x}))'$. In such special case, the estimator $\hat{\theta}(\mathsf{x})$ corresponds to one of the estimators introduced in \cite{Cattaneo-Jansson-Ma_2020_JASA}. But, as further discussed below, other choices of $R(\cdot)$ and/or $\theta(\mathsf{\cdot})$ can be attractive, and as a consequence we take \eqref{cdf approximation, generic} as the starting point for our analysis. Section \ref{section:Extensions} discusses other extensions and generalization of the basic local regression distribution estimator $\hat{\theta}(\mathsf{x})$ in \eqref{Local regression estimator}.

The class of estimators defined in \eqref{Local regression estimator} is motivated by standard local polynomial regression methods \citep{Fan-Gijbels_1996_Book}. However, well-known results for local polynomial regression are not applicable to the local regression distribution estimator, $\hat{\theta}(\mathsf{x})$, because the empirical distribution function estimator, $\hat{F}_{i}$, which plays the role of the ``dependent'' variable in the construction, depends on not only $x_i$ but also all of the ``independent'' observations $x_1,x_2,\dots,x_n$. This implies that, unlike the case of standard local polynomial regression, $\hat{\theta}(\mathsf{x})$ cannot be studied by conditioning on the ``covariates'' $x_{1},x_{2},\dots,x_{n}$. Instead, we employ U-statistic methods for analyzing the statistical properties of $\hat{\theta}(\mathsf{x})$. This observation explains the quite different asymptotic variance of our estimator: see Section \ref{subsection:Efficiency} for details. Furthermore, as discussed in Section \ref{subsection:Reweighted Distribution}, when a weighted distribution function is used in place of $\hat{F}_i$ in \eqref{Local regression estimator}, the resulting (weighted) local regression distribution estimators are consistent for a density-weighted regression function, as opposed to being consistent for the regression function itself (as it is the case for standard local polynomial regression methods). Finally, the SA highlights other technical differences between the two types of local regression estimators.

\section{Pointwise Distribution Theory}\label{section:Pointwise Distribution Theory}

This section discusses the large sample properties of the estimator $\hat{\theta}(\mathsf{x}),$ pointwise in $\mathsf{x}\in\mathcal{X}$. We first establish asymptotic normality, and then discuss asymptotic efficiency. Other results are reported in the SA to conserve space. We drop the dependence on the evaluation point $\mathsf{x}$ whenever possible. 

\subsection{Assumptions}\label{subsection:Assumptions}

We impose the following assumption throughout this section. We do not restrict the support of $\mathcal{X}$, which can be a compact set or unbounded, because our estimator automatically adapts to boundary evaluation points.

\begin{Assumption}
\label{ass:AN}$x_{1},\dots,x_{n}$ is a random sample from a distribution $F(\cdot)$ supported on $\mathcal{X}\subseteq\mathbb{R}$, and $\mathsf{x}\in\mathcal{X}$.

(i) For some $\delta>0,$ $F(\cdot)$ is absolutely continuous on $[\mathsf{x} -\delta,\mathsf{x}+\delta]$ with a density $f(\cdot)$ admitting constants $f(\mathsf{x}-)$, $\dot{f}(\mathsf{x}-)$, $f(\mathsf{x}+)$, and $\dot{f}(\mathsf{x}+)$ such that
\[ \sup_{u\in\lbrack-\delta,0)}\frac{|f(\mathsf{x}+u)-f(\mathsf{x}-)-\dot{f}(\mathsf{x}-)u|}{|u|^{2}}
  +\sup_{u\in(0,\delta]}\frac{|f(\mathsf{x}+u)-f(\mathsf{x}+)-\dot{f}(\mathsf{x}+)u|}{|u|^{2}}<\infty.
\]

(ii)\ $K(\cdot )$ is nonnegative, symmetric, and continuous on its support $[-1,1]$, and integrates to 1. 

(iii)\ $R(\cdot)$ is locally bounded, and there exists a positive-definite diagonal matrix $\Upsilon_{h}$ for each $h>0$, such that $\Upsilon_{h}R(u)=R(u/h)$.

(iv) Let $\mathcal{X}_{h,\mathsf{x}}=\frac{\mathcal{X}-\mathsf{x}}{h}$. For all $h$ sufficiently small, the minimum eigenvalues of $\Gamma_{h,\mathsf{x}}$ and $h^{-1}\Sigma_{h,\mathsf{x}}$ are bounded away from zero, where
\begin{align*}
\Gamma_{h,\mathsf{x}} &= \int\nolimits_{\mathcal{X}_{h,\mathsf{x}}}R(u)R(u)'K(u)f(\mathsf{x}+hu)\diff u,\\
\Sigma_{h,\mathsf{x}} &= \int_{\mathcal{X}_{h,\mathsf{x}}}\int_{\mathcal{X}_{h,\mathsf{x}}} R(u)R(v)' \Big[ F(\mathsf{x}+h \min\{u,v\}) - F(\mathsf{x}+hu)F(\mathsf{x}+hv) \Big]\\
&\qquad\qquad\qquad\qquad\cdot K(u)K(v) f(\mathsf{x}+hu)f(\mathsf{x}+hv)\diff u\diff v.
\end{align*}
\end{Assumption}

Part (i) imposes smoothness conditions on the distribution function $F(\cdot)$, separately for the two regions on the left and on the right of the evaluation point $\mathsf{x}$. In most applications, the distribution function will also be smooth at the evaluation point, in which case $f(\mathsf{x}-)=f(\mathsf{x}+)$ and $\dot f(\mathsf{x}-)=\dot f(\mathsf{x}+)$. However, there are important situations where $F(\cdot)$ only has one-sided derivatives, such as at boundary or kink evaluation points. Part (ii) imposes standard restrictions on the kernel function, which allows for all commonly used (compactly supported) second-order kernel functions. Part (iii) requires that the local basis $R(\cdot)$ can be stabilized by a suitable normalization. Parts (iv) give assumptions on two (non-random) matrices which will feature in the asymptotic distribution.
 
The error of the approximation in \eqref{cdf approximation, generic} depends on the choice of $R(\cdot)$ and $\theta$, and is quantified by $\varrho(h)$, where we suppress the dependence on the evaluation point $\mathsf{x}$ to save notation. The approximation error will be required to be ``small'' in the sense that $n\varrho(h)^{2}/h\to0$. In the cases of main interest (i.e., when $R(\cdot)$ is polynomial), we have either $\varrho(h)=O(h^{p+1})$ or $\varrho(h)=o(h^{p})$ for some $p$. The condition can therefore be stated as $nh^{2p+1}\to0$ and $nh^{2p-1}=O(1)$, respectively, in those cases.

We do not discuss how to choose the bandwidth $h$, or the order $p$ if $R(\cdot)$ contains polynomials, as both choices can be developed following standard ideas in the local polynomial literature. We focus instead on distributional approximation (Section \ref{subsection:Asymptotic Normality}) and asymptotic variance minimization (Section \ref{subsection:Efficiency}), given a choice of bandwidth sequence and polynomial order. Bandwidth selection can be developed by extending the results in \citet*{Cattaneo-Jansson-Ma_2020_JASA} and polynomial order selection can be developed following \citet[Section 3.3]{Fan-Gijbels_1996_Book}. In particular, a larger $p$ can lead to more bias reduction whenever the target population function is smooth enough at the expense of a larger asymptotic variance. We discuss this trade-off explicitly in our efficiency calculations (Section \ref{subsection:Efficiency}).

\subsection{Asymptotic Normality}\label{subsection:Asymptotic Normality}

We show that, under regularity conditions and if $h$ vanishes at a suitable rate as $n\to\infty$, then
\begin{equation}\label{AN, studentized estimator}
\hat{\Omega}^{-1/2}(\hat{\theta}-\theta)\rightsquigarrow\mathcal{N}(0,I),\qquad\hat{\Omega}=\hat{\Gamma}^{-1}\hat{\Sigma}\hat{\Gamma}^{-1},
\end{equation}
where
\[\hat{\Gamma}=\frac{1}{n}\sum_{i=1}^nW_{i}R_{i}R_{i}',\qquad
  \hat{\Sigma}=\frac{1}{n^2}\sum_{i=1}^n\hat{\psi}_{i}\hat{\psi}_{i}',\qquad \hat{\psi}_{i}=\frac{1}{n}\sum_{j=1}^nW_{j}R_{j}(\I(  x_{i}\leq x_{j}) - \hat{F}_j).
\]
It follows from this result that inference on $\theta$ can be based on $\hat{\theta}$ by employing the (pointwise) distributional approximation $\hat{\theta}\overset{a}{\sim}\mathcal{N}(\theta,\hat{\Omega})$. The three matrices, $\hat\Gamma$, $\hat\Sigma$ and $\hat{\Omega}$, depend on the evaluation point $\mathsf{x}$, but such dependence is again suppressed for simplicity. This distributional result will rely on the ``small'' bias condition $n\varrho(h)^{2}/h\to0$ mentioned above, which makes the asymptotic approximation (or smoothing) bias of $\hat{\theta}$ negligible relative to the standard error. From an inference perspective, such bias condition can be achieved by employing undersmoothing or robust bias correction: see \cite{Calonico-Cattaneo-Farrell_2018_JASA,Calonico-Cattaneo-Farrell_2020_CEOptimal} for discussion and background references. The SA includes more details on the bias of the estimator. 

To provide some insight into the distributional approximation \eqref{AN, studentized estimator}, and to see why it cannot be established using standard results for local polynomial regression, first observe that
\[\hat{\theta}-\theta=\hat{\Gamma}^{-1}S,\qquad S=\frac{1}{n}\sum_{i=1}^nW_{i}R_{i}(\hat{F}_{i}-R_{i}'\theta),\]
assuming $\hat{\Gamma}$ is invertible with probability approaching one. The statistic $S$ can be written as
\begin{equation}
	S=U + B, \qquad U = \frac{1}{n(n-1)}\sum_{i,j=1,i\neq j}^nW_{j}R_{j} \Big(\I(  x_{i}\leq x_{j})-F(x_j)\Big),\label{eq:U-stat representation}
\end{equation}
where $B$ consists of a leave-in bias term and a smoothing bias term. Since $S$ is approximately a second-order $U$-statistic, result (\ref{AN, studentized estimator}) should follow from a central limit theorem for ($n$-varying) $U$-statistics under suitable regularity conditions, including conditions ensuring that the approximation errors are negligible. More specifically, result (\ref{AN, studentized estimator}) follows if $U$ is asymptotically mean-zero Gaussian $\mathbb{V}[U]^{-1/2}U\rightsquigarrow\mathcal{N}(0,I)$, where $\mathbb{V}[U]$ denotes the variance of $U$, $\mathbb{V}[U]^{-1/2}B\to_{\mathbb{P}} 0$, and if the variance estimator $\hat{\Sigma}$ is consistent in the sense that $\mathbb{V}[U]^{-1}(\hat{\Sigma}-\mathbb{V}[U])\to_{\mathbb{P}}0$.  Moreover, the projection theorem for $U$-statistics implies that, under appropriate regularity conditions,
\[\mathbb{V}[U]\approx\frac{1}{n}\mathbb{E}[\psi_i\psi_i'],\qquad\psi_i=\mathbb{E}[W_{j}R_{j}\I(x_{i}\leq x_{j})-F(x_j)|x_{i}],\]
which motivates the functional form of the variance estimator $\hat{\Sigma}$ used to form $\hat{\Omega}$.

The following theorem formalizes the above intuition with precise sufficient conditions.

\begin{theorem}[Pointwise Asymptotic Normality]
\label{thm:AN}Suppose Assumption \ref{ass:AN} holds. If $n\varrho(h)^{2}/h\to0$ and $nh^{2}\to\infty$, then (\ref{AN, studentized estimator}) holds.
\end{theorem}

This theorem establishes a (pointwise) Gaussian distributional approximation for the Studentized statistic $\hat{\Omega}^{-1/2}(\hat{\theta}-\theta)$, which is valid for each evaluation point $\mathsf{x}\in\mathcal{X}$. For example, letting $c$ be a vector of conformable dimension and $\alpha\in(0,1)$, this result justifies the standard $100(1-\alpha)\%$ confidence interval
\[
\text{CI}_{\alpha}(\mathsf{x})=\left[c'\hat{\theta}(\mathsf{x})-\mathfrak{q}_{1-\alpha/2}\sqrt{c'\hat{\Omega}(\mathsf{x})c}~,~
                                     c'\hat{\theta}(\mathsf{x})-\mathfrak{q}_{\alpha/2}\sqrt{c'\hat{\Omega}(\mathsf{x})c}\right],
\]
where $\mathfrak{q}_{a}=\inf\{u\in\mathbb{R}:\mathbb{P}\left[  \mathcal{N}(0,1)\leq u\right]  \geq a\}$.
The above confidence interval is asymptotically valid for each evaluation point $\mathsf{x}$, which is reflected by the notation $\text{CI}_{\alpha}(\mathsf{x})$. That is, 
\[\lim_{n\to\infty}\mathbb{P}\Big[ c' \theta(\mathsf{x})\in\text{CI}_{\alpha}(\mathsf{x})\Big]  =1-\alpha,\qquad\text{for all }\mathsf{x}\in\mathcal{X}.\] Section \ref{section:Uniform Distribution Theory} develops asymptotically valid confidence bands, which will be denoted by $\text{CI}_{\alpha}(\mathcal{I})$ for some region $\mathcal{I}\subseteq \mathcal{X}$.

\subsection{Efficiency}\label{subsection:Efficiency}

As it is well known in the literature \citep{Fan-Gijbels_1996_Book}, the standard local polynomial regression estimator of $\mathbb{E}[y|x=\mathsf{x}]$, for dependent variable $y$ and independent variable $x$, has a limiting asymptotic variance of the ``sandwich form'' $e_0'\Gamma^{-1}A\Gamma^{-1}e_0$, where $e_\ell$ denotes the $(\ell + 1)$th standard basis vector, and
\[\Gamma = f(\mathsf{x})\int_{-1}^1 R(u)R(u)'K(u)\diff u, \quad A = \mathbb{V}[y|x=\mathsf{x}]f(\mathsf{x}) \int_{-1}^1 R(u)R(u)'K(u)^{2}\diff u.\]
This variance structure implies that setting $K(\cdot)$ to be the uniform kernel makes $\Gamma$ proportional to $A$ (i.e., $K(u)=K(u)^2$ whenever $K(u)=\I(|u|\leq1)$), and hence minimizes the above asymptotic variance, at least in the sense that $\Gamma^{-1}A\Gamma^{-1}\geq A^{-1}$. See also \citet*{Granovsky-Muller_1991_ISR} for a more general discussion on the optimality of the uniform kernel for kernel-based estimation.

Unlike the case of the asymptotic variance of local polynomial regression, however, our local regression distribution estimators exhibit a more complex and uneven asymptotic variance formula due to their construction. As a result, employing the uniform kernel may not exhaust the potential efficiency gains. For example, in the case of local polynomial density estimation \citep*{Cattaneo-Jansson-Ma_2020_JASA}, $R(u)$ is polynomial of order $p\geq1$ and the asymptotic variance of the density estimator $\hat{f}(\mathsf{x})=e_1'\hat{\theta}(\mathsf{x})$ takes the form $e_1'\Gamma^{-1}\Sigma\Gamma^{-1}e_1$ with
\[\Sigma = f(\mathsf{x})^3 \int_{-1}^1\int_{-1}^1 \min\{u,v\}R(u)R(v)'K(u)K(v)\diff u\diff v,\]
which implies that $\Gamma$ is no longer proportional to $\Sigma$ even when the kernel function is uniform. (To show this result, one first recognizes that the asymptotic variance of $\hat{f}(\mathsf{x})$ is $h^{-1}e_1'\Gamma^{-1}_{h}\Sigma_{h}\Gamma^{-1}_{h}e_1$, where the matrices are defined in Assumption \ref{ass:AN}. Then the expression reduces to $e_1'\Gamma^{-1}\Sigma\Gamma^{-1}e_1$ after taking the limit $h\to 0$, provided that $\mathsf{x}$ is an interior evaluation point. See the SA for omitted details.) This observation applies to the general case where the local basis function $R(\cdot)$ needs not to be of polynomial form, or when higher-order derivatives are of interest. See the SA for further discussion and detailed formulas.

In this section we employ a minimum distance approach to develop a lower bound on the asymptotic variance of the local regression distribution estimators, and also propose more efficient estimators based on the observation that their asymptotic variance is of the sandwich form $\Gamma^{-1}\Sigma\Gamma^{-1}$ but with $\Gamma$ not proportional to $\Sigma$ even when the uniform kernel is used. 

To motivate our approach, notice that in many cases it is possible to specify $R(\cdot)$ in such a way that $\theta$ can be partitioned as $\theta=(\theta_{1}',\theta_{2}')',$ where $\theta_{2}=0$. In such cases several distinct estimators of $\theta_{1}$ are available. To describe some leading candidates and their salient properties, partition $\hat{\theta}$, $\hat{\Gamma}$, $\hat{\Sigma}$, and $\hat{\Omega}$ conformable with $\theta$ as $\hat{\theta}=(\hat{\theta}_{1}',\hat{\theta}_{2}')'$ and
\[
\hat{\Gamma}=\left(
\begin{array}[c]{cc}
\hat{\Gamma}_{11} & \hat{\Gamma}_{12}\\
\hat{\Gamma}_{21} & \hat{\Gamma}_{22}
\end{array}\right)  ,\qquad\hat{\Sigma}=\left(
\begin{array}[c]{cc}
\hat{\Sigma}_{11} & \hat{\Sigma}_{12}\\
\hat{\Sigma}_{21} & \hat{\Sigma}_{22}
\end{array}\right)  ,\qquad\hat{\Omega}=\left(
\begin{array}[c]{cc}
\hat{\Omega}_{11} & \hat{\Omega}_{12}\\
\hat{\Omega}_{21} & \hat{\Omega}_{22}
\end{array}\right)  .
\]
The ``short'' regression counterpart of $\hat{\theta}_{1}$ obtained by dropping $R_{2}(\cdot)$ from $R(\cdot)=(R_{1}(\cdot)',R_{2}(\cdot)')'$ is given by
\[\hat{\theta}_{\mathtt{R},1}=\hat{\theta}_{1}+\hat{\Gamma}_{11}^{-1}\hat{\Gamma}_{12}\hat{\theta}_{2},\]
while an optimal minimum distance estimator of $\theta_{1}$ is given by
\begin{equation}\label{eq:minimum distance estimator}
\hat{\theta}_{\mathtt{MD},1}=\operatorname*{argmin}_{\theta_{1}}\binom{\hat{\theta}_{1}-\theta_{1}}{\hat{\theta}_{2}}'\hat{\Omega}^{-1}\binom{\hat{\theta}_{1}-\theta_{1}}{\hat{\theta}_{2}}
                            =\hat{\theta}_{1}-\hat{\Omega}_{12}\hat{\Omega}_{22}^{-1}\hat{\theta}_{2}.
\end{equation}
As a by-product of results obtained when establishing (\ref{AN, studentized estimator}), it follows that
\begin{align*}
\hat{\Omega}_{11}^{-1/2}(\hat{\theta}_{1}-\theta_{1})&\rightsquigarrow\mathcal{N}(0,I),\\
\hat{\Omega}_{\mathtt{R},11}^{-1/2}(\hat{\theta}_{\mathtt{R},1}-\theta_{1})&\rightsquigarrow\mathcal{N}(0,I),\qquad\hat{\Omega}_{\mathtt{R},11}
=\hat{\Gamma}_{11}^{-1}\hat{\Sigma}_{11}\hat{\Gamma}_{11}^{-1},\\
\text{and}\qquad\hat{\Omega}_{\mathtt{MD},11}^{-1/2}(\hat{\theta}_{\mathtt{MD},1}-\theta_{1})&\rightsquigarrow\mathcal{N}(0,I),\qquad\hat{\Omega}_{\mathtt{MD},11}
=\hat{\Omega}_{11}-\hat{\Omega}_{12}\hat{\Omega}_{22}^{-1}\hat{\Omega}_{21},
\end{align*}
under regularity conditions. Since $\hat{\Omega}$ is of ``sandwich'' form, the estimators $\hat{\theta}_{1}$ and $\hat{\theta}_{\mathtt{R},1}$ cannot be ranked in terms of (asymptotic) efficiency in general. On the other hand, $\hat{\theta}_{\mathtt{MD},1}$ will always be (weakly) superior to both $\hat{\theta}_{1}$ and $\hat{\theta}_{\mathtt{R},1}$. In fact, because
\[\hat{\theta}_{1}=\operatorname*{argmin}_{\theta_{1}}\binom{\hat{\theta}_{1}-\theta_{1}}{\hat{\theta}_{2}}'\left(\begin{array}
[c]{cc}
\hat{\Omega}_{11}^{-1} & 0\\
0 & \hat{\Omega}_{22}^{-1}
\end{array}
\right)  \binom{\hat{\theta}_{1}-\theta_{1}}{\hat{\theta}_{2}},
\]
and
\[\hat{\theta}_{\mathtt{R},1}=\operatorname*{argmin}_{\theta_{1}}\binom{\hat{\theta}_{1}-\theta_{1}}{\hat{\theta}_{2}}'\hat{\Gamma}\binom{\hat{\theta}_{1}-\theta_{1}}{\hat{\theta}_{2}},\]
each estimator admits a minimum distance interpretation, but only $\hat{\theta}_{\mathtt{MD},1}$ can be interpreted as an optimal minimum distance estimator based on $\hat{\theta}$. See \citet*{Newey-McFadden_1994_Handbook} for more discussion on minimum distance estimation.

As a consequence, we investigate whether an appropriately implemented $\hat{\theta}_{\mathtt{MD},1}$ can lead to asymptotic efficiency gains relative to $\hat{\theta}_{1}$ and $\hat{\theta}_{\mathtt{R},1}$. More generally, as a by-product, we obtain an efficiency bound among minimum distance estimators and show that this bound coincides with those known in the literature for kernel-based density estimation at interior points \citep*{Granovsky-Muller_1991_ISR,Cheng-Fan-Marron_1997_AoS}.

In the remaining of this section we focus on the case of local polynomial density estimation at an interior point for concreteness, but the SA presents more general results. Consequently, we assume that $F(\cdot)$ is $p$-times continuously differentiable in a neighborhood of $\mathsf{x}$. Then, (\ref{cdf approximation, interior point}) is satisfied and a natural choice of
$R(\cdot)$ is
\begin{equation}\label{eq:polynomial long regression basis}
R(u) = \Big(R_{1}(u)', R_{2}(u)'\Big)' = \Big(1, P(u)', Q(u)'\Big)',
\end{equation}
where $P(u)=(u,u^2/2,\cdots,u^p/p!)'$ is a polynomial basis, and $Q(\cdot)$ represent redundant regressors. Therefore, in our minimum distance construction, the parameters are
\begin{equation}\label{eq:polynomial long regression parameter}
\theta = \Big( \underbrace{F(\mathsf{x})}_{\text{intercept}},\ \underbrace{f(\mathsf{x}),\ \cdots,\ f^{(p-1)}(\mathsf{x})}_{\text{slope, }P(\cdot)},\ \underbrace{0,\ \cdots,\ 0}_{\text{redundant, }Q(\cdot)} \Big)',
\end{equation}
with smoothing error of order $\varrho(h)=o(h^{p})$. 

With \eqref{eq:polynomial long regression basis} and \eqref{eq:polynomial long regression parameter}, we define the minimum distance density estimator as $\hat{f}_{\mathtt{MD}}(\mathsf{x})=e_{1}'\hat{\theta}_{\mathtt{MD, 1}}$. Similarly, we have $\hat{f}(\mathsf{x})=e_{1}'\hat{\theta}_{1}$ and $\hat{f}_{\mathtt{R}}(\mathsf{x})=e_{1}'\hat{\theta}_{\mathtt{R}, 1}$. Of course, if it is known a priori that the distribution function is $p+q$ times continuously differentiable, then one can specify $Q(\cdot)$ to include higher order polynomials: $Q(u)=(u^{p+1}/(p+1)!,\cdots,u^{p+q}/(p+q)!)'$. By redefining the parameters as $\theta = ( F(\mathsf{x}),\ f(\mathsf{x}),\ \cdots,\ f^{(p+q-1)}(\mathsf{x}))'$, the smoothing error will be of order $\varrho(h)=o(h^{p+q})$. Notice that, in this case, $\hat{f}(\mathsf{x})$ and $\hat{f}_{\mathtt{R}}(\mathsf{x})$ correspond to the density estimator introduced in
\cite{Cattaneo-Jansson-Ma_2020_JASA} implemented with $R(u)=(1,u,\cdots,u^{p+q}/(p+q)!)'$ and $R(u)=(1,u,\cdots,u^{p}/p!)'$, respectively. Since the purpose of this section is to investigate the efficiency gains of incorporating additional redundant regressors, we do not exploit the extra smoothness condition, and we will treat $Q(\cdot)$ as redundant regressors even if $Q(\cdot)$ contains higher order polynomials. 

As both $\hat{f}(\mathsf{x})$ and $\hat{f}_{\mathtt{R}}(\mathsf{x})$ are (weakly) asymptotically inefficient relative to $\hat{f}_{\mathtt{MD}}(\mathsf{x})$ for any choice of $Q(\cdot)$, we consider the asymptotic variance of the minimum distance estimator, which can be obtained by establishing asymptotic counterparts of $\hat\Gamma$ and $\hat\Sigma$ after suitable scaling. Under regularity conditions (e.g., lack of perfect collinearity between $P$ and $Q$), the asymptotic variance of the minimum distance $\ell$-th derivative density estimator, $\hat{f}^{(\ell)}_{\mathtt{MD}}(\mathsf{x})=e_{\ell+1}'\hat{\theta}_{\mathtt{MD, 1}}$ with $0\leq\ell\leq p-1$, is
\[\mathsf{AsyVar}[\hat{f}^{(\ell)}_{\mathtt{MD}}(\mathsf{x})] = e_{\ell}'\left[\Omega_{PP}-\Omega_{PQ}\Omega_{QQ}^{-1}\Omega_{QP}\right]  e_{\ell},\]
where 
\[
\left(
\begin{array}
[c]{ccc}
\Omega_{11} & \Omega_{1P} & \Omega_{1Q}\\
\Omega_{P1} & \Omega_{PP} & \Omega_{PQ}\\
\Omega_{Q1} & \Omega_{QP} & \Omega_{QQ}
\end{array}
\right)=\Gamma^{-1}\Sigma\Gamma^{-1}.
\]

Therefore, the objective is to find a function $Q(\cdot)$ that minimizes the asymptotic variance $\mathsf{AsyVar}[\hat{f}^{(\ell)}_{\mathtt{MD}}(\mathsf{x})]$. Taking $Q(\cdot)$ scalar and properly orthogonalized, without loss of generality, we have $\int_{-1}^{1}P(u)K(u)\diff u=0$ and $\int_{-1}^{1}(1,P(u)')'Q(u)K(u)\diff u=0$. It follows that the problem of selecting an optimal $Q(\cdot)$ to minimize $\mathsf{AsyVar}[\hat{f}^{(\ell)}_{\mathtt{MD}}(\mathsf{x})]$ is equivalent to the following variational problem:
\begin{equation}
\sup_{Q\in\mathcal{Q}}\frac{\left[\int_{-1}^{1}\int_{-1}^{1}P_{\ell}(u)Q(v)\min\{u,v\}K(u)K(v)\diff u\diff v\right]^{2}}
                               {\int_{-1}^{1}\int_{-1}^{1}Q(u)Q(v)\min\{u,v\}K(u)K(v)\diff u\diff v}\label{eq:OptQ-1}
\end{equation}
where
\[\mathcal{Q} = \left\{Q(\cdot):\int_{-1}^{1}Q(u)K(u)\diff u=0,\quad \int_{-1}^{1}P(u)Q(u)K(u)\diff u=0\right\},\]
with $P_{\ell}(u)=e_{\ell}'\Big(\int_{-1}^{1}P(u)P(u)'K(u)\diff u\Big)^{-1}P(u)$ and $\ell=1,2,\dots,p-1$. The objective function is obtained from the fact that, after proper orthogonalization, the matrix $\Gamma$ becomes block diagonal. See the SA for all other omitted details.

The following theorem characterizes a lower bound for the asymptotic variance of the minimum distance density estimator among all possible choices of redundant regressors. 

\begin{theorem}[Efficiency: Local Polynomial Density Estimator at Interior Points]\label{thm: optimal kernel}Suppose the conditions of Theorem
\ref{thm:AN} hold. If $\mathsf{x}\in\mathcal{X}$ is an interior point, then
\[\inf_{Q\in\mathcal{Q}}\mathsf{AsyVar}[\hat{f}^{(\ell)}_{\mathtt{MD}}(\mathsf{x})] \geq \nu_\ell, \qquad 
  \nu_\ell = f(\mathsf{x})e_{\ell}'\left(\int_{-1}^{1}\dot{P}(u)\dot{P}(u)'\diff u\right)^{-1}e_{\ell}, \qquad
  0\leq\ell\leq p-1,\]
  where $\dot{P}(u)=(1,u,\cdots,u^{p-1}/(p-1)!)'$ is the derivative of $P(u)$. 
\end{theorem}

This theorem establishes a lower bound among minimum distance estimators. Importantly, it is shown in the SA that this bound coincides with the variance bound of all kernel-type density (and derivatives thereof) estimators employing the same order of the (induced) kernel function \citep{Granovsky-Muller_1991_ISR}. Therefore, our minimum distance approach sheds new light on minimum variance results for nonparametric kernel-based estimators of the density function and its derivatives.

This lower bound can be (approximately) achieved by setting the redundant regressor $Q(\cdot)$ to include a certain higher order polynomial function. By direct calculation for each $p=1,2,\dots,10$, it is also shown in the SA that $\lim_{j\to\infty}\mathsf{AsyVar}[\hat{f}^{(\ell)}_{\mathtt{MD},j}(\mathsf{x})]=\nu_\ell$,
where the minimum distance estimator $\hat{f}^{(\ell)}_{\mathtt{MD},j}(\mathsf{x})=e_{\ell}'\hat{\theta}_{\mathtt{MD},j}$ is constructed with
\[Q(u)=u^{2j+1} - P(u)'\left(\int_{-1}^1 P(u)P(u)'\diff u\right)^{-1}\int_{-1}^1 P(u)u^{2j+1}\diff u,\quad \text{for }\ell=0,\ 2,\ 4,\ \cdots,\]
or
\[Q(u)=u^{2j+2} - P(u)'\left(\int_{-1}^1 P(u)P(u)'\diff u\right)^{-1}\int_{-1}^1 P(u)u^{2j+2}\diff u,\quad \text{for }\ell=1,\ 3,\ 5,\ \cdots,\]
and $K(\cdot)$ being the uniform kernel. While we found that other kernel shapes can also be used, we chose the uniform kernel in this construction for three reasons. First, this choice is intuitive and coincides with the optimal choice in standard local polynomial regression settings. Second, when $p\geq3$ the other allowed kernel shapes overweight observations near the boundary of the kernel's support. Third, the uniform kernel makes the asymptotic variance calculation more tractable. See the SA for further details. 

The resulting recipe for implementation is simple: it proposes a specific choice of $Q(\cdot)$ so that the corresponding minimum distance estimator approximately achieves the variance bound for $j$ large enough. Interestingly, $Q(\cdot)$ is scalar and known, but the larger $j$ the closer the asymptotic variance of the minimum distance density estimator will be to the efficiency bound. We assume $Q(\cdot)$ is orthogonal to $P(\cdot)$ for theoretical convenience. To implement this estimator, one only needs to run a local polynomial regression of the empirical distribution function on a constant, the polynomial basis $P(\cdot)$, and one additional regressor, either $u^{2j+1}$ or $u^{2j+2}$ (depending on the choice of $\ell$), and then apply \eqref{eq:minimum distance estimator} with the corresponding estimated variance-covariance matrix.

\begin{figure}[!tb]
\centering
\resizebox{0.6\textwidth}{!}{\includegraphics{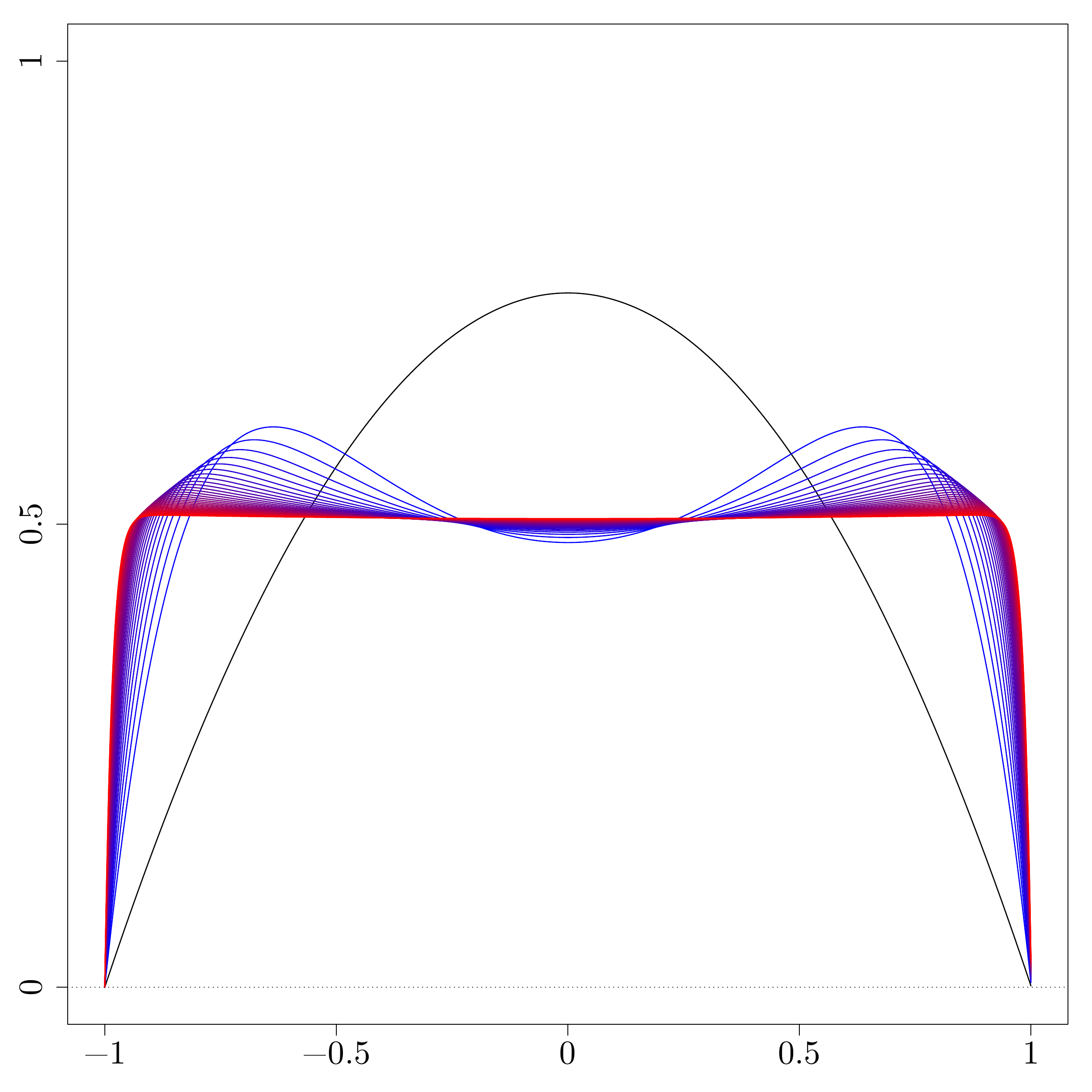}}
\caption{Equivalent Kernels of the Minimum Distance Density Estimators.\label{fig:equiv kernels}}
\begin{flushleft}
\footnotesize\textit{Notes}. We set $P(u)=u$ or $P(u)=(u,u^2/2)'$, and $K$ uniform. The redundant regressor is $Q(u)=u^{2j+1}$ for $j=1,2,\cdots,30$. The initial equivalent kernel is quadratic (black solid line), and the minimum variance kernel is uniform (red solid line).
\end{flushleft}
\end{figure}

In Figure \ref{fig:equiv kernels}, we consider the local linear/quadratic density estimator ($\ell=0$) with the redundant regressor being a higher order polynomial (i.e., $P(u)=u$ or $P(u)=(u,u^2/2)'$, and $Q(u)=u^{2j+1}$), and plot the corresponding equivalent kernel of our minimum distance density estimator for $j=1,2,\dots,30$. As $j$ increases, the equivalent kernel converges to the uniform kernel, which is well-known to minimize the (asymptotic) variance among all density estimators employing second order kernels \citep{Granovsky-Muller_1991_ISR}. The asymptotic variance of our proposed minimum distance density estimator converges to the optimal asymptotic variance as $j\to\infty$.

Finally, in this paper we focus on minimizing the asymptotic variance of the estimator $\hat{\theta}$ and its variants because our main goal is inference. However, our results could be modified and extended to optimize the asymptotic mean square error (MSE). We do not pursue point estimation optimality further for brevity, but we do note that in the case of local polynomial density estimation \citep*{Cattaneo-Jansson-Ma_2020_JASA}, the resulting estimator is automatically MSE-optimal at interior points when $p\leq2$, because the induced equivalent kernel coincides with the Epanechnikov kernel \citep*{Granovsky-Muller_1991_ISR,Cheng-Fan-Marron_1997_AoS}.

\section{Uniform Distribution Theory}\label{section:Uniform Distribution Theory}

The distributional result presented in Theorem \ref{thm:AN} is valid pointwise for $\mathsf{x}\in\mathcal{X}$. We now develop an uniform distributional approximation for the Studentized process
\[\left\{T(\mathsf{x})=\frac{c'\hat{\theta}(\mathsf{x})-c'\theta(\mathsf{x})}{\sqrt{c'\hat{\Omega}(\mathsf{x})c}}\ :\ \mathsf{x}\in\mathcal{I}\right\},\]
using the notation in \eqref{AN, studentized estimator}, where $c$ is a conformable vector and $\mathcal{I}\subseteq \mathcal{X}$ is some prespecified region. This stochastic process is not asymptotically tight, and hence does not converge in distribution. Our approximation proceeds in two steps. First, for an positive (vanishing) sequence, $r_{\mathtt{L},n}$, we establish a uniform ``linearization'' of the process $T(\cdot)$ of the form:
\begin{equation}\label{eq:UnifLin}
\sup_{\mathsf{x}\in\mathcal{I}}\left\vert T(\mathsf{x})-\mathfrak{T}(\mathsf{x})\right\vert = O_\mathbb{P}(r_{\mathtt{L},n}),
\end{equation}
where
\[\left\{\mathfrak{T}(\mathsf{x}) = \frac{1}{\sqrt{n}}\sum_{i=1}^n \mathcal{K}_{h,\mathsf{x}}(x_i) \ :\ \mathsf{x}\in\mathcal{I}\right\}\] with
\[\mathcal{K}_{h,\mathsf{x}}(x_i)
= \frac{c'\Upsilon_{h}\Gamma_{h,\mathsf{x}}^{-1}  \int_{\mathcal{X}_h} R(u)\Big[\I(x_i\leq \mathsf{x} + hu) - F(\mathsf{x} + hu)\Big] K\left(u\right) f(\mathsf{x}+hu)\mathrm{d} u}
{\sqrt{c'\Upsilon_{h}\Omega_{h,\mathsf{x}}\Upsilon_{h}c}},\]
and $\Omega_{h,\mathsf{x}}=\Gamma_{h,\mathsf{x}}^{-1}\Sigma_{h,\mathsf{x}}\Gamma_{h,\mathsf{x}}^{-1}$. In words, we show that the Studentized process $T(\cdot)$, which involves various pre-asymptotic estimated quantities, is uniformly close to the linearized process $\mathfrak{T}(\cdot)$, which is a sample average of independent observations. To obtain \eqref{eq:UnifLin}, we develop new uniform approximations with precise convergence rates, which may be of independent interest in semiparametric estimation and inference settings. See the SA for more details.

Second, in a possibly enlarged probability space, we show that there exists a copy of $\mathfrak{T}(\cdot)$, denoted by $\tilde{\mathfrak{T}}(\cdot)$, and a centered Gaussian process $\{\mathfrak{B}(\mathsf{x}):\mathsf{x}\in\mathcal{I}\}$, with a suitable variance-covariance structure, such that
\begin{equation}
\sup_{\mathsf{x}\in\mathcal{I}}\left\vert \tilde{\mathfrak{T}}(\mathsf{x})-\mathfrak{B}(\mathsf{x})\right\vert = O_\mathbb{P}(r_{\mathtt{G},n}),\label{eq:SA}
\end{equation}
where $r_{\mathtt{G},n}$ is another positive (vanishing) sequence. This type of strong approximation result, when established with suitably fast rate $r_{\mathtt{G},n}\to0$, can be used to deduce distributional approximations for statistics such as $\sup_{\mathsf{x}\in\mathcal{I}}|T(\mathsf{x})|$, which are useful for constructing confidence bands or for conducting hypothesis tests about shape or other restrictions on the function of interest. To obtain (\ref{eq:SA}), we employ a result established by \cite{Rio_1994_PTRF}, and later extended in \cite{Gine-Koltchinskii-Sakhanenko_2004_PTRF}; see also \cite[Proof of their Proposition 5]{Gine-Nickl_2010_AoS}.

In this section we consider a fixed linear combination $c$ for ease of exposition, but in the SA we discuss the more general case where $c$ can depend on both the evaluation point $\mathsf{x}$ and the tuning parameter $h$, which is necessary to establish uniform distribution approximations for the minimum distance estimator introduced in Section \ref{subsection:Efficiency}. All the results reported in this section apply to the latter class of estimators as well.

\subsection{Assumptions}\label{subsection:Assumptions Uniform}

In addition to Assumption \ref{ass:AN}, we impose the following conditions on the data generating process. In the sequel, continuity and differentiability conditions at boundary points should be interpreted as one-sided statements (i.e., as in part (i) of Assumption \ref{ass:AN}).

\begin{Assumption} \label{ass:SA}
	Let $\mathcal{I}\subseteq\mathcal{X}$ be a compact interval.

(i) The density function $f(\mathsf{x})$ is twice continuously differentiable and bounded away from zero on $\mathcal{I}$. 

(ii) There exists some $\delta>0$ and compactly supported kernel functions $K^{\dag}(\cdot)$ and $\{K^{\ddag,d}(\cdot) \}_{d\leq \delta}$, such that (ii.1) $\sup_{u\in\mathbb{R}}| K^{\dag}(u) |+ \sup_{d\leq \delta,u\in\mathbb{R}} | K^{\ddag,d}(u) |<\infty$, (ii.2) the support of $K^{\ddag,d}(\cdot)$ has Lebesgue measure bounded by $Cd$, where $C$ is independent of $d$; and (ii.3) for all $u$ and $v$ such that $|u-v|\leq \delta$, 
\[|K(u)-K(v)| \leq |u-v|\cdot K^{\dag}(u)  + K^{\ddag,|u-v|}(u).\]

(iii) The basis function $R(\cdot)$ is Lipschitz continuous in $[-1,1]$. 

(iv) For all $h$ sufficiently small, the minimum eigenvalues of $\Gamma_{h,\mathsf{x}}$ and $h^{-1}\Sigma_{h,\mathsf{x}}$ are bounded away from zero uniformly for $\mathsf{x}\in\mathcal{I}$. 
\end{Assumption}

The above strengthens and expands Assumption \ref{ass:AN}. Part (i) requires the density function to be reasonably smooth uniformly in $\mathcal{I}$. Part (ii) imposes additional requirements on the kernel function. Although seemingly technical, it permits a decomposition of the difference $|K(u)-K(v)|$ into two parts. The first part, $|u-v|\cdot K^{\dag}(u)$, is a kernel function which vanishes uniformly as $|u-v|$ becomes small. Note that this will be the case for all piecewise smooth kernel functions, such as the triangular or the Epanechnikov kernel. However, difference of discontinuous kernels, such as the uniform kernel, cannot be made uniformly close to zero. This motivates the second term in the above decomposition. Part (iii) requires the basis function to be reasonably smooth. Together, parts (i)--(iii) imply that the estimator $\hat\theta(\mathsf{x})$ will be ``smooth'' in $\mathsf{x}$, which is important to control the complexity of the process $T(\cdot)$. Finally, part (iv) implies that the matrices $\Gamma_{h,\mathsf{x}}$ and $\Sigma_{h,\mathsf{x}}$ are well-behaved uniformly for $\mathsf{x}\in \mathcal{I}$. 

\subsection{Strong Approximation}\label{subsection:Strong Approximation}

We first discuss the covariance of the process $\mathfrak{T}(\cdot)$. It is straightforward to show that \[\mathbb{C}\mathrm{ov}[\mathfrak{T}(\mathsf{x}),\mathfrak{T}(\mathsf{y})]= \frac{c'\Upsilon_{h}\Omega_{h,\mathsf{x},\mathsf{y}}\Upsilon_{h}c }{ \sqrt{c'\Upsilon_{h}\Omega_{h,\mathsf{x}}\Upsilon_{h}c}\sqrt{c'\Upsilon_{h}\Omega_{h,\mathsf{y}}\Upsilon_{h}c}}, \qquad \Omega_{h,\mathsf{x},\mathsf{y}} = \Gamma_{h,\mathsf{x}}^{-1}\Sigma_{h,\mathsf{x},\mathsf{y}}\Gamma_{h,\mathsf{y}}^{-1},\]
where
\begin{align*}
\Sigma_{h,\mathsf{x},\mathsf{y}}
= \int_{\mathcal{X}_{h,\mathsf{y}}}\int_{\mathcal{X}_{h,\mathsf{x}}} & R(u)R(v)'  \Big[ F(\min\{\mathsf{x}+hu,\mathsf{y}+hv\}) - F(\mathsf{x}+hu)F(\mathsf{y}+hv) \Big]\\
&\cdot K(u)K(v)f(\mathsf{x}+hu)f(\mathsf{y}+hv)\diff u\diff v,
\end{align*}
and $\Sigma_{h,\mathsf{x},\mathsf{x}}=\Sigma_{h,\mathsf{x}}$.

Now we state the second main distributional result of this paper in the following theorem. 

\begin{theorem}[Strong Approximation]\label{thm:SA-KS distance}
	Suppose Assumptions \ref{ass:AN} and \ref{ass:SA} hold, and that $h\to 0$ and $nh^2/\log(n)\to\infty$.
	
	\begin{enumerate}
		\item \eqref{eq:UnifLin} holds with 
		\[r_{\mathtt{L},n} = \sqrt{\frac{n}{h}}\sup_{\mathsf{x}\in\mathcal{I}} \varrho(h,\mathsf{x}) + \frac{\log(n)}{\sqrt{nh^2}}.\]
		
		\item On a possibly enlarged probability space, there exists a copy $\tilde{\mathfrak{T}}(\cdot)$ of $\mathfrak{T}(\cdot)$, and a centered Gaussian process, $\{\mathfrak{B}(\mathsf{x}),\mathsf{x}\in\mathcal{I}\}$, defined with the same covariance as $\mathfrak{T}(\cdot)$, such that \eqref{eq:SA} holds with 
		\[r_{\mathtt{G},n} = \frac{\log(n)}{\sqrt{nh}}.\]
	\end{enumerate}
\end{theorem}

The first part of this theorem gives conditions such that the feasible Studentized process $T(\cdot)$ is well approximated by the infeasible (linear) process $\mathfrak{T}(\cdot)$, uniformly for $\mathsf{x}\in\mathcal{I}$. The latter process is mean zero, and takes a kernel-based form. However, standard strong approximation results for kernel-type estimators do not apply directly to the process $\mathfrak{T}(\cdot)$, as the implied (equivalent, Studentized) kernel $\mathcal{K}_{h,\mathsf{x}}(\cdot)$ depends not only on the bandwidth but also on the evaluation point in a non-standard way. That is, due to the boundary adaptive feature of the local regression distribution estimators, the shape of the implied kernel automatically changes for different evaluation points depending on whether they are interior or boundary points.

Putting the two results together, it follows that the distribution of $T(\cdot)$ is approximated by that of $\mathfrak{B}(\cdot)$, provided the following condition holds:
\begin{equation*}
\sqrt{\frac{n}{h}}\sup_{\mathsf{x}\in\mathcal{I}}\varrho(h,\mathsf{x}) + \frac{\log(n)}{\sqrt{nh^2}} \to0.
\end{equation*}
To facilitate understanding of this rate restriction, we consider the local polynomial density estimation setting of \cite{Cattaneo-Jansson-Ma_2020_JASA}, where the basis function takes the form $R(u)=(1,u,u^2/2,\cdots,u^p/p!)'$ for some $p\geq 1$, and the second element of $\hat\theta(\mathsf{x})$ estimates the density $f(\mathsf{x})$. That is, $e_1'\hat\theta(\mathsf{x}) = \hat{f}(\mathsf{x})\to_\mathbb{P} f(\mathsf{x})$ under Assumption \ref{ass:AN}, where $c=e_1$. By a Taylor expansion argument, it is easy to see that the smoothing bias has order $h^{p+1}$ as long as the distribution function $F(\cdot)$ is suitably smooth. Then, the above rate restriction reduces to $\sqrt{nh^{2p+1}} + {\frac{\log(n)}{\sqrt{nh^2}}} \to 0$.

Finally, if the goal is to approximate the distribution of $\sup_{\mathsf{x}\in\mathcal{I}}|T(\mathsf{x})|$, then an extra $\sqrt{\log(n)}$ factor is needed in the rate restriction, as discussed in \cite{Chernozhukov-Chetverikov-Kato_2014b_AoS}. A formal statement of such result is given below, after we discuss how we can further approximate the infeasible Gaussian process $\mathfrak{B}(\cdot)$. 

\subsection{Confidence Bands}\label{subsection:Confidence Bands}

Feasible inference cannot be based on the Gaussian process $\mathfrak{B}(\cdot)$, as its covariance structure is unknown and has to be estimated in practice. For estimation, first recall from Sections \ref{section:Setup} and \ref{section:Pointwise Distribution Theory} that $W_i(\mathsf{x}) = K((x_i-\mathsf{x})/h)/h$, $R_i(\mathsf{x})=R(x_i-\mathsf{x})$, and $\hat{\Gamma}(\mathsf{x})=\frac{1}{n}\sum_{i=1}^nW_{i}(\mathsf{x})R_{i}(\mathsf{x})R_{i}(\mathsf{x})'$. Then, we construct the plug-in estimator of $\Omega_{h,\mathsf{x},\mathsf{y}}$ as follows:
\[\hat\Omega_{h,\mathsf{x},\mathsf{y}} = n\Upsilon_{h}^{-1} \hat{\Gamma}(\mathsf{x})^{-1}\hat{\Sigma}(\mathsf{x},\mathsf{y})\hat{\Gamma}(\mathsf{y})^{-1}\Upsilon_{h}^{-1},\qquad
\hat{\Sigma}(\mathsf{x},\mathsf{y})=\frac{1}{n^2}\sum_{i=1}^n\hat{\psi}_{i}(\mathsf{x})\hat{\psi}_{i}(\mathsf{y})'
\]
where
\[
\hat{\psi}_{i}(\mathsf{x})=\frac{1}{n}\sum_{j=1}^nW_{j}(\mathsf{x})R_{j}(\mathsf{x})(\I(x_{i}\leq x_{j}) - \hat{F}_j).
\]

The following theorem combines previous results, and justifies the uniform confidence band constructed using critical values from $\sup_{\mathsf{x}\in\mathcal{I}}|\hat{\mathfrak{B}}(\mathsf{x})|$. Let $X_n=(x_1,x_2,\dots,x_n)'$.

\begin{theorem}[Kolmogorov-Smirnov Distance]\label{thm:confidence band}
Suppose Assumptions \ref{ass:AN} and \ref{ass:SA} hold, and that $n\sup_{\mathsf{x}\in\mathcal{I}}\varrho(h,\mathsf{x})^2\log(n)/h + \log(n)^5 / (nh^2)\to 0$.
Then, conditional on $X_n$, there exists a centered Gaussian process $\{ \hat{\mathfrak{B}}(\mathsf{x}), \mathsf{x}\in\mathcal{I} \}$ with covariance
\[
\mathbb{C}\mathrm{ov}\left[\left.\hat{\mathfrak{B}}(\mathsf{x}),\hat{\mathfrak{B}}(\mathsf{y})\right|X_n \right] =  \frac{c'\Upsilon_{h}\hat{\Omega}_{h,\mathsf{x},\mathsf{y}}\Upsilon_{h}c}{\sqrt{c'\Upsilon_{h}\hat{\Omega}_{h,\mathsf{x}}\Upsilon_{h}c}\sqrt{c'\Upsilon_{h}\hat{\Omega}_{h,\mathsf{y}}\Upsilon_{h}c}},
\]
such that
\[\sup_{u\in\mathbb{R}}\left|\mathbb{P}\Big[ \sup_{\mathsf{x}\in\mathcal{I}}|T(\mathsf{x})| \leq u \Big] 
  - \mathbb{P}\Big[  \sup_{\mathsf{x}\in\mathcal{I}}|\hat{\mathfrak{B}}(\mathsf{x})| \leq u\Big| X_n \Big]\right| = o_{\mathbb{P}}(1).
\]
\end{theorem}

From Theorem \ref{thm:confidence band}, an asymptotically valid $100(1-\alpha)\%$ confidence band for $\{c'{\theta}(\mathsf{x}):\mathsf{x}\in\mathcal{I}\}$ is given by
\[\text{CI}_{\alpha}(\mathcal{I})=\left\{\left[c'\hat{\theta}(\mathsf{x})-\mathfrak{q}_{1-\alpha}\sqrt{c'\hat{\Omega}(\mathsf{x})c}
                                               ~,~ c'\hat{\theta}(\mathsf{x})+\mathfrak{q}_{1-\alpha}\sqrt{c'\hat{\Omega}(\mathsf{x})c}\right],\quad\mathsf{x}\in\mathcal{I}\right\},
\]
where $\mathfrak{q}_{1-\alpha}$ is the $1-\alpha$ quantile of $\sup_{\mathsf{x}\in\mathcal{I}}|\hat{\mathfrak{B}}(\mathsf{x})|$, conditional on the data. That is,
\[\mathfrak{q}_{a}=\inf\left\{u\in\mathbb{R}:\mathbb{P}\left[\sup_{\mathsf{x}\in\mathcal{I}}|\hat{\mathfrak{B}}(\mathsf{x})| \leq u\Big| X_n \right]  \geq a \right\},\]
which can be obtained by simulating the process $\hat{\mathfrak{B}}(\cdot)$ on a dense grid.

As an alternative to analytic estimation of the covariance kernel, it is possible to consider resampling methods as in \citet*{Chernozhukov-Chetverikov-Kato_2014b_AoS}, \citet*{Cheng-Chen_2019_EJS}, \citet*{Cattaneo-Farrell-Feng_2020_AoS}, and references therein. We relegate resampling-based inference for future research.

\section{Extensions and Other Applications}\label{section:Extensions}

We briefly outline some extensions of our main results. First, we introduce a re-weighted version of $\hat{\theta}$, which is useful in applications as illustrated in Section \ref{section:Applications}. Second, we discuss a new class of local regression estimators based on a non-random least-squares loss function, which has some interesting theoretical properties and may be of interest in some semiparametric settings. Finally, we discuss how to incorporate restrictions in the estimation procedure, employing the generic structure of the local basis $R(\cdot)$.

\subsection{Re-weighted Distribution Estimator}\label{subsection:Reweighted Distribution}

Suppose $(x_{1},w_{1}),(x_{2},w_{2}),\cdots,(x_{n},w_{n})$ is a random sample, where $x_{i}$ is a continuous random variable with a smooth cumulative distribution function, but now $w_{i}$ is an additional ``weighting'' variable, possibly random and involving unknown parameters. We consider the generic weighted distribution parameter
\[H(\mathsf{x})=\mathbb{E}[w_{i}\I(x_{i}\leq\mathsf{x})],\]
whose practical interpretation depends on the specific choice of $w_{i}$.

We discuss some examples. If $w_{i}=1$, $H(\cdot)$ becomes the distribution function $F(\cdot)$, and hence the results above apply. If $w_{i}$ is set to be a certain ratio of propensity scores for subpopulation membership, then the derivative $\diff H(\mathsf{x})/\diff \mathsf{x}$ becomes a counterfactual density function, as in \cite{DiNardo-Fortin-Lemieux_1996_ECMA}; see Section \ref{subsection:DiNardo-Fortin-Lemieux} below. If $w_{i}$ is set to be a combination of the treatment assignment and treatment status variables, then the resulting derivative can be used to conduct specification testing in IV models, or if $w_{i}$ is set to be a certain ratio of propensity scores for a binary instrument, then the derivative can be used to identify distributions of compliers, as in \cite{Imbens-Rubin_1997_RESTUD}, \cite{Abadie_2003_JOE}, and \cite{Kitagawa_2015_ECMA}; see Section \ref{subsection:IV Specification and Heterogeneity} below. Other examples of applicability of this extension include bunching, missing data, measurement error, data combination, and treatment effect settings.

More generally, when weights are allowed for, there is another potentially interesting connection between the estimand $\diff H(\mathsf{x})/\diff \mathsf{x}$ and classical weighted averages featuring prominently in econometrics because $\diff H(\mathsf{x})/\diff \mathsf{x}=\mathbb{E}[w_{i}|x_{i}=\mathsf{x}]f(\mathsf{x})$, which is useful in the context of partial means and related problems as in \cite{Newey_1994_ET}.

Our main results extend immediately to allow for $\sqrt{n}$-consistent estimated weights $w_{i}$ or, more generally, to estimated weights that converge sufficiently fast. Specifically, we let $\hat{F}_{w,i}(x)=\frac{1}{n}\sum_{j=1}^{n}w_{j}\I(x_{j}\leq x)$ in place of $\hat{F}_i$, and investigate the large sample properties of our proposed estimator in (\ref{Local regression estimator}) when $w_{i}$ is replaced by $\hat{w_{i}}=w_{i}(\hat{\beta})$ with $\hat{\beta}$ an $a_n$-consistent estimator, for some $a_n\to\infty$, and $w_{i}(\cdot)$ a known function of the data. That is, when estimated weights are used to construct the weighted empirical distribution function $\hat{F}_{w,i}(x)$. Provided that $a_n^{-1}\to 0$ sufficiently fast, this extra estimation step will not affect the asymptotic properties of our estimator of the density function or its  derivatives (which will be true, for example, in parametric estimation cases,  where $a_n=\sqrt{n}$ under regularity conditions). All the results reported in the previous sections apply to this extension, which we illustrate empirically below.

\subsection{Local $L^2$ Distribution Estimators}\label{subsection:Local Projection Distribution Estimators}

The local regression distribution estimator is obtained from a least squares projection of the empirical distribution function onto a local basis, where the projection puts equal weights at all observations. That is, \eqref{Local regression estimator} employs an $L^2(\hat{F})$-projection
\[
\hat{\theta}(\mathsf{x})=\operatorname*{argmin}_{\theta}\int\Big(\hat
{F}(u)-R(u-\mathsf{x})'\theta\Big)^{2}K\left(  \frac{u-\mathsf{x}}{h}\right)
\diff\hat{F}(u).
\]
This representation motivates a general class of local $L^2$ distribution estimators given by
\[
\hat{\theta}_{G}(\mathsf{x})=\operatorname*{argmin}_{\theta}\int\Big(\hat
{F}(u)-R(u-\mathsf{x})'\theta\Big)^{2}K\left(  \frac{u-\mathsf{x}}{h}\right)
\diff G(u)
\]
for some measure $G$. We show in the SA that all our theoretical results continue to hold for $\hat{\theta}_{G}$, provided that $G$ is absolutely continuous with respect to the Lebesgue measure and the Radon-Nikodym derivative is reasonably smooth. (Note that $G$ does not need to be a proper distribution function.)

The estimator $\hat\theta_{G}$ involves only one average, while the local regression estimator $\hat\theta$ has two layers of averages (one from the construction of the empirical distribution function, and the other from the $L^2(\hat{F})$-projection/regression). As a result, with suitable centering and scaling, the local $L^2$ distribution estimator, $\hat\theta_G$, can be written as the sum of a mean-zero influence function and a smoothing bias term. Since $\hat\theta_G$ no longer involves a second order U-statistic (c.f. \eqref{eq:U-stat representation}), or a leave-in bias, pointwise asymptotic normality can be established under weaker conditions: it is no longer needed to assume $nh^2\to \infty$ (Theorem \ref{thm:AN}), and $nh\to\infty$ will suffice. Similarly, for the strong approximation results we only need to restrict $\log(n)/\sqrt{nh}$ as opposed to $\log(n)/\sqrt{nh^2}$ (part 1 of Theorem \ref{thm:SA-KS distance}).

In addition, the local $L^2$ distribution estimator $\hat{\theta}_G$ is robust to ``low'' density. To see this, recall that the local regression estimator $\hat\theta$ involves regressing the empirical distribution on a local basis, which means that this estimator can be numerically unstable if there are only a few observations near the evaluation point. More precisely, the matrix $\hat\Gamma$ will be close to singular if the effective sample size is small.

Although the local $L^2$ distribution estimator $\hat{\theta}_G$ takes a simpler form, is robust to low density, and its large sample properties can be established under weaker bandwidth conditions, it does have one drawback: it requires knowledge of the support $\mathcal{X}$. To be more precise, let $G$ be the Lebesgue measure, then the local $L^2$ distribution estimator may be biased at or near boundaries of $\mathcal{X}$ if it is compact. In contrast, the local regression distribution estimator is fully boundary adaptive, even in cases where the location of the boundary is unknown. See \citet*{Cattaneo-Jansson-Ma_2020_JASA} for further discussion for the case of density estimation. 

\subsection{Incorporating Restrictions}\label{subsection:otherbasis}

The formulation (\ref{Local regression estimator}) is general enough to allow for some interesting extensions in the definition of the local regression distribution estimator. The key observation is that the estimator has a weighted least squares representation with a generic local basis function $R(\cdot)$, which allows for deploying well-know results from linear regression models. We briefly illustrate this idea with three examples.

First, consider the case where the local basis $R(u)$ incorporates specific restrictions, such as continuity or lack thereof, on the distribution function, density function or higher-order derivatives at the evaluation point $\mathsf{x}$. To give a concrete example, suppose that $F(\mathsf{x})$ and $f(\mathsf{x})$ are known to be continuous at some interior point $\mathsf{x}\in\mathcal{X}$, while no information is available for the higher-order derivatives. Then, these restriction can be effortlessly incorporated to the local regression distribution estimator by considering the local basis function
\[R(u)=\left(1, u, \frac{u^2}{2}\I(u< \mathsf{x}), \frac{u^2}{2}\I(u\geq \mathsf{x}), \frac{u^3}{6}\I(u< \mathsf{x}), \frac{u^3}{6}\I(u\geq \mathsf{x}), \cdots, \frac{u^p}{p!}\I(u< \mathsf{x}), \frac{u^p}{p!}\I(u\geq \mathsf{x})\right)'.\]
It follows that $\hat{f}(\mathsf{x})=e_1'\hat{\theta}(\mathsf{x})$ consistently estimates the density $f(\mathsf{x})$ at the kink point $\mathsf{x}$, while $e_2'\hat{\theta}(\mathsf{x})$ and $e_3'\hat{\theta}(\mathsf{x})$ are consistent estimators of the left and the right derivatives of the density function, respectively (and similarly for other higher-order one-sided derivatives). In this example, the generalized formulation not only reduces the bias of $\hat{f}(\mathsf{x})=e_1'\hat{\theta}(\mathsf{x})$ even in the absence of continuity of higher-order derivatives, but also provides the basis for testing procedures for continuity of higher-order derivatives; e.g., by considering an statistic based on $(e_2-e_3)'\hat{\theta}(\mathsf{x})$. This provides a concrete illustration of the advantages of allowing for generic local basis. A distinct example was developed in \citet*{Cattaneo-Jansson-Ma_2018_Stata,Cattaneo-Jansson-Ma_2020_JASA} for density discontinuity testing in regression discontinuity designs.

As a second example, consider imposing shape constraints, such as positivity or monotonicity, in the construction of the local regression distribution estimator. Such constraints amount to specific restrictions on the parameter space of $\theta$, which naturally leads to restricted weighted least squares estimation in the context of our estimator. To be concrete, consider constructing a local polynomial density estimator which is non-negative, in which case $R(u)$ is a polynomial basis of order $p\geq1$ and (\ref{Local regression estimator}) is extended to:
\begin{align*}
\hat{\theta}(\mathsf{x}) = &\operatorname*{argmin}_{\theta}\sum_{i=1}^n W_{i}(\hat{F}_{i}-R_{i}'\theta)^{2}\qquad\text{subject to } T\theta \geq 0,
\end{align*}
where $T$ denotes a matrix of restrictions; in this example, $T=e_1'$ to ensure that $\hat{f}(\mathsf{x})=e_1'\hat{\theta}(\mathsf{x})\geq0$. This example showcases the advantages of the weighted least squares formulation of our estimator. Local monotonicity constraints, for instance, could also be easily incorporated in a similar fashion.

The final example of extensions of our basic local regression distribution estimation approach concerns non-identity link functions, leading to a non-linear least squares formulation. Specifically, (\ref{Local regression estimator}) can be generalized to $\hat{\theta}(\mathsf{x}) = \operatorname*{argmin}_{\theta}\sum_{i=1}^n W_{i}(\hat{F}_{i}-\Lambda(R_{i}'\theta))^{2}$ for some known link function $\Lambda(\cdot)$. For instance, such extension may be useful to model distributions with large support or to impose specific local shape constraints.

All of the examples above, as well as many others, can be analyzed using the large sample results developed in this paper and proper extensions thereof. We plan to investigate these and other extensions in future research.

\section{Applications}\label{section:Applications}

We discuss two applications of our main results in the context of program evaluation \citep[see][and references therein]{Abadie-Cattaneo_2018_ARE}.

\subsection{Counterfactual Densities}\label{subsection:DiNardo-Fortin-Lemieux}

In this first application, the objects of interest are density functions over their entire support, including boundaries and near-boundary regions, which are estimated using estimated weighting schemes, as this is a key feature needed for counterfactual analysis (and many other applications). Our general estimation strategy is specialized to the counterfactual density approach originally proposed by \cite{DiNardo-Fortin-Lemieux_1996_ECMA}. We focus on density estimation, and we refer readers to \cite{Chernozhukov-FernandezVal-Melly_2013_ECMA} for related methods based on distribution functions as well as for an overview of the literature on counterfactual analysis.

To construct a counterfactual density or, more generally, re-weighted density estimators, we simply need to set the weights $(w_{1},w_{2},\cdots,w_{n})$ appropriately. In most applications, this also requires constructing preliminary consistent estimators of these weights, as we illustrate in this section. Suppose the observed data is $(x_{i},t_{i},{z}_{i}')'$, $i=1,2,\dots, n$, where $x_{i}$ continues to be the main outcome variable, ${z}_{i}$ collects other covariates, and $t_{i}$ is a binary variable indicating to which group unit $i$ belongs. For concreteness, we call these two groups control and treatment, though our discussion does not need to bear any causal interpretation.

The marginal distribution of the outcome variable $x_{i}$ for the full sample can be easily estimated without weights (that is, $w_{i}=1$). In addition, two conditional densities, one for each group, can be estimated using $w^{1}_{i} = t_{i}/\mathbb{P}[t_{i}=1]$ for the treatment group and $w^{0}_{i} = (1 -t_{i})/\mathbb{P}[t_{i}=0]$ for the control group, and are denote by $\hat{f}_{1}(x)$ and $\hat{f}_{0}(x)$, respectively. For example, in the context of randomized controlled trials, these density estimators can be useful to depict the distribution of the outcome variables for control and treatment units.

A more challenging question is: what would the outcome distribution have been, had the treated units had the same covariates distribution as the control units? The resulting density is called the counterfactual density for the treated, which is denoted by $f_{1\rhd0}(x)$. Knowledge about this distribution is important for understanding differences between $f_{1}(x)$ and $f_{0}(x)$, as the outcome distribution is affected by both group status and covariates distribution. Furthermore, the counterfactual distribution has another useful interpretation: Assume the outcome variable is generated from potential outcomes, $x_{i} = t_{i}x_{i}(1) + (1-t_{i})x_{i}(0)$, then under unconfoundedness, that is, assuming $t_{i}$ is independent of $(x_{i}(0),x_{i}(1))'$ conditional on the covariates ${z}_{i}$, $f_{1\rhd0}(x)$ is the counterfactual distribution for the control group: it is the density function associated with the distribution of $x_{i}(1)$ conditional on $t_{i}=0$.

Regardless of the interpretation taken, $f_{1\rhd0 }(x)$ is of interest and can be estimated using our generic density estimator $\hat{f}(x)$ with the following weights:
\[w^{1\rhd0}_{i} = t_{i}\cdot\frac{\mathbb{P}[t_{i}=0|{z}_{i}]}{\mathbb{P}[t_{i}=1|{z}_{i}]} \frac{\mathbb{P}[t_{i}=1]}{\mathbb{P}[t_{i}=0]}.\]
In practice, this choice of weighting scheme is unknown because the conditional probability $\mathbb{P}[t_{i}=1|{z}_{i}]$, a.k.a. the propensity score, is not observed. Thus, researchers estimate this quantity using a flexible parametric model, such as Probit or Logit. Our technical results allow for these estimated weights to form counterfactual density estimators after replacing the theoretical weights by their estimated counterparts, provided the estimated weights converge sufficiently fast to their population counterparts.

To be more precise, we can model $\mathbb{P}[t_{i}=1|{z}_{i}] = G(b(z_i)'\beta)$ for some known link function $G(\cdot)$, such as Logit or Probit, and $K$-dimensional basis expansion $b(z_i)$, such as power series or B-splines. If the model is correctly specified for some fixed $K$ and basis function $b(\cdot)$, then $\max_{1\leq i \leq n} |w_i - \hat{w_{i}}|=O_{\mathbb{P}}(a^{-1}_n)$ with $a_n=\sqrt{n}$ under mild regularity conditions, and all our results carry over to the setting with estimated weights mentioned in Section \ref{subsection:Reweighted Distribution}. Alternatively, from a nonparametric perspective, if $K\to\infty$ as $n\to\infty$, and for appropriate basis function $b(\cdot)$ and regularity conditions, $\max_{1\leq i \leq n} |w_i - \hat{w_{i}}|=O_{\mathbb{P}}(a^{-1}_n)$ with $a_n$ depending on both $K$ and $n$. Then, as in the parametric case, our main results carry over if $a_n^{-1}\to0$ fast enough. The exact rate requirements can be deduced from the main theorems above.

\subsubsection{Empirical Illustration}\label{subsubsection:Empirical Illustration DiNardo-Fortin-Lemieux}

We demonstrate empirically how marginal, conditional, and counterfactual densities can be estimated with our proposed method. We consider the effect of education on earnings using a subsample of the data in
\cite{Abadie-Angrist-Imbens_2002_ECMA}. The data consists of individuals who did not enroll in the Job Training Partnership Act (JTPA). The main outcome variable is the sum of earnings in a 30-month period, and individuals are split into two groups according to their education attainment: $t_{i}=1$ for those with high school degree or GED, and $t_{i}=0$ otherwise. Also available are demographic characteristics, including gender, ethnicity, age, marital status, AFDC receipt (for women), and a dummy indicating whether the individual worked at least 12 weeks during a one-year period. The sample size is $5,447$, with $3,927$ being either high school graduates or GED. Summary statistics are available as the fourth column in Table \ref{table:jtpa}. We leave further details on the JTPA program to Section \ref{subsection:IV Specification and Heterogeneity}, where we utilize a larger sample and conduct distribution estimation in a randomized controlled (intention-to-treat) and instrumental variables (imperfect compliance) setting.

\begin{table}[!tb]
\centering
\renewcommand{\arraystretch}{1.2}
\caption{Summary Statistics for the JTPA data.\label
{table:jtpa}}\vspace{-.1in}
\resizebox{.7\textwidth}{!}{
\begin{tabular}{lrcrrcrr}
\hline\hline
\multicolumn{1}{l}{\bfseries }&\multicolumn{1}{c}{\bfseries Full}&\multicolumn{1}{c}{\bfseries }&\multicolumn{2}{c}{\bfseries JTPA Offer}&\multicolumn{1}{c}{\bfseries }&\multicolumn{2}{c}{\bfseries JTPA Enrollment}\tabularnewline
 \cline{4-5} \cline{7-8}
\multicolumn{1}{l}{}&\multicolumn{1}{c}{}&\multicolumn{1}{c}{}&\multicolumn{1}{c}{N}&\multicolumn{1}{c}{Y}&\multicolumn{1}{c}{}&\multicolumn{1}{c}{N}&\multicolumn{1}{c}{Y}\tabularnewline
\hline
{\bfseries Income}&$17949.20$&&$17191.13$&$18321.59$&&$17015.58$&$19098.44$\tabularnewline

{\bfseries HS or GED}&$    0.72$&&$    0.71$&$    0.72$&&$    0.70$&$    0.74$\tabularnewline

{\bfseries Male}&$    0.46$&&$    0.47$&$    0.46$&&$    0.48$&$    0.45$\tabularnewline

{\bfseries Nonwhite}&$    0.36$&&$    0.36$&$    0.36$&&$    0.36$&$    0.37$\tabularnewline

{\bfseries Married}&$    0.28$&&$    0.27$&$    0.29$&&$    0.27$&$    0.29$\tabularnewline

{\bfseries Work $\leq$ 12}&$    0.44$&&$    0.43$&$    0.44$&&$    0.44$&$    0.44$\tabularnewline

{\bfseries AFDC}&$    0.17$&&$    0.17$&$    0.17$&&$    0.16$&$    0.19$\tabularnewline

{\bfseries Age}&&&&&&&\tabularnewline
~~22-25&$    0.24$&&$    0.25$&$    0.24$&&$    0.24$&$    0.25$\tabularnewline
~~26-29&$    0.21$&&$    0.20$&$    0.21$&&$    0.21$&$    0.21$\tabularnewline
~~30-35&$    0.24$&&$    0.25$&$    0.24$&&$    0.24$&$    0.25$\tabularnewline
~~36-44&$    0.19$&&$    0.19$&$    0.19$&&$    0.20$&$    0.19$\tabularnewline
~~45-54&$    0.08$&&$    0.08$&$    0.08$&&$    0.08$&$    0.07$\tabularnewline
\hline
{ Sample Size}&$ 9872$&&$ 3252$&$ 6620$&&$ 5447$&$ 4425$\tabularnewline
\hline
\end{tabular}
}
\begin{flushleft}
\footnotesize\textit{Columns}
: (i) Full: full sample; (ii) JTPA Offer: whether offered JTPA services; (iii) JTPA Enrollment: whether enrolled in JTPA.
\textit{Rows}
: (i) Income: cumulative income over 30-month period post random selection; (ii) HS or GED: whether has high school degree or GED; (iii) Male: gender being male; (iv) Nonwhite: black or Hispanic; (v) Married: whether married; (vi) Work $\leq
$ 12: worked less than 12 weeks during one year period prior to random assignment; (vii) Age: age groups.
\end{flushleft}
\end{table}

\begin{figure}[!tb]
\centering
\subfloat[Marginal Distributions]{\resizebox{0.49\textwidth}{!}{\includegraphics{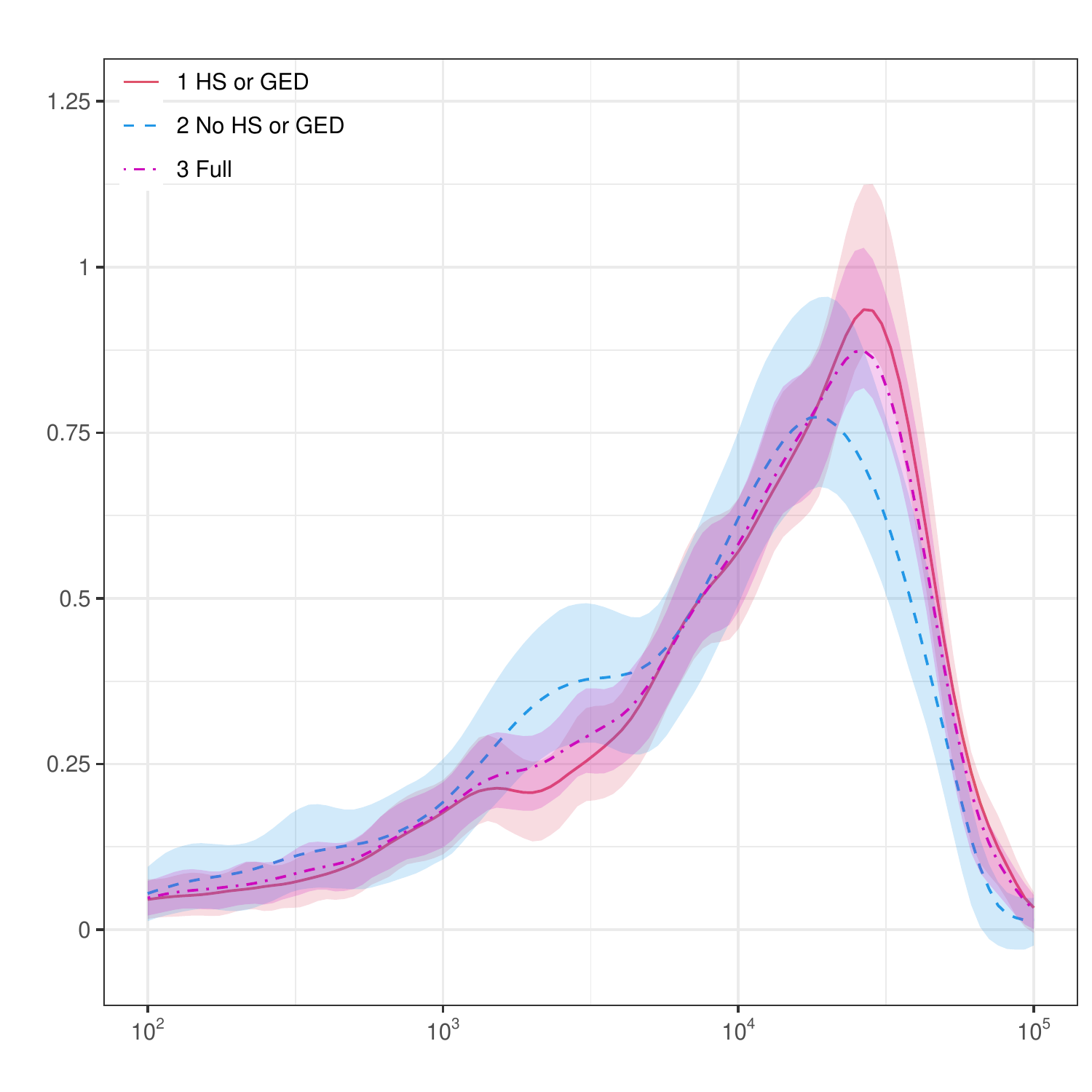}}}
\subfloat[Counterfactual Distribution]{\resizebox{0.49\textwidth}{!}{\includegraphics{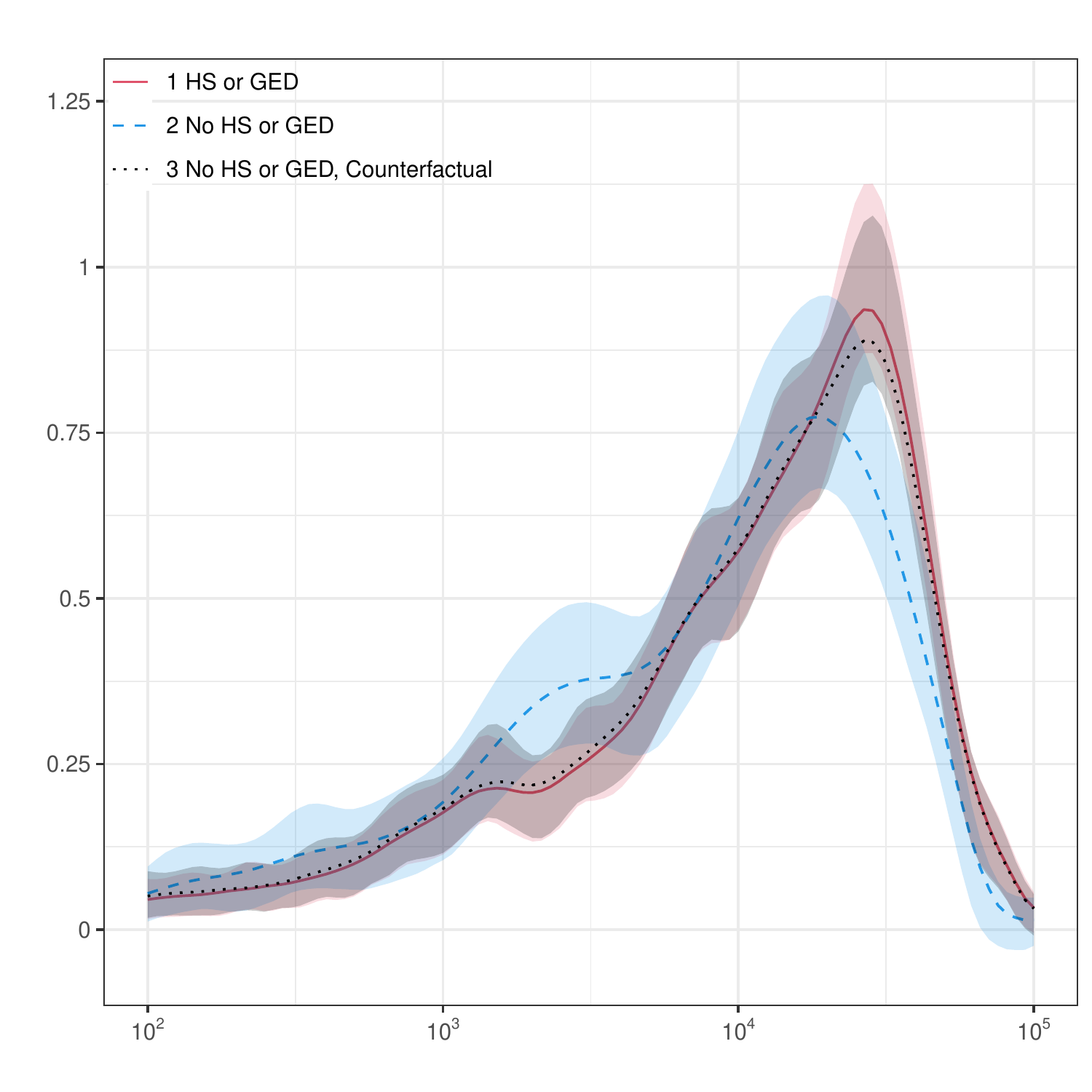}}}
\caption{Earning Distributions by Education, JTPA.\label{fig:jtpa edu}}
\begin{flushleft}
\footnotesize\textit{Notes}: (i) Full: earning distribution for the full sample ($n=5,447$); (ii) HS or GED: earning distributions for subgroups with and without high school degree or GED ($n=1,520$ and $3,927$, respectively); (iii) No HS or GED, Counterfactual: counterfactual earning distribution. Point estimates are obtained by using local polynomial regression with order 2, and robust confidence intervals are obtained with local polynomial of order 3. Bandwidths are chosen by minimizing integrated mean squared errors. All estimates are obtained using companion \texttt{R} (and \texttt{Stata}) package described in \cite{Cattaneo-Jansson-Ma_2021_JSS}.
\end{flushleft}
\end{figure}

It is well-known that education has significant impact on labor income, and we first plot earning distributions separately for subsamples with and without high school degree or GED. The two estimates, $\hat{f}_{1}(x)$ and $\hat{f}_{0}(x)$, are plotted in panel (a) of Figure \ref{fig:jtpa edu}. There, it is apparent that the earning distribution for high school graduates is very different compared to those without high school degree. More specifically, both the mean and median of $\hat{f}_{1}(x)$ are higher than $\hat{f}_{0}(x)$, and $\hat{f}_{1}(x)$ seems to have much thinner left tail and thicker right tail.

As mentioned earlier, direct comparison between $\hat{f}_{1}(x)$ and $\hat{f}_{0}(x)$ does not reveal the impact of having high school degree on earning, since the difference is confounded by the fact that individuals with high school degree can have very different characteristics (measured by covariates) compared to those without. We employ covariates adjustments, and ask the following question: what would the earning distribution have been for high school graduates, had they had the same characteristics as those without such degree?

We estimate the counterfactual distribution $f_{1\rhd0}(x)$ by our proposed method, and is shown in panel (b) of Figure \ref{fig:jtpa edu}. The difference between $\hat{f}_{1\rhd0}(x)$ and $\hat{f}_{1}(x)$ is not very profound, although it seems $\hat{f}_{1\rhd0}(x)$ has smaller mean and median. On the other hand, difference between $\hat{f}_{0}(x)$ and $\hat{f}_{1\rhd0}(x)$ remains highly nontrivial. Our empirical finding is compatible with existing literature on return to education: it is generally believed that education leads to significant accumulation of human capital, hence increase in labor income. As a result, educational attainment is usually one of the most important ``explanatory variables'' for differences in income.

\subsection{IV Specification and Heterogeneity}\label{subsection:IV Specification and Heterogeneity}

Self-selection and treatment effect heterogeneity are important concerns in causal inference and studies of socioeconomic programs. It is now well understood that classical treatment parameters, such as the average treatment effect or the treatment effect on the treated, are not identifiable even when treatment assignment is fully randomized due to imperfect compliance. Indeed, what can be recovered is either an intention-to-treat parameter or, using the instrumental variables method, some other more local treatment effect, specific to a subpopulation: the ``compliers.'' See \cite{Imbens-Rubin_2015_Book} and references therein for further discussion. Practically, this poses two issues for empirical work employing instrumental variables methods focusing on local average treatment effects. First, since compliers are usually not identified, it is crucial to understand how different their characteristics are compared to the population as a whole. Second, it is often desirable to have a thorough estimate of the distribution of potential outcomes, which provides information not only on the mean or median, but also its dispersion, overall shape, or local curvatures.

Motivated by these observations, and to illustrate the applicability of our density estimation methods, we now consider two related problems. First, we investigate specification testing in the context of local average treatment effects based on comparison of two (rescaled) densities as discussed by \cite{Kitagawa_2015_ECMA}. This method requires estimating two densities nonparametrically. Second, we consider estimating the density of potential outcomes for compliers in the IV setting of \cite{Abadie_2003_JOE}, which allows for conditioning on covariates. The resulting density plots not only provide visual guides on treatment effects, but also can be used for further analysis to construct a rich set of summary statistics or as inputs for semiparametric procedures. Both methods require estimated weights.

We first introduce the notation and the potential outcomes framework. For each individual there is a binary indicator of treatment assignment (a.k.a. the instrument), denoted by $d_{i}$. The actual treatment (takeup), however, can be different, due to imperfect compliance. More specifically, let $t_{i}(0) $ and $t_{i}(1)$ be the two potential treatments, corresponding to $d_{i}=0$ and $1$, then the observed binary treatment indicator is $t_{i} = d_{i}t_{i}(1) + (1-d_{i})t_{i}(0)$. We also have a pair of potential outcomes, $x_{i}(0)$ and $x_{i}(1)$, associated with $t_{i}=0$ and $1$, and what is observed is $x_{i} = t_{i}x_{i}(1) + (1-t_{i})x_{i}(0)$. Finally, also available are some covariates, collected in ${z}_{i}$. We assume that the observed data is a random sample $\{ (x_{i},t_{i}, d_{i},{z}_{i}')':1\leq i\leq n \}$.

There are three important assumptions for identification. First, the instrument has to be exogenous, meaning that conditional on covariates, it is independent of the potential treatments and outcomes. Second, the instrument has to be relevant, meaning that conditional on covariates, the instrument should be able to induce changes in treatment takeups. Third, there are no defiers (a.k.a. the monotonicity assumption). We do not reproduce the exact details of those assumptions and other technical requirements for identification; see the references given for more details.

Building on \cite{Imbens-Rubin_1997_RESTUD}, \citet{Kitagawa_2015_ECMA} discusses interesting testable implications in this IV setting, which can be easily adapted to test instrument validity using our density estimator. In the current context, the testable implications take the following form: for any (measurable) set $\mathcal{I}\subset\mathbb{R}$,
\begin{align*}
                 & \mathbb{P}[x_{i}\in \mathcal{I},\ t_{i}=1|d_{i}=1]\geq\mathbb{P}[x_{i}\in \mathcal{I},\ t_{i}=1|d_{i}=0],\\
\text{and}\qquad & \mathbb{P}[x_{i}\in \mathcal{I},\ t_{i}=0|d_{i}=0]\geq \mathbb{P}[x_{i}\in \mathcal{I},\ t_{i}=0|d_{i}=1].
\end{align*}
The first requirement holds trivially in the JTPA context, since the program does not allow enrollment without being offered (that is, $\mathbb{P}[t_{i}=1|d_{i}=0]=0$). Therefore we demonstrate the second with our density estimator. Let $f_{d=0,t=0}(x)$ be the earning density for the subsample $d_{i}=0$ and $t_{i}=0$, that is, for individuals without JTPA offer and not enrolled. Similarly let $f_{d=1,t=0}(x)$ be the earning density for individuals offered JTPA but not enrolled. Then the second inequality in the above display is equivalent to, for all $x\in\mathbb{R}$,
\[\mathbb{P}[t_{i}=0|d_{i}=0]\cdot f_{d=0,t=0}(x)\geq\mathbb{P}[t_{i}=0|d_{i}=1]\cdot f_{d=1,t=0}(x).\]
Thus, our density estimator can be used directly, where $f_{d=0,t=0}(x)$ is consistently estimated with weights $w_{i}^{d=0,t=0}=(1-d_{i})(1-t_{i})/\mathbb{P}[d_{i}=0,t_{i}=0]$, and $f_{d=1,t=0}(x)$ is consistently estimated with $w_{i}^{d=1,t=0}=d_{i}(1-t_{i})/\mathbb{P}[d_{i}=1,t_{i}=0]$. 

\cite{Abadie_2003_JOE} showed that the distributional characteristics of compliers are identified, and can be expressed as re-weighted marginal quantities. We focus on three distributional parameters here. The first one is the distribution of the observed outcome variable, $x_{i}$, for compliers, which is denoted by $f_{c}$. This parameter is important for understanding the overall characteristics of compliers, and how different it is from the populations. The other two parameters are distributions of the potential outcomes, $x_{i}(0)$ and $x_{i}(1)$, for compliers, since the difference thereof reveals the effect of treatment for this subgroup. They are denoted by $f_{c,0}$ and $f_{c,1}$, respectively. The three density functions can also be estimated using our proposed local polynomial density estimator $\hat{f}(x)$ with, respectively, the following weights:
\begin{align*}
w^{c}_{i}   & = \frac{1}{\mathbb{P}[t_{i}(1)>t_{i}(0)]}\cdot\left( 1 - \frac{t_{i}(1-d_{i})}{\mathbb{P}[d_{i}=0|{z}_{i}]} - \frac{(1-t_{i})d_{i}}{\mathbb{P}[d_{i}=1|{z}_{i}]}\right),\\
w^{c,0}_{i} & = \frac{1}{\mathbb{P}[t_{i}(1)>t_{i}(0)]}\cdot(1-t_{i})\cdot\frac{1 - d_{i} - \mathbb{P}[d_{i}=0|{z}_{i}]}{\mathbb{P}[d_{i}=0|{z}_{i}]\mathbb{P}[d_{i}=1|{z}_{i}]},\\
w^{c,1}_{i} & = \frac{1}{\mathbb{P}[t_{i}(1)>t_{i}(0)]}\cdot t_{i}\cdot\frac{d_{i} - \mathbb{P}[d_{i}=1|{z}_{i}]}{\mathbb{P}[d_{i}=0|{z}_{i}]\mathbb{P}[d_{i}=1|{z}_{i}]}.
\end{align*}
Here, the weights need to be estimated in practice, unless precise knowledge about the treatment assignment mechanism is available. As mentioned previously, our results allow for estimated weights such as those obtained by fitting a flexible Logit or Probit model to approximate the propensity score $\mathbb{P}[d_{i}=1|{z}_{i}]$ so long as they converge sufficiently fast to their population counterparts. 

\subsubsection{Empirical Illustration}\label{subsubsection:Empirical Illustration IV Specification and Heterogeneity}

The JTPA is a large publicly funded job training program targeting at individuals who are economically disadvantaged and/or facing significant barriers to employment. Individuals were randomly offered JTPA training, the treatment take-up, however, was only about 67\% among those who were offered. Therefore the JTPA offer provides valid instrument to study the impact of the job training program. We continue to use the same data as \cite{Abadie-Angrist-Imbens_2002_ECMA}, who analyzed quantile treatment effects on earning distributions.

Besides the main outcome variable and covariates already introduced in Section \ref{subsection:DiNardo-Fortin-Lemieux}, also available are the treatment take-up (JTPA enrollment) and the instrument (JTPA Offer). See Table \ref{table:jtpa} for summary statistics for the full sample and separately for subgroups. As the JTPA offers were randomly assigned, it is possible to estimate the intent-to-treat effect by mean comparison. Indeed, individuals who are offered JTPA services earned, on average, $\$1,130$ more than those not offered. On the other hand, due to imperfect compliance, it is in general not possible to estimate the effect of job training (i.e. the effect of JTPA enrollment), unless one is willing to impose strong assumptions such as constant treatment effect.

\begin{figure}[!tb]
\centering
\resizebox{0.60\textwidth}{!}{\includegraphics{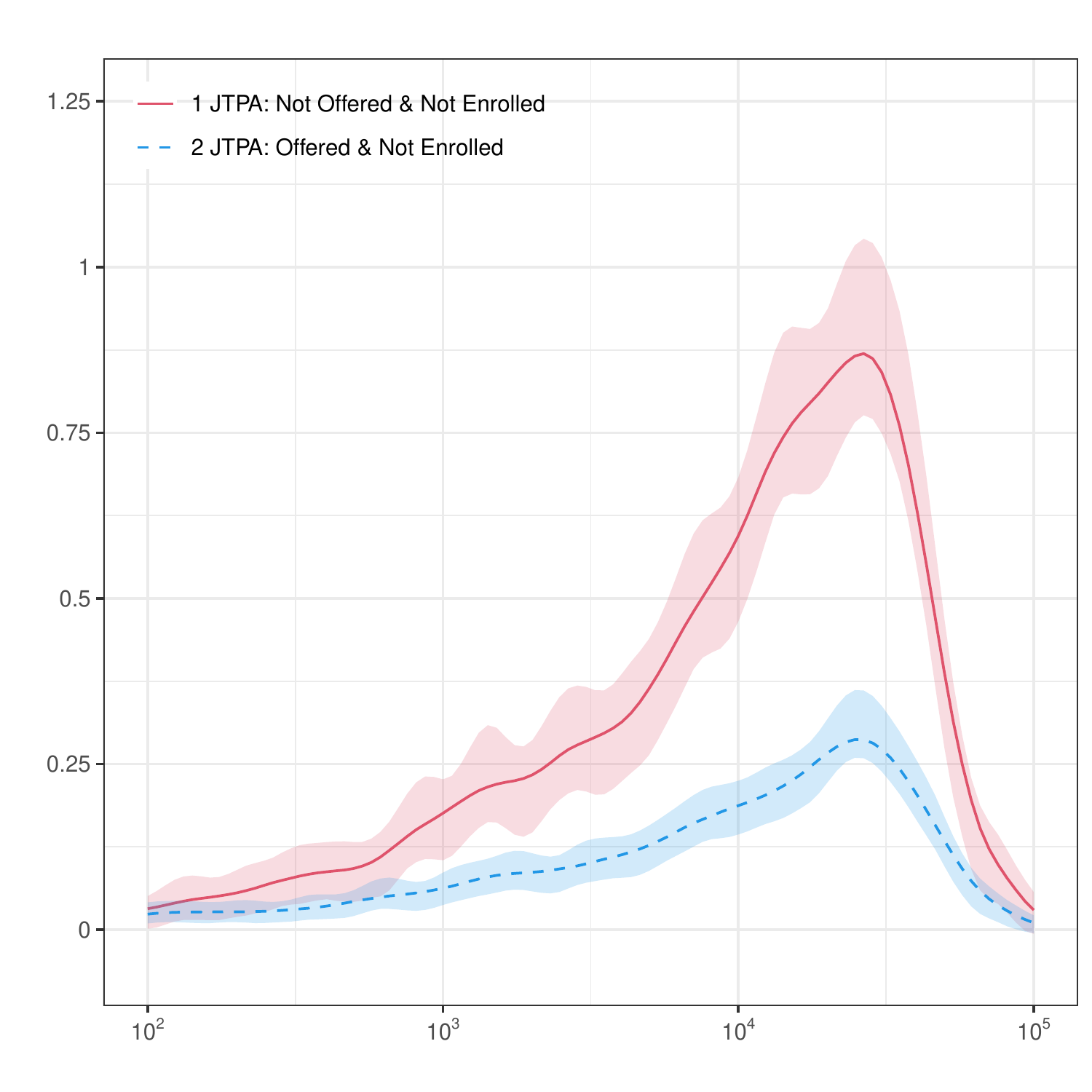}}
\caption{Testing Validity of Instruments, JTPA.\label{fig:jtpa IV validity}}
\begin{flushleft}
\footnotesize\textit{Notes}
: (i) JTPA: Not Offered \& Not Enrolled: the scaled density estimate $\frac{\sum_i \I(t_i=0,d_i=0)}{\sum_i \I(d_i=0)} \hat{f}_{d=0,t=0}(x)$; (ii) JTPA: Offered \& Not Enrolled: the scaled density estimate $\frac{\sum_i \I(t_i=0,d_i=1)}{\sum_i \I(d_i=1)} \hat{f}_{d=1,t=0}(x)$. Point estimates are obtained by using local polynomial regression with order 2, and robust confidence bands are obtained with local polynomial of order 3. Bandwidths are chosen by minimizing integrated mean squared errors. All estimates are obtained using companion \texttt{R} (and \texttt{Stata}) package described in \cite{Cattaneo-Jansson-Ma_2021_JSS}.
\end{flushleft}
\end{figure}

We first implement the IV specification test, which is straightforward using our density estimator $\hat{f}(x)$. We plot the two estimated (rescaled) densities in Figure \ref{fig:jtpa IV validity}. A simple eyeball test suggests no evidence against instrumental variable validity. A formal hypothesis test, justified using our theoretical results, confirms this finding.

\begin{figure}[!tb]
\centering
\subfloat[Marginal Distributions]{\resizebox{0.49\textwidth}{!}{\includegraphics{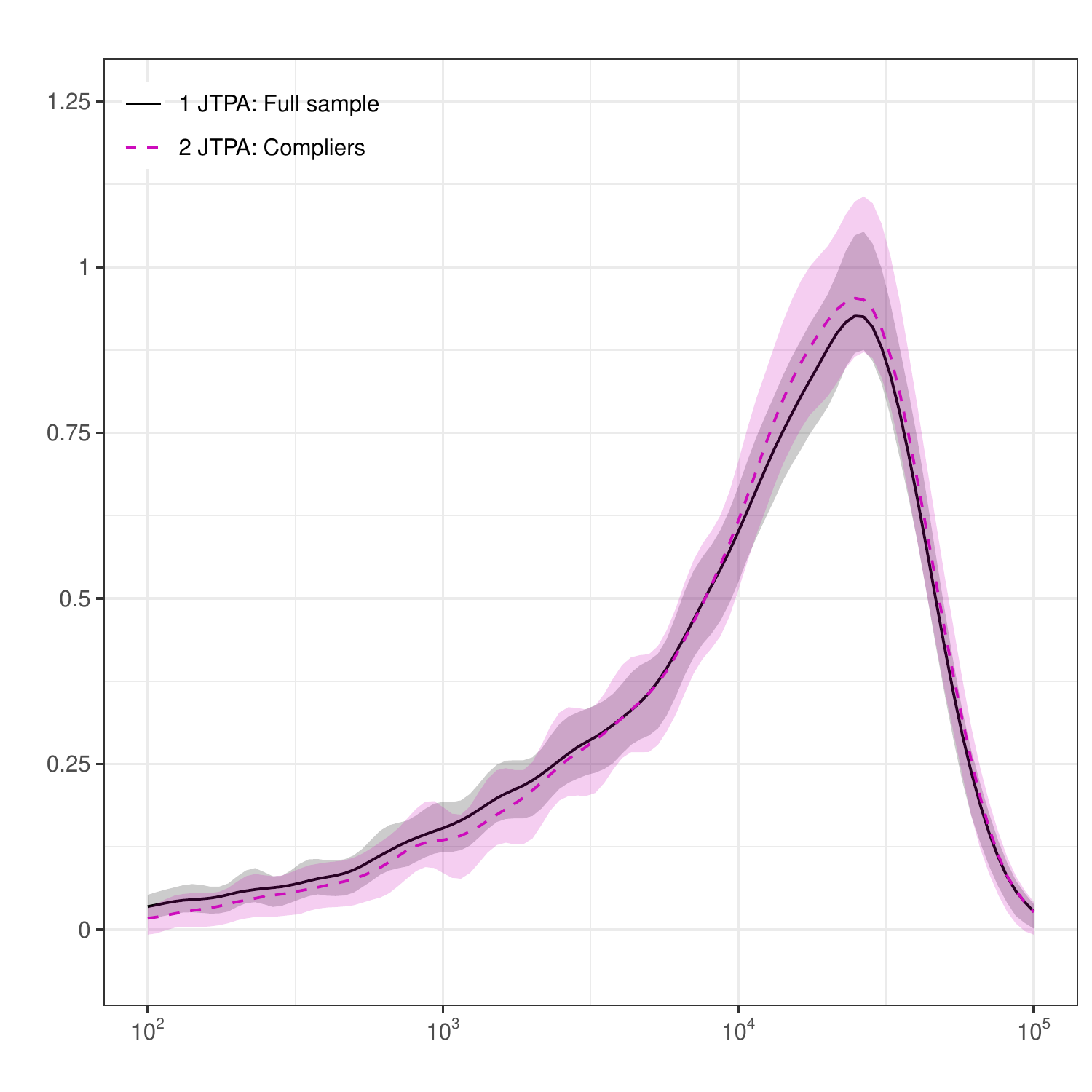}}}
\subfloat[Marginal Distributions]{\resizebox{0.49\textwidth}{!}{\includegraphics{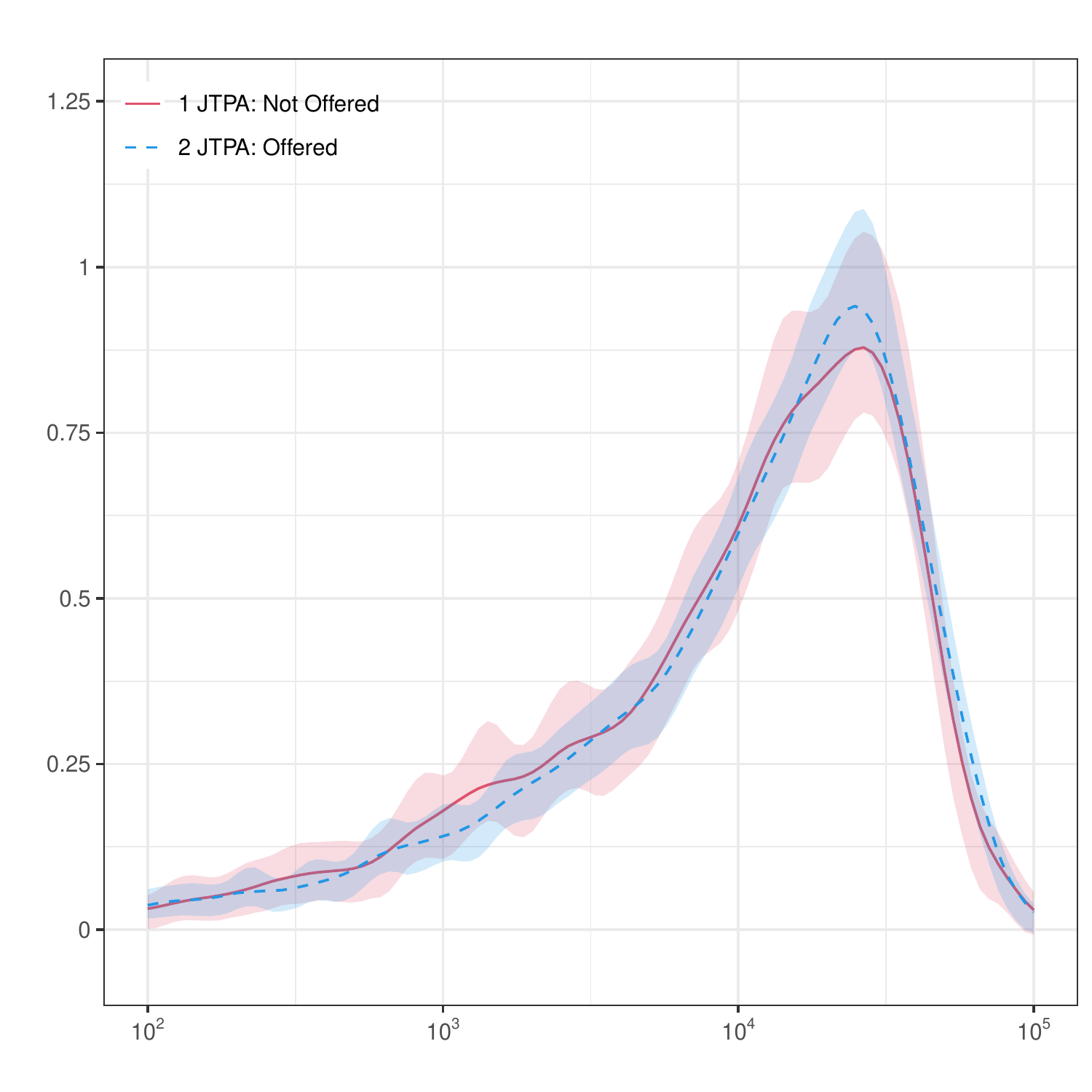}}}\\
\subfloat[Marginal Distributions]{\resizebox{0.49\textwidth}{!}{\includegraphics{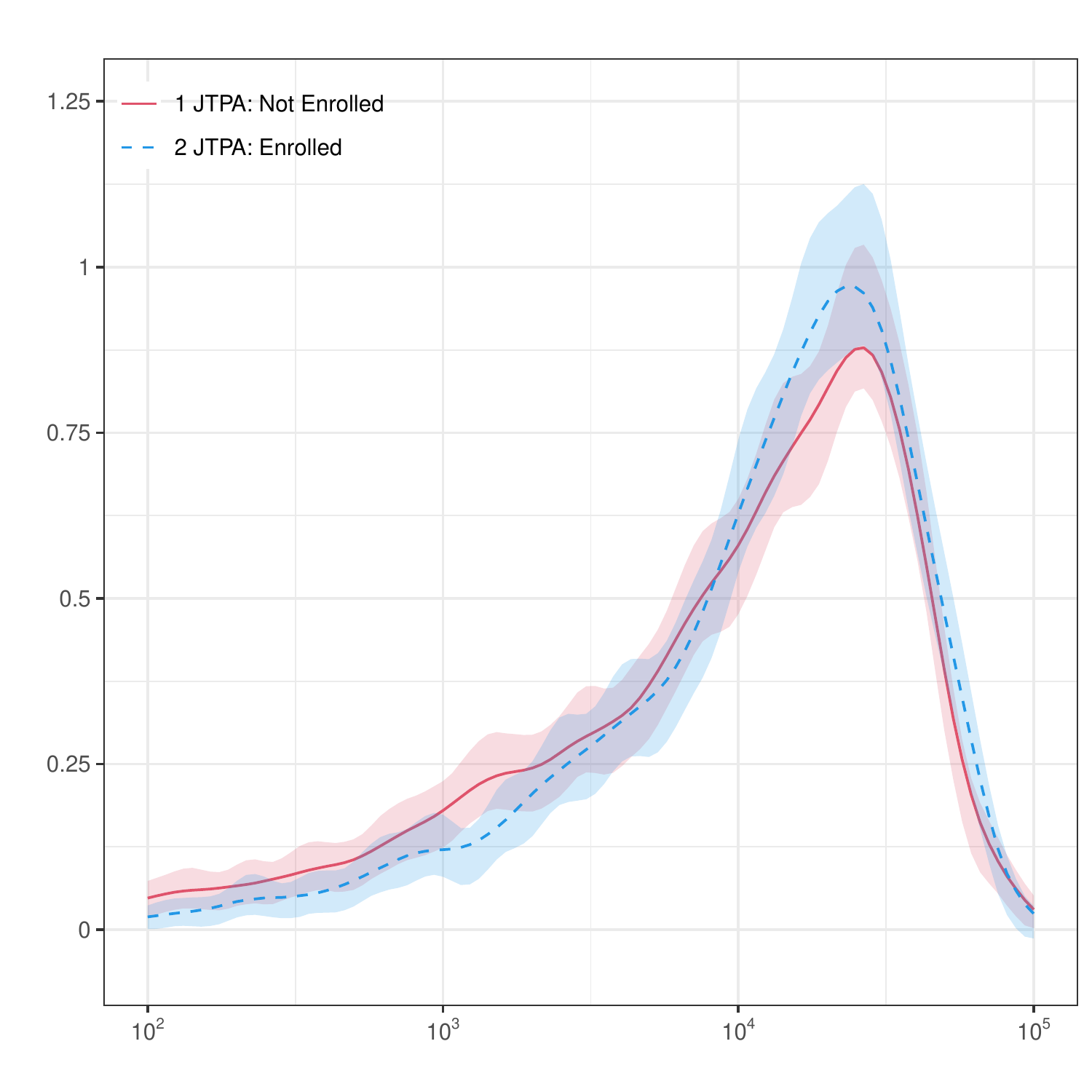}}}
\subfloat[Potential Outcome Distributions]{\resizebox{0.49\textwidth}{!}{\includegraphics{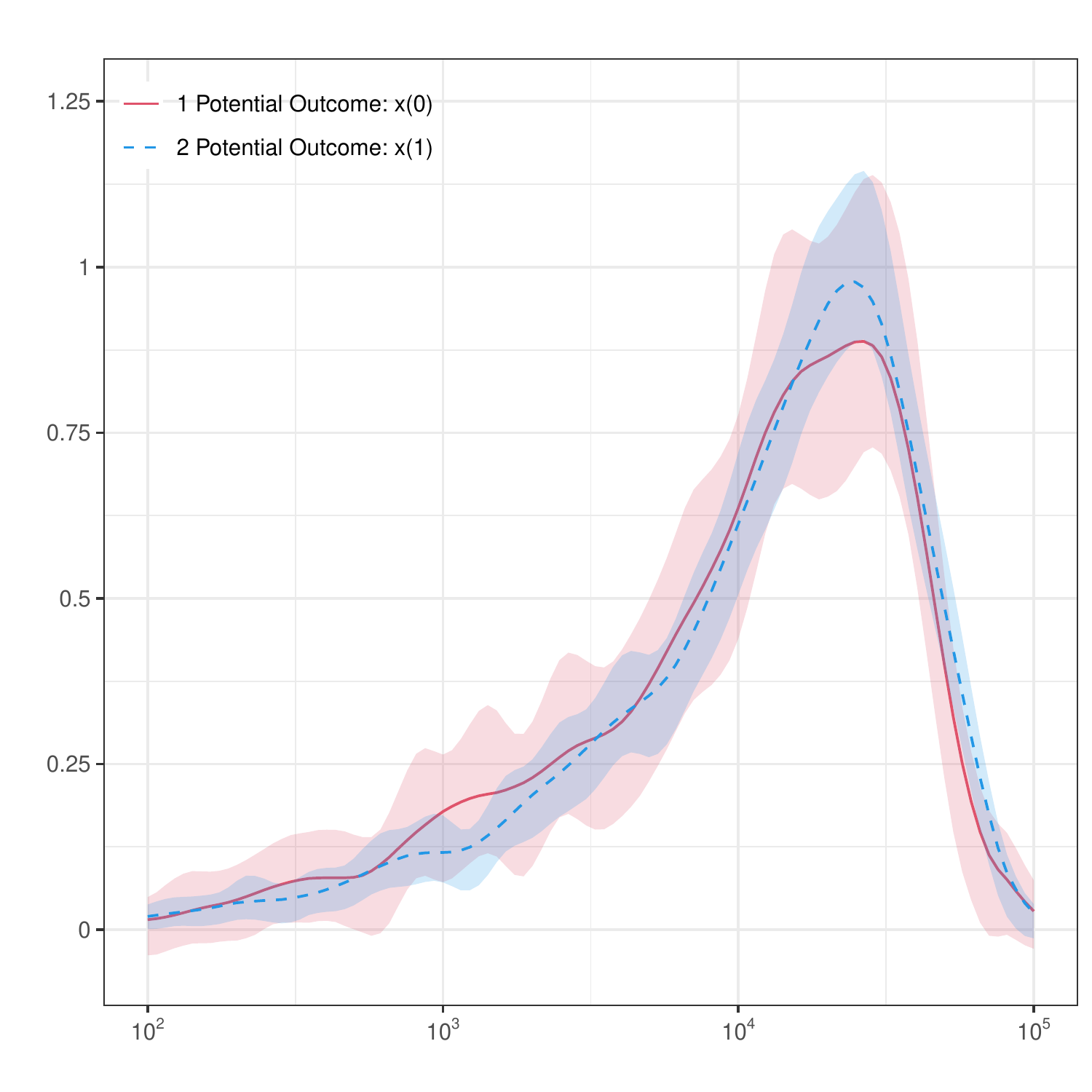}}}

\caption{Earning Distributions, JTPA.\label{fig:jtpa complier}}
\begin{flushleft}
\footnotesize\textit{Notes}: (a) earning distributions in the full sample and for compliers; (b) earning distributions by JTPA offer; (c) earning distributions by JTPA enrollment; (d) distributions of potential outcomes for compliers. Point estimates are obtained by using local polynomial regression with order 2, and robust confidence bands are obtained with local polynomial of order 3. Bandwidths are chosen by minimizing integrated mean squared errors. All estimates are obtained using companion \texttt{R} (and \texttt{Stata}) package described in \cite{Cattaneo-Jansson-Ma_2021_JSS}.
\end{flushleft}

\end{figure}

Second, we estimate the density of the potential outcomes for compliers. In panel (a) of Figure \ref{fig:jtpa complier}, we plot earning distributions for the full sample and that for the compliers, where the second is estimated using the weights $w_{i}^{c}$, introduced earlier. The two distributions seem quite similar, while compliers tend to have higher mean and thinner left tail in the earning distribution. Next we consider the intent-to-treat effect, as the difference in earning distributions for subgroups with and without JTPA offer (a.k.a. the reduced form estimate in the 2SLS context). This is given in panel (b) of Figure \ref{fig:jtpa complier}. The effect is significant, albeit not very large. We also plot earning distributions for individuals enrolled (and not) in JTPA in panel (c). Not surprisingly, the difference is much larger. Simple mean comparison implies that enrolling in JTPA is associated with $\$2,083$ more income.

Unfortunately, neither panel (b) nor (c) reveals information on distribution of potential outcomes. To see the reason, note that in panel (b) earning distributions are estimated according to treatment assignment, but potential outcomes are defined according to treatment takeup. And panel (c) does not give potential outcome distributions since treatment takeup is not randomly assigned. In panel (d) of Figure \ref{fig:jtpa complier}, we use weighting schemes $w_{i}^{c,0}$ and $w_{i}^{c,1}$ to construct potential earning distributions for compliers, which estimates the identified distributional treatment effect in this IV setting. Indeed, treatment effect on compliers is larger than the intent-to-treat effect, but is smaller than that in panel (c). The result is compatible with the fact that JTPA has positive and nontrivial effect on earning. Moreover, it demonstrates the presence of self-selection: those who participated in JTPA on average would benefit the most, followed by compliers who are regarded as ``on the margin of indifference.''

\section{Conclusion}\label{section:Conclusion}

We introduced a new class of local regression distribution estimator, which can be used to construct distribution, density, and higher-order derivatives estimators. We established valid large sample distributional approximations, both pointwise and uniform over their support. Pointwise on the evaluation point, we characterized a minimum distance implementation based on redundant regressors leading to asymptotic efficiency improvements, and gave precise results in terms of (tight) lower bounds for interior points. Uniformly over the evaluation points, we obtained valid linearizations and strong approximations, and constructed confidence bands. Finally, we discussed several extensions of our work.

Although beyond the scope of this paper, it would be useful to generalize our results to the case of multivariate regressors $x_{i}\in\mathbb{R}^{d}$. Boundary adaptation is substantially more difficult in multiple dimensions, and hence our proposed methods are potentially very useful in such setting. In addition, multidimensional density estimation can be used to construct new conditional distribution, density and higher derivative estimators in a straightforward way. These new estimators would be useful in several areas of economics, including for instance estimation of auction models.

\bibliographystyle{econometrica}
\bibliography{Cattaneo-Jansson-Ma_2021_JoE}

\newpage

\end{document}


\title{ 
Local Regression Distribution Estimators\thanks{Cattaneo gratefully acknowledges financial support from the National Science Foundation through grant SES-1947805, and Jansson gratefully acknowledges financial support from the National Science Foundation through grant SES-1947662 and the research support of CREATES.}\bigskip\\ Supplemental Appendix\bigskip}
\author{
Matias D. Cattaneo\thanks{Department of Operations Research and Financial Engineering, Princeton University.} 
\and 
Michael Jansson\thanks{Department of Economics, UC Berkeley and CREATES.}
\and 
Xinwei Ma\thanks{Department of Economics, UC San Diego.}
}
\maketitle

\begin{abstract}
\noindent This Supplemental Appendix contains general theoretical results encompassing those discussed in the main paper, includes proofs of those general results, and discusses additional methodological and technical results.
\end{abstract}

\vfill
\thispagestyle{empty}
\clearpage

\tableofcontents

\clearpage

\onehalfspacing
\pagestyle{plain}

\section{Setup}\label{section:setup}

Suppose $x_1,x_2,\cdots,x_n$ is a random sample from a univariate distribution with cumulative distribution function $F(\cdot)$. Also assume the distribution function admits a (sufficiently accurate) linear-in-parameters local approximation near an evaluation point $\Evapointx$:
\begin{align*}
\varrho(h,\Evapointx) := \sup_{|x-\Evapointx|\leq h}\left|F(x) - R(x-\Evapointx)^\Trans \theta(\Evapointx)\right|\ \text{is small for $h$ small}, 
\end{align*}
where $R(\cdot)$ is a known basis function. The parameter $\theta(\Evapointx)$ can be estimated by the following local $L^2$ method:
\begin{align}\label{eq:local L2 estimator}
\hat\theta_{G} &= \argmin_{\theta} \int_{\mathcal{X}}\left( \hat{F}(u) - R(u-\Evapointx)^\Trans\theta \right)^2 \frac{1}{h}K\left(\frac{u-\Evapointx}{h}\right)\diff G(u),\qquad \hat{F}(u) = \frac{1}{n}\sum_{i=1}^n \Indicator(x_i\leq u),
\end{align}
where $K(\cdot)$ is a kernel function, $\mathcal{X}$ is the support of $F(\cdot)$, and $G(\cdot)$ is a known weighting function to be specified later. The local $L^2$ estimator \eqref{eq:local L2 estimator} is closely related to another estimator, which is constructed by local regression:
\begin{align}\label{eq:local regression estimator}
\hat\theta &= \argmin_{\theta} \sum_{i=1}^n \left( \hat{F}(x_i) - R(x_i-\Evapointx)^\Trans\theta \right)^2 \frac{1}{h}K\left(\frac{x_i-\Evapointx}{h}\right). 
\end{align}
The local regression estimator can be equivalently expressed as $\hat\theta_{\hat{F}}$, meaning that it can be viewed as a special case of the local $L^2$ estimator, with $G(\cdot)$ in \eqref{eq:local L2 estimator} replaced by the empirical distribution function $\hat{F}(\cdot)$. 

For future reference, we first discuss some of the notation we use in the main paper and this Supplemental Appendix (SA). For a function $g(\cdot)$, we denote its $j$-th derivative as $g^{(j)}(\cdot)$. For simplicity, we also use the ``dot'' notation to denote the first derivative: $\dot{g}(\cdot) = g^{(1)}(\cdot)$. Assume $g(\cdot)$ is suitably smooth on $[\Evapointx-\delta,\Evapointx)\cup(\Evapointx,\Evapointx+\delta]$ for some $\delta>0$, but not necessarily continuous or differentiable at $\Evapointx$. If $g(\cdot)$ and its one-sided derivatives can be continuously extended to $\Evapointx$ from the two sides, we adopt the following special notation: 
\begin{align*}
g^{(j)}_u &= \Indicator(u<0)g^{(j)}(\Evapointx-) + \Indicator(u\geq 0)g^{(j)}(\Evapointx+). 
\end{align*}
With $j=0$, the above is simply $g_u = \Indicator(u<0)g(\Evapointx-)+\Indicator(u\geq 0)g(\Evapointx+)$. Also for $j=1$, we use $\dot{g}_u = g^{(1)}_u$. Convergence in probability and in distribution are denoted by $\toProb$ and $\rightsquigarrow$, respectively, and limits are taken with respect to the sample size $n$ going to infinity unless otherwise specified. We use $|\cdot|$ to denote the Euclidean norm. 

The following matrices will feature in asymptotic expansions of our estimators:
\begin{align*}
\Gamma_{h,\Evapointx} &= \int_{\Intregion} R(u)R(u)^\Trans  K\left(u\right) g(\Evapointx + hv)\diff u = \int_{\Intregion} R(u)R(u)^\Trans  K\left(u\right) g_u\diff u + O(h) = \Gamma_{1h,\Evapointx} + O(h),
\end{align*}
and 
\begin{align*}
	\Sigma_{h,\Evapointx} &= \iint_{\Intregion} R(u)R(v)^\Trans\Big[ F(\Evapointx+h(u\wedge v)) - F(\Evapointx+hu)F(\Evapointx+hv) \Big] K\left(u\right)K\left(v\right) g(\Evapointx+hu)g(\Evapointx+hv)\diff u\diff v\\
	&= F(\Evapointx)(1-F(\Evapointx))\left(\int_{\Intregion} R(u) K(u)  g_u\diff u\right)\left(\int_{\Intregion} R(u) K(u)  g_u\diff u\right)^\Trans\\
	&\quad + h\iint_{\Intregion} R(u)R(v)^\Trans K(u)K(v)\Big[ - F(\Evapointx)(uf_u+vf_v)g_ug_v + F(\Evapointx)(1-F(\Evapointx))(u\dot{g}_ug_v+v\dot{g}_vg_u)\Big]  \diff u\diff v\\
	&\quad + h\iint_{\Intregion} R(u)R(v)^\Trans K(u)K(v)(u\wedge v)f_{u\wedge v}g_ug_v   \diff u\diff v + O(h^2)\\
	&= \Sigma_{1h,\Evapointx} + h\Sigma_{2h,\Evapointx} + O(h^2).
\end{align*}
Now we list the main assumptions. 

\begin{assumption}\label{assumption:dgp pointwise}
$x_{1},\dots,x_{n}$ is a random sample from a distribution $F(\cdot)$ supported on $\mathcal{X}\subseteq\mathbb{R}$, and $\mathsf{x}\in\mathcal{X}$.

(i) For some $\delta>0$, $F(\cdot)$ is absolutely continuous on $[\Evapointx-\delta,\Evapointx+\delta]$ with a density $f(\cdot)$ admitting constants $f(\Evapointx-)$, $f(\Evapointx+)$, $\dot{f}(\Evapointx-)$, and $\dot{f}(\Evapointx+)$, such that 
\begin{align*}
\sup_{u\in[-\delta,0)} \frac{f(\Evapointx+u) - f(\Evapointx-) - u\dot{f}(\Evapointx-)}{u^2} + \sup_{u\in(0,\delta]} \frac{f(\Evapointx+u) - f(\Evapointx+) - u\dot{f}(\Evapointx+)}{u^2} < \infty.
\end{align*}

(ii) $K(\cdot)$ is nonnegative, symmetric, and continuous on its support $[-1,1]$, and integrates to 1. 

(iii) $R(\cdot)$ is locally bounded, and there exists a positive-definite diagonal matrix $\Upsilon_{h}$ for each $h>0$, such that $\Upsilon_{h}R(u)=R(u/h)$

(iv) For all $h$ sufficiently small, the minimum eigenvalues of $\Gamma_{h,\Evapointx}$ and $h^{-1}\Sigma_{h,\Evapointx}$ are bounded away from zero. 
\qed
\end{assumption} 

\begin{assumption}\label{assumption:design pointwise}
For some $\delta>0$, $G(\cdot)$ is absolutely continuous on $[\Evapointx-\delta,\Evapointx+\delta]$ with a derivative $g(\cdot)\geq 0$ admitting constants $g(\Evapointx-)$, $g(\Evapointx+)$, $\dot{g}(\Evapointx-)$, and $\dot{g}(\Evapointx+)$, such that 
\begin{align*}
\sup_{u\in[-\delta,0)} \frac{g(\Evapointx+u) - g(\Evapointx-) - u\dot{g}(\Evapointx-)}{u^2} + \sup_{u\in(0,\delta]} \frac{g(\Evapointx+u) - g(\Evapointx+) - u\dot{g}(\Evapointx+)}{u^2} < \infty.
\end{align*}
\vskip-2em\qed
\end{assumption}

\begin{example}[Local Polynomial Estimator]\label{example:local polynomial density estimator}
Before closing this section, we briefly introduce the local polynomial estimator of \cite*{Cattaneo-Jansson-Ma_2020_JASA}, which is a special case of our local regression distribution estimator. The local polynomial estimator employs the following polynomial basis:
\[ R(u) = \Big(1,\ u,\ \frac{1}{2}u^2,\ \cdots,\ \frac{1}{p!}u^p\Big)^\Trans,\]
for some $p\in \mathbb{N}$. As a result, it estimates the distribution function, the density function, and derivatives thereof. To be precise, 
\[\theta(\Evapointx) = \Big( F(\Evapointx),\ f(\Evapointx),\ f^{(1)}(\Evapointx),\ \cdots,\ f^{(p-1)}(\Evapointx) \Big)^\Trans.\]
With $R(\cdot)$ being a polynomial basis, it is straightforward to characterize the approximation bias $\varrho(h,\Evapointx)$. Assuming the distribution function $F(\cdot)$ is at least $p+1$ times continuously differentiable in a neighborhood of $\Evapointx$, one can employ a Taylor expansion argument and show that $\varrho(h,\Evapointx) = O(h^{p+1})$. We will revisit this local polynomial estimator below as a leading example when we discuss pointwise and uniform asymptotic properties of our local distribution estimator. 
\qed
\end{example}

\section{Pointwise Distribution Theory}\label{section:pointwise distribution theory}

We discuss pointwise (i.e., for a fixed evaluation point $\Evapointx\in\mathcal{X}$) large-sample properties of the local $L^2$ estimator \eqref{eq:local L2 estimator}, and that of the local regression estimator \eqref{eq:local regression estimator}. For ease of exposition, we suppress the dependence on the evaluation point $\Evapointx$ whenever possible. 

\subsection{Local $L^2$ Distribution Estimation}\label{subsection:local projection: pointwise distribution theory}

With simple algebra, the local $L^2$ estimator in \eqref{eq:local L2 estimator} takes the following form
\begin{align*}
\hat\theta_G &= \left( \int_{\mathcal{X}} R(u-\Evapointx)R(u-\Evapointx)^\Trans \frac{1}{h}K\left(\frac{u-\Evapointx}{h}\right) \diff G(u) \right)^{-1}\left( \int_{\mathcal{X}} R(u-\Evapointx)\hat{F}(u) \frac{1}{h}K\left(\frac{u-\Evapointx}{h}\right) \diff G(u) \right). 
\end{align*}
We can further simplify the above. First note that the ``denominator'' can be rewritten as
\begin{align*}
&\ \int_{\mathcal{X}} R(u-\Evapointx)R(u-\Evapointx)^\Trans \frac{1}{h}K\left(\frac{u-\Evapointx}{h}\right) \diff G(u)\\
&= \Upsilon_h^{-1}\left(\int_{\mathcal{X}} \Upsilon_h R(u-\Evapointx)R(u-\Evapointx)^\Trans \Upsilon_h \frac{1}{h}K\left(\frac{u-\Evapointx}{h}\right) g(u)\diff u\right) \Upsilon_h^{-1}= \Upsilon_h^{-1}\Gamma_h \Upsilon_h^{-1}.
\end{align*}
The same technique can be applied to the ``numerator'', which leads to
\begin{align}
\nonumber &\ \hat\theta_G - \theta = \Upsilon_h\Gamma_h^{-1}  \left(\int_{\Intregion} R(u)\hat{F}(\Evapointx + hu) K\left(u\right) g(\Evapointx+hu)\diff u\right) - \theta\\
\label{eq:approximation bias, projection}=& \Upsilon_h\Gamma_h^{-1}\int_{\Intregion} R(u)\Big[F(\Evapointx + hu) - \theta^\Trans R(u)\Upsilon_h^{-1}\Big] K\left(u\right) g(\Evapointx+hu)\diff u\\
\label{eq:linear variance, projection} &\qquad +\Upsilon_h \frac{1}{n}\sum_{i=1}^n\Gamma_h^{-1}  \int_{\Intregion} R(u)\Big[\Indicator(x_i\leq \Evapointx + hu) - F(\Evapointx + hu)\Big] K\left(u\right) g(\Evapointx+hu)\diff u.
\end{align}
The above provides a further expansion of the local $L^2$ estimator into a term that contributes as bias, and another term that contributes asymptotically to the variance. 

The large-sample properties of the local $L^2$ estimator \eqref{eq:local L2 estimator} are as follows. 

\begin{thm}[Local $L^2$ Distribution Estimation: Asymptotic Normality]\label{thm:local projection: asymptotic normality}
Assume Assumptions \ref{assumption:dgp pointwise} and \ref{assumption:design pointwise} hold, and that $h\to 0$, $nh\to \infty$ and $n\varrho(h)^2/h\to 0$. Then\\
(i) \eqref{eq:approximation bias, projection} satisfies
\begin{align*}
\left|\int_{\Intregion} R(u)\Big[F(\Evapointx + hu) - \theta^\Trans R(u)\Upsilon_h^{-1}\Big] K\left(u\right) \diff u\right| = O(\varrho(h)). 
\end{align*}
(ii) \eqref{eq:linear variance, projection} satisfies
\begin{align*}
\Var\left[ \int_{\Intregion} R(u)\Big[\Indicator(x_i\leq \Evapointx + hu) - F(\Evapointx + hu)\Big] K\left(u\right) g(\Evapointx+hu)\diff u \right] &= \Sigma_h,
\end{align*}
and
\begin{align*}
\Sigma_h^{-1/2} \left(\frac{1}{\sqrt{n}}\sum_{i=1}^n\int_{\Intregion} R(u)\Big[\Indicator(x_i\leq \Evapointx + hu) - F(\Evapointx + hu)\Big] K\left(u\right) g(\Evapointx+hu)\diff u\right) \rightsquigarrow \mathcal{N}(0,I). 
\end{align*}
(iii) The local $L^2$ estimator is asymptotically normally distributed
\begin{align*}
\sqrt{n}\left(\Gamma_h^{-1}\Sigma_h\Gamma_h^{-1}\right)^{-1/2}\Upsilon_h^{-1} (\hat\theta_G - \theta) \rightsquigarrow \mathcal{N}(0, I).
\end{align*}
\vskip-2em\qed
\end{thm} 

For valid inference, one needs to construct standard errors. To start, note that $\Gamma_{h}$ is known, and hence we only need to estimate $\Sigma_{h}$. Consider the following:
\begin{align}
\nonumber\hat\Sigma_{h} &= \frac{1}{n}\sum_{i=1}^n\iint_{\Intregion} R(u)R(v)^\Trans\Big[\Indicator(x_i\leq \Evapointx + hu) - \hat{F}(\Evapointx + hu)\Big]\Big[\Indicator(x_i\leq \Evapointx + hv) - \hat{F}(\Evapointx + hv)\Big] \\
\label{eq:local projection: standard error pointwise}&\qquad \qquad \qquad \qquad K(u)K(v) g(\Evapointx+hu)g(\Evapointx+hv)\diff u\diff v,
\end{align}
where $\hat{F}(\cdot)$ is the empirical distribution function. The following theorem shows that standard errors constructed using $\hat{\Sigma}_h$ are consistent. 

\begin{thm}[Local $L^2$ Distribution Estimation: Standard Errors]\label{thm:local projection: standard error pointwise}
Assume Assumptions \ref{assumption:dgp pointwise} and \ref{assumption:design pointwise} hold, and that $h\to 0$ and $nh\to \infty$. Let $c$ be a nonzero vector of suitable dimension, then
\begin{align*}
\left|\frac{c^\Trans \hat{\Sigma}_h c}{c^\Trans \Sigma_{h} c} - 1\right| &= \Op\left(  \sqrt{\frac{1}{nh}} \right).
\end{align*}
If, in addition that $n\varrho(h)^2/h\to 0$, then
\begin{align*}
\frac{c^\Trans(\hat\theta_G - \theta)}{\sqrt{c^\Trans \Upsilon_{h}\Gamma_h^{-1}\hat\Sigma_{h} \Gamma_h^{-1}\Upsilon_{h} c/n}} \rightsquigarrow \mathcal{N}(0,1).
\end{align*}
\vskip-2em\qed
\end{thm}

\subsection{Local Regression Distribution Estimation}\label{subsection:local regression: pointwise distribution theory}

The local regression distribution estimator \eqref{eq:local regression estimator} can be understood as a special case of the local $L^2$ estimator by setting $G=\hat{F}$ (i.e., using the empirical distribution as the design). However, the empirical measure $\hat{F}$ is not smooth, so that large-sample properties of the local regression estimator cannot be deduced directly from Theorem \ref{thm:local projection: asymptotic normality}. In this subsection, we will show that estimates obtained by the two approaches, \eqref{eq:local L2 estimator} and \eqref{eq:local regression estimator}, are asymptotically equivalent under suitable regularity conditions. To be precise, we establish the equivalence of the local regression distribution estimator, $\hat\theta$, and the (infeasible) local $L^2$ distribution estimator, $\hat{\theta}_F$ (i.e., using $F$ as the design weighting in \eqref{eq:local L2 estimator}). As before, we suppress the dependence on the evaluation point $\Evapointx$.

First, the local regression estimator can be written as 
\begin{align*}
\hat\theta - \theta &= \left( \frac{1}{n}\sum_{i=1}^n R(x_i-\Evapointx)R(x_i-\Evapointx)^\Trans \frac{1}{h}K\left(\frac{x_i-\Evapointx}{h}\right)\right)^{-1}\\
&\qquad \qquad \left( \frac{1}{n}\sum_{i=1}^n R(x_i-\Evapointx)\Big[ \hat{F}(x_i) -R(x_i-\Evapointx)^\Trans\theta  \Big] \frac{1}{h}K\left(\frac{x_i-\Evapointx}{h}\right)\right)\\
&= \Upsilon_h\hat{\Gamma}_h^{-1}\Gamma_h\Gamma_h^{-1}\left( \frac{1}{n}\sum_{i=1}^n \Upsilon_hR(x_i-\Evapointx)\Big[ \hat{F}(x_i) -R(x_i-\Evapointx)^\Trans\theta  \Big] \frac{1}{h}K\left(\frac{x_i-\Evapointx}{h}\right)\right),
\end{align*}
where 
\begin{align*}
\hat{\Gamma}_h &= \frac{1}{n}\sum_{i=1}^n \Upsilon_hR(x_i-\Evapointx)R(x_i-\Evapointx)^\Trans\Upsilon_h \frac{1}{h}K\left(\frac{x_i-\Evapointx}{h}\right),
\end{align*}
and $\Gamma_h$ is defined as before with $G=F$. 

To proceed, we further expand as follows
\begin{align}
\nonumber&\ \frac{1}{n}\sum_{i=1}^n \Upsilon_hR(x_i-\Evapointx)\Big[ \hat{F}(x_i) -R(x_i-\Evapointx)^\Trans\theta  \Big] \frac{1}{h}K\left(\frac{x_i-\Evapointx}{h}\right)\\
\nonumber=&\ \frac{1}{n^2}\sum_{i,j=1,i\neq j}^n \Upsilon_hR(x_j-\Evapointx)\Big[ \Indicator(x_i\leq x_j) -F(x_j)  \Big] \frac{1}{h}K\left(\frac{x_j-\Evapointx}{h}\right)\\
\label{eq:leave-in bias}&\ + \frac{1}{n^2}\sum_{j=1}^n \Upsilon_hR(x_j-\Evapointx)\Big[ 1 -F(x_j)  \Big] \frac{1}{h}K\left(\frac{x_j-\Evapointx}{h}\right)\\
\label{eq:approximation bias}&\ + \frac{1}{n}\sum_{j=1}^n \Upsilon_hR(x_j-\Evapointx)\Big[ F(x_j) -R(x_j-\Evapointx)^\Trans\theta  \Big] \frac{1}{h}K\left(\frac{x_j-\Evapointx}{h}\right).
\end{align}
The last two terms correspond to the leave-in bias and the approximation bias, respectively. We further decompose the first term with conditional expectation:
\begin{align}
\nonumber&\ \frac{1}{n^2}\sum_{i,j=1,i\neq j}^n \Upsilon_hR(x_j-\Evapointx)\Big[ \Indicator(x_i\leq x_j) -F(x_j)  \Big] \frac{1}{h}K\left(\frac{x_j-\Evapointx}{h}\right)\\
\nonumber=&\ \frac{1}{n^2}\sum_{i,j=1,i\neq j}^n \Expectation\left[\left.\Upsilon_hR(x_j-\Evapointx)\Big[ \Indicator(x_i\leq x_j) -F(x_j)  \Big] \frac{1}{h}K\left(\frac{x_j-\Evapointx}{h}\right)\right| x_i\right]\\
\nonumber&\ + \frac{1}{n^2}\sum_{i,j=1,i\neq j}^n \Upsilon_hR(x_j-\Evapointx)\Big[ \Indicator(x_i\leq x_j) -F(x_j)  \Big] \frac{1}{h}K\left(\frac{x_j-\Evapointx}{h}\right) \\
\nonumber&\qquad \qquad \qquad - \Expectation\left[\left.\Upsilon_hR(x_j-\Evapointx)\Big[ \Indicator(x_i\leq x_j) -F(x_j)  \Big] \frac{1}{h}K\left(\frac{x_j-\Evapointx}{h}\right)\right| x_i\right]\\
\label{eq:linear variance}=&\ \frac{n-1}{n^2}\sum_{i=1}^n \int_{\Intregion} R(u)\Big[\Indicator(x_i\leq \Evapointx + hu) - F(\Evapointx + hu)\Big] K\left(u\right) f(\Evapointx + hu) \diff u\\
\nonumber&\ + \frac{1}{n^2}\sum_{i,j=1,i\neq j}^n \Upsilon_hR(x_j-\Evapointx)\Big[ \Indicator(x_i\leq x_j) -F(x_j)  \Big] \frac{1}{h}K\left(\frac{x_j-\Evapointx}{h}\right) \\
\label{eq:quadratic variance}&\qquad \qquad \qquad- \int_{\Intregion} R(u)\Big[\Indicator(x_i\leq \Evapointx + hu) - F(\Evapointx + hu)\Big] K\left(u\right) f(\Evapointx + hu)\diff u.
\end{align}

The following theorem studies the large-sample properties of each term in the above decomposition, and shows that the local regression distribution estimator is asymptotically equivalent to the local $L^2$ estimator by setting $G=F$, and hence it is asymptotically normally distributed. 

\begin{thm}[Local Regression Distribution Estimation: Asymptotic Normality]\label{thm:local regression: asymptotic normality}
Assume Assumption \ref{assumption:dgp pointwise} holds, and that $h\to 0$, $nh^2\to \infty$ and $n\varrho(h)^2/h\to 0$. Then\\
(i) $\hat{\Gamma}_h$ satisfies
\begin{align*}
\left|\hat{\Gamma}_h - \Gamma_h\right| = \Op\left( \sqrt{\frac{1}{nh}} \right).
\end{align*}
(ii) \eqref{eq:leave-in bias} and \eqref{eq:approximation bias} satisfy
\begin{align*}
\text{\eqref{eq:leave-in bias}} &= \Op\left(\frac{1}{n}\right),\qquad \text{\eqref{eq:approximation bias}}= \Op(\varrho(h)).
\end{align*}
(iii) \eqref{eq:quadratic variance} satisfies
\begin{align*}
\text{\eqref{eq:quadratic variance}} &= \Op\left( \sqrt{\frac{1}{n^2h}} \right).
\end{align*}
(iv) The local regression distribution estimator \eqref{eq:local regression estimator} satisfies
\begin{align*}
\sqrt{n}\left(\Gamma_h^{-1}\Sigma_h\Gamma_h^{-1}\right)^{-1/2}\Upsilon_h^{-1} (\hat\theta - \theta) &= \sqrt{n}\left(\Gamma_h^{-1}\Sigma_h\Gamma_h^{-1}\right)^{-1/2}\Upsilon_h^{-1} (\hat\theta_F - \theta)  + \op(1) 
\rightsquigarrow \mathcal{N}(0,I).
\end{align*}
\vskip-2.5em\qed
\end{thm} 

We now discuss how to construct standard errors in the local regression framework. Note that $\Gamma_{h}$ can be estimated by $\hat{\Gamma}_h$, whose properties have already been studied in Theorem \ref{thm:local regression: asymptotic normality}(i). To estimate $\Sigma_{h}$, we propose the following
\begin{align*}
\hat\Sigma_{h} &= \frac{1}{n}\sum_{i=1}^n \left[\frac{1}{n}\sum_{j=1}^n \Upsilon_hR(x_j-\Evapointx)\Big[ \Indicator(x_i\leq x_j) -\hat{F}(x_j)  \Big] \frac{1}{h}K\left(\frac{x_j-\Evapointx}{h}\right)\right]\\
&\qquad \qquad \qquad \cdot\left[\frac{1}{n}\sum_{j=1}^n \Upsilon_hR(x_j-\Evapointx)\Big[ \Indicator(x_i\leq x_j) -\hat{F}(x_j)  \Big] \frac{1}{h}K\left(\frac{x_j-\Evapointx}{h}\right)\right]^\Trans. 
\end{align*}
where $\hat{F}(\cdot)$ is the empirical distribution function. The following theorem shows that standard errors constructed using $\hat{\Sigma}_h$ are consistent. 

\begin{thm}[Local Regression Distribution Estimation: Standard Errors]\label{thm:local regression: standard error pointwise}
Assume Assumption \ref{assumption:dgp pointwise} holds. In addition, assume $h\to 0$ and $nh^2\to \infty$. Let $c$ be a nonzero vector of suitable dimension. Then
\begin{align*}
\left|\frac{c^\Trans \hat\Gamma_h^{-1} \hat{\Sigma}_h \hat\Gamma_h^{-1} c}{c^\Trans \Gamma_h^{-1} \Sigma_{h} \Gamma_h^{-1} c} - 1\right| &= \Op\left(   \sqrt{\frac{1}{nh^2}} \right).
\end{align*}
If, in addition that $n\varrho(h)^2/h\to 0$, one has
\begin{align*}
\frac{c^\Trans(\hat\theta - \theta)}{\sqrt{c^\Trans \Upsilon_{h}\hat{\Gamma}_h^{-1}\hat\Sigma_{h} \hat{\Gamma}_h^{-1}\Upsilon_{h} c/n}} \rightsquigarrow \mathcal{N}(0,1).
\end{align*}
\vskip-2em\qed
\end{thm}

\section{Efficiency}\label{section:Efficiency}

For ease of presentation, we focus on the (infeasible) local $L^2$ distribution estimator $\hat\theta_F$, 
\begin{align}\label{eq:local polynomial projection estimator}
\hat\theta_{F} &= \argmin_{\theta} \int_{\mathcal{X}}\left( \hat{F}(u) - R(u-\Evapointx)^\Trans\theta \right)^2 \frac{1}{h}K\left(\frac{u-\Evapointx}{h}\right)\diff F(u),
\end{align}
but all the results in this section are applicable to the local regression distribution estimator $\hat\theta$, as we showed earlier that it is asymptotically equivalent to $\hat\theta_F$. In addition, we consider a specific basis:
\begin{align}\label{eq:polynomial basis}
R(u) &= \left(1,\ P(u)^\Trans,\ Q(u)\right)^\Trans,
\end{align}
where $P(u)$ is a polynomial basis of order $p$:
\begin{align*}
P(u) = \left(u,\ \frac{1}{2}u^2,\ \cdots,\ \frac{1}{p!}u^p\right)^\Trans,
\end{align*}
and $Q(u)$ is a scalar function, and hence is a ``redundant regressor.'' Without $Q(\cdot)$, the above reduces to the local polynomial estimator of \cite*{Cattaneo-Jansson-Ma_2020_JASA}. See Section \ref{section:setup} and Example \ref{example:local polynomial density estimator} for an introduction. 

We consider additional regressors because they may help improve efficiency (i.e., reduce the asymptotic variance). Following Assumption \ref{assumption:dgp pointwise}, we assume there exists a scalar $\upsilon_h$ (depending on $h$) such that $\upsilon_hQ(u)=Q(u/h)$. Therefore, $\Upsilon_h$ is a diagonal matrix containing $1,h^{-1},h^{-2},\cdots,h^{-p},\upsilon_h$. As we consider a (local) polynomial basis, it is natural to impose smoothness assumptions on $F(\cdot)$. In particular, 

\begin{assumption}\label{assumption:smoothness}
For some $\delta>0$, $F(\cdot)$ is $(p+1)$-times continuously differentiable in $\mathcal{X}\cap[\Evapointx-\delta,\Evapointx+\delta]$ for some $p\geq 1$, and $G(\cdot)$ is twice continuously differentiable in $\mathcal{X}\cap[\Evapointx-\delta,\Evapointx+\delta]$. 
\qed
\end{assumption}

Under the above assumption, the approximation error satisfies $\varrho(h)=O(h^{p+1})$, and the parameter $\theta$ can be partitioned into the following:
\begin{align*}
\theta = \Big(\theta_{1},\ \theta_{P}^\Trans,\ \theta_{Q} \Big)^\Trans = \Big( F(\Evapointx),\quad f(\Evapointx),\ \cdots,\ f^{(p-1)}(\Evapointx),\quad 0 \Big)^\Trans.
\end{align*} 

We first state a corollary, which specializes Theorem \ref{thm:local projection: asymptotic normality} to the polynomial basis \eqref{eq:polynomial basis}.

\begin{coro}[Local Polynomial $L^2$ Distribution Estimation: Asymptotic Normality]\label{coro:asy normal loc pol projection estimator}
Assume Assumptions \ref{assumption:dgp pointwise} and \ref{assumption:smoothness} hold, and that $h\to 0$, $nh\to \infty$, and $n\varrho(h)^2/h\to 0$. Then the local polynomial $L^2$ distribution estimator in \eqref{eq:local polynomial projection estimator} satisfies
\begin{align*}
\sqrt{n}\left(\Gamma_h^{-1}\Sigma_h\Gamma_h^{-1}\right)^{-1/2}\Upsilon_h^{-1} (\hat\theta_F - \theta) \rightsquigarrow \mathcal{N}(0, I).
\end{align*}
\vskip-2em\qed
\end{coro} 

\subsection{Effect of Orthogonalization}

To start, consider the following (sequentially) orthogonalized basis:
\begin{align}\label{eq:orthogonalized polynomial basis}
R^\perp(u) &= \left(1,\ P^\perp(u)^\Trans,\ Q^\perp(u)\right)^\Trans,
\end{align}
where 
\begin{align*}
P^\perp(u) &= P^\perp(u) - \int_{\Intregion} K(u) P(u) \diff u,\\
Q^\perp(u) &= Q(u) - \Big(1, P(v)^\Trans\Big)\left( \int_{\Intregion} K(v)\Big(1, P(v)^\Trans\Big)^\Trans\Big(1, P(v)^\Trans\Big) \diff v \right)^{-1}\left( \int_{\Intregion} K(v)\Big(1, P(v)^\Trans\Big)^\Trans Q(v) \diff v \right).
\end{align*}
The above transformation can be represented by the following:
\begin{align*}
R^\perp(u) &= \Lambda_h^\Trans R(u),
\end{align*}
where $\Lambda_h$ is a nonsingular upper triangular matrix. (Note that the matrix $\Lambda_h$ depends on the bandwidth only because we would like to handle both interior and boundary evaluation points. If, for example, we fix the evaluation point to be in the interior of the support of the data, then $\Lambda_h$ is a fixed matrix and no longer depends on $h$. Alternatively, one could also use the notation ``$\Lambda_\Evapointx$'' to denote such dependence.) Now consider the following orthogonalized local polynomial $L^2$  estimator
\begin{align}\label{eq:local orthogonal polynomial projection estimator}
\hat\theta_{F}^\perp &= \argmin_{\theta} \int_{\mathcal{X}}\left( \hat{F}(u) - \Lambda_h^\Trans R(u-\Evapointx)^\Trans\theta \right)^2 \frac{1}{h}K\left(\frac{u-\Evapointx}{h}\right)\diff F(u).
\end{align}

To discuss its properties, we partition the estimator and the target parameter as
\begin{align*}
\hat\theta_F^\perp &= \Big( \hat\theta_{1,F}^\perp,\ (\hat\theta_{P,F}^\perp)^\Trans,\ \hat\theta_{Q,F}^\perp \Big)^\Trans,
\end{align*}
where $\hat\theta_{1,F}^\perp$ is the first element of $\hat\theta_F^\perp$ and $\hat\theta_{Q,F}^\perp$ is the last element of $\hat\theta_F^\perp$. Similarly, we can partition the target parameter, 
\begin{align*}
\theta^\perp &= \Lambda_h^{-1}\theta = \Big( \theta^\perp_{1},\ (\theta^\perp_{P})^\Trans,\ \theta^\perp_{Q} \Big)^\Trans,
\end{align*}
so that $\theta^\perp_{1}$ is the first element of $\Lambda_h^{-1}\theta$ and $\theta^\perp_{Q}$ is the last element of $\Lambda_h^{-1}\theta$. As $\theta_{Q}=0$, simple least squares algebra implies
\begin{align*}
\theta^\perp &= \Big( \theta^\perp_{1},\ \theta_{P}^\Trans,\ 0 \Big)^\Trans = \Big( \theta^\perp_{1},\ f(\Evapointx),\ f^{(1)}(\Evapointx),\ \cdots,\ f^{(p-1)}(\Evapointx),\ 0 \Big)^\Trans.
\end{align*}
Note that, in general, $\theta^\perp_{1}\neq \theta_{1}$, meaning that after orthogonalization, the intercept of the local polynomial estimator no longer estimates the distribution function $F(\Evapointx)$. 

The following corollary gives the large-sample properties of the orthogonalized local polynomial estimator, excluding the intercept. 

\begin{coro}[Orthogonalized Local Polynomial $L^2$ Distribution Estimation: Asymptotic Normality]\label{coro:asy normal ortho loc pol projection estimator}
Assume Assumptions \ref{assumption:dgp pointwise} and \ref{assumption:smoothness} hold, and that $h\to 0$, $nh\to \infty$, and $n\varrho(h)^2/h\to 0$. Then the orthogonalized local polynomial $L^2$ distribution estimator in \eqref{eq:local orthogonal polynomial projection estimator} satisfies
\begin{align*}
\begin{bmatrix}
(\Gamma_{P,h}^\perp)^{-1} \Sigma_{PP,h}^\perp (\Gamma_{P,h}^\perp)^{-1} & (\Gamma_{P,h}^\perp)^{-1} \Sigma_{PQ,h}^\perp (\Gamma_{Q,h}^\perp)^{-1} \\
(\Gamma_{Q,h}^\perp)^{-1} \Sigma_{QP,h}^\perp (\Gamma_{P,h}^\perp)^{-1} & (\Gamma_{Q,h}^\perp)^{-1} \Sigma_{QQ,h}^\perp (\Gamma_{Q,h}^\perp)^{-1}
\end{bmatrix}^{-1/2} \sqrt{\frac{n}{hf(\Evapointx)}}\Upsilon_{-1, h}^{-1}\begin{bmatrix}
\hat\theta^\perp_{P,F} - \theta_{P} \\
\hat\theta^\perp_{Q,F}
\end{bmatrix} \rightsquigarrow \mathcal{N}(0, I),
\end{align*}
where
\begin{align*}
&\ \Gamma_{P,h}^\perp = \int_{\Intregion} P^\perp(u)P^\perp(u)^\Trans K(u)\diff u,
\quad \Gamma_{Q,h}^\perp = \int_{\Intregion} Q^\perp(u)^2 K(u)\diff u,\\
&\ \Sigma_{PP,h}^\perp = \iint_{\Intregion} K(u)K(v)P^\perp(u)P^\perp(v)^\Trans (u\wedge v) \diff u\diff v,\\
&\ \Sigma_{QQ,h}^\perp = \iint_{\Intregion} K(u)K(v)Q^\perp(u)Q^\perp(v) (u\wedge v) \diff u\diff v,\\ 
&\ \Sigma_{PQ,h}^\perp = (\Sigma_{QP,h}^\perp)^\Trans = \iint_{\Intregion} K(u)K(v)P^\perp(u)Q^\perp(v) (u\wedge v) \diff u\diff v,
\end{align*}
and $\Upsilon_{-1, h}$ is a diagonal matrix containing $h^{-1},h^{-2},\cdots,h^{-p},\upsilon_h$.
\qed
\end{coro}

\subsection{Optimal $Q$}

Now we discuss the optimal choice of $Q$, which minimizes the asymptotic variance of the minimum distance estimator. Recall from the main paper that, with orthogonalized basis, the minimum distance estimator of ${f}^{(\ell)}(\Evapointx)$, for $0\leq \ell\leq p-1$, has an asymptotic variance 
\begin{align*}
f(\Evapointx)\Big[e_\ell^\Trans(\Gamma_{P,h}^\perp)^{-1} \Sigma_{PP,h}^\perp (\Gamma_{P,h}^\perp)^{-1}e_\ell - e_\ell^\Trans(\Gamma_{P,h}^\perp)^{-1} \Sigma_{PQ,h}^\perp 
 (\Sigma_{QQ,h}^{\perp} )^{-1}
 \Sigma_{QP,h}^\perp (\Gamma_{P,h}^\perp)^{-1}e_\ell\Big],
\end{align*}
where $e_\ell$ is the $(\ell+1)$-th standard basis vector. In subsequent analysis, we drop the multiplicative factor $f(\Evapointx)$. 

Let $p_\ell(u)$ be defined as
\begin{align*}
p_\ell(u) &= e_\ell^\Trans(\Gamma_{P,h}^\perp)^{-1} P^\perp(u),
\end{align*}
then the objective is to maximize 
\begin{align*}
\left( \iint_{\Intregion} K(u)K(v)Q^\perp(u)Q^\perp(v)(u\wedge v) \diff u\diff v \right)^{-1}\left( \iint_{\Intregion} K(u)K(v)p_\ell(u)Q^\perp(v)(u\wedge v) \diff u\diff v \right)^2.
\end{align*}
Alternatively, we would like to solve (recall that $Q(u)$ is a scaler function)
\begin{align*}
\text{maximize}&\ \frac{\left(\iint_{\Intregion} K(u)K(v)p_\ell(u)q(v)(u\wedge v) \diff u\diff v\right)^2}{\iint_{\Intregion} K(u)K(v)q(u)q(v)(u\wedge v) \diff u\diff v},\quad 
\text{subject to}\ \int_{\Intregion} K(u)q(u)(1, P(u)^\Trans)\diff u = 0.
\end{align*}
To proceed, define the following transformation for a function $g(\cdot)$:
\begin{align*}
\mathcal{H}(g)(u) &= \int_{\Intregion} \Indicator(v\geq u)K(v)g(v)\diff v.
\end{align*}
This transformation satisfies two important properties, which are summarized in the following lemma. 

\begin{lem}[$\mathcal{H}$-transformation]\label{lem:survival transformation}\ 

(i) If $g_1(\cdot)$ and $g_2(\cdot)$ are bounded, and that either $\int_{\Intregion}K(u)g_1(u)\diff u$ or $\int_{\Intregion}K(u)g_2(u)\diff u$ is zero, then
\begin{align*}
\int_{\Intregion\cap[-1,1]} \mathcal{H}(g_1)(u)\mathcal{H}(g_2)(u) \diff u &= \iint_{\Intregion} K(u)K(v)g_1(u)g_1(v)(u\wedge v)\diff u\diff v.
\end{align*}

(ii) If $g_1(\cdot)$ and $g_2(\cdot)$ are bounded, $g_2(\cdot)$ is continuously differentiable with a bounded derivative, and that either $\int_{\Intregion}K(u)g_1(u)\diff u$ or $\int_{\Intregion}K(u)g_2(u)\diff u$ is zero, then 
\begin{align*}
\int_{\Intregion\cap[-1,1]} \mathcal{H}(g_1)(u)\dot{g}_2(u) \diff u &= \int_{\Intregion}K(u)g_1(u)  g_2(u) \diff u.
\end{align*}
\vskip-2.3em\qed
\end{lem}

With the previous lemma, we can rewrite the maximization problem as
\begin{align}
\nonumber\text{maximize}&\ \qquad \frac{\left(\int_{\Intregion\cap[-1,1]} \mathcal{H}(p_\ell)(u)\mathcal{H}(q)(u)\diff u\right)^2}{\int_{\Intregion\cap[-1,1]} \mathcal{H}(q)(u)^2 \diff u}\\
\label{eq:maximization}\text{subject to}&\ \qquad \int_{\Intregion\cap[-1,1]} \dot{P}(u)\mathcal{H}(q)(u)\diff u = 0,\qquad \mathcal{H}(q)\left( \frac{\inf \mathcal{X}-\Evapointx}{h}\vee (-1) \right) =0.
\end{align}

\begin{thm}[Variance Bound of the Minimum Distance Estimator]\label{thm:variance bound}
An upper bound of the maximization problem \eqref{eq:maximization} is
\begin{align*}
e_\ell^\Trans(\Gamma_{P,h}^\perp)^{-1} \Sigma_{PP,h}^\perp (\Gamma_{P,h}^\perp)^{-1}e_\ell - e_\ell^\Trans \left(\int_{\Intregion\cap[-1,1]} \dot{P}(u)\dot{P}(u)^\Trans \diff u\right)^{-1} e_{\ell}. 
\end{align*}
Therefore, the asymptotic variance of the minimum distance estimator is bounded below by
\begin{align*}
f(\Evapointx)e_\ell^\Trans \left(\int_{\Intregion\cap[-1,1]} \dot{P}(u)\dot{P}(u)^\Trans \diff u\right)^{-1} e_{\ell},
\end{align*}
where $\dot{P}(u)=(1,\ u,\ u^2/2,\ u^3/3!,\ \cdots,\ u^{p-1}/(p-1)!)^\Trans$.
\qed
\end{thm}

\begin{figure}[!t]
\centering

\includegraphics[width=0.45\textwidth]{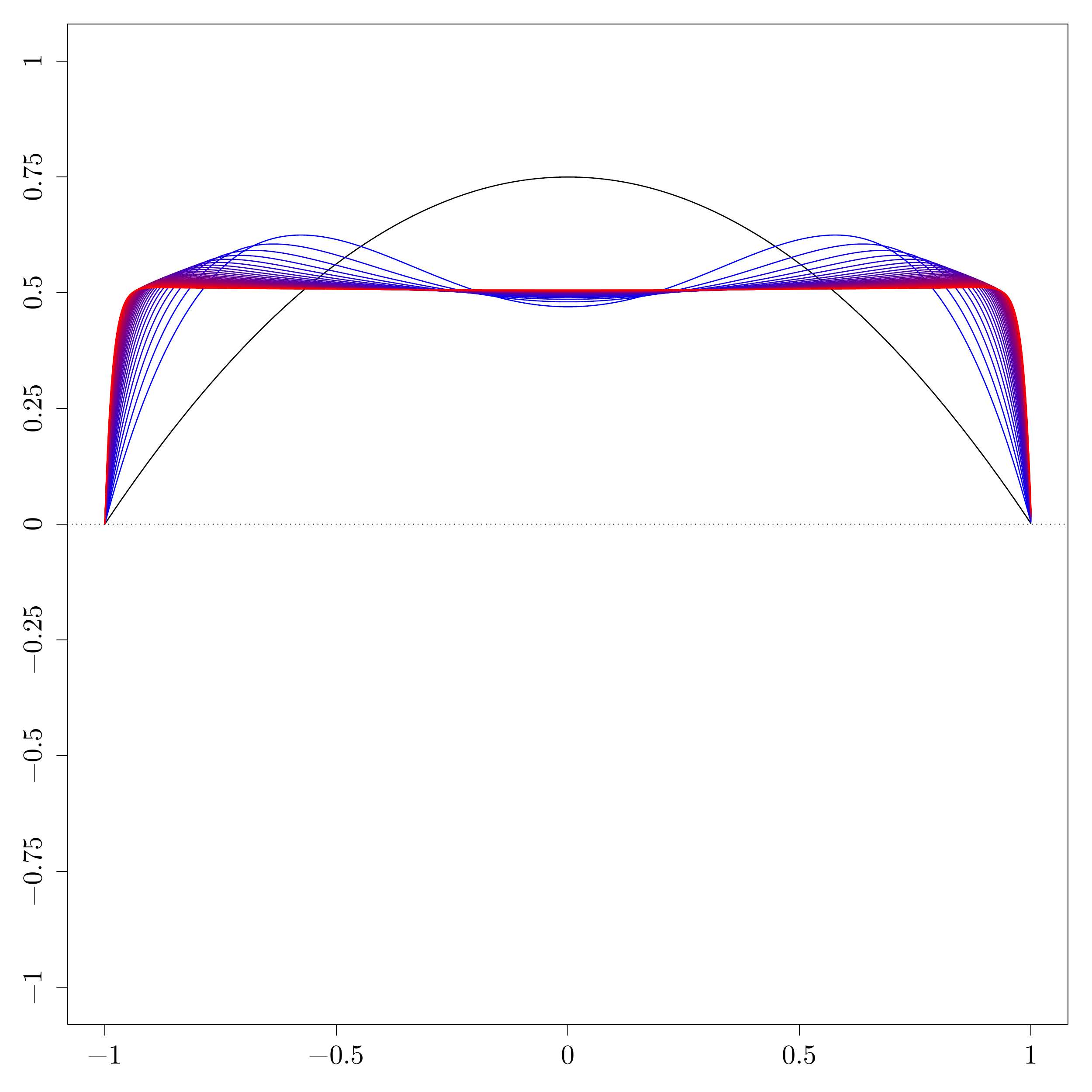}

\caption{Equivalent Kernel of the Local Linear Minimum Distance Density Estimator.} \label{fig:equi kerenl p=1 v=1}

\begin{flushleft}
\footnotesize
\textit{Notes}: The basis function $R(u)$ consists of an intercept, a linear term $u$ (i.e., local linear regression), and an odd higher-order polynomial term $u^{2j+1}$ for $j=1, 2, \cdots,30$. Without the higher-order polynomial regressor, the local linear density estimator using the uniform kernel is equivalent to the kernel density estimator using the Epanechnikov kernel (black line). Including a higher-order redundant regressor leads to an equivalent kernel that approaches the uniform kernel as $j$ tends to infinity (red).
\end{flushleft} 
\end{figure}

\begin{example}[Local Linear/Quadratic Minimum Distance Density Estimation]\label{example:p=1 v=1 interior}
Consider a simple example where $\ell=0$ and $P(u)=u$, which means we focus on the asymptotic variance of the estimated density in a local linear regression. Also assume we employ a uniform kernel: $K(u)=\frac{1}{2}\Indicator(|u|\leq 1)$, and that the integration region is $\Intregion=\mathbb{R}$ (i.e., $\Evapointx$ is an interior evaluation point). Note that this example also applies to local quadratic regressions, as $u$ and $u^2$ are orthogonal for interior evaluation points. 

Taking $P(u) = u$, the variance bound in Theorem \ref{thm:variance bound} is easily found to be
\begin{align*}
 f(\Evapointx)\left(\int_{-1}^1 \dot{P}(u)\dot{P}(u)^\Trans \diff u\right)^{-1}  =  f(\Evapointx)\frac{1}{2}. 
\end{align*}

We now calculate the asymptotic variance of the minimum distance estimator. To be specific, we choose $Q(u) = u^{2j+1}$, which is a higher-order polynomial function. With tedious calculation, one can show that the minimum distance estimator has the following asymptotic variance
\begin{align*}
\mathrm{Asy}\Var[\hat{f}_{\mathtt{MD}}(\Evapointx)] &=  f(\Evapointx)\frac{11+4j}{20+8j}, 
\end{align*}
which asymptotes to $f(\Evapointx)/2$ as $j\to \infty$. As a result, it is possible to achieve the maximum amount of efficiency gain by including one higher-order polynomial and using our minimum distance estimator. 

In Figure \ref{fig:equi kerenl p=1 v=1}, we plot the equivalent kernel of the local linear minimum distance density estimator using a uniform kernel. Without the redundant regressor, it is equivalent to the kernel density estimator using the Epanechnikov kernel. As $j$ gets larger, however, the equivalent kernel of the minimum distance estimator becomes closer to the uniform kernel, which is why, as $j\to \infty$, the minimum distance estimator has an asymptotic variance the same as the kernel density estimator using the uniform kernel. 
\qed
\end{example}

\begin{example}[Local Cubic Minimum Distance Estimation]\label{example:p=3 v=1 interior}
We adopt the same setting in Example \ref{example:p=1 v=1 interior}, i.e., local polynomial density estimation with the uniform kernel at an interior evaluation point. The difference is that we now consider a local cubic regression: $P(u) = (u, \frac{1}{2}u^2, \frac{1}{3!}u^3)^\Trans$. 

As before, the variance bound in Theorem \ref{thm:variance bound} is easily found to be
\begin{align*}
 f(\Evapointx)\left(\int_{-1}^1 \dot{P}(u)\dot{P}(u)^\Trans \diff u\right)^{-1}  =  f(\Evapointx)
 \begin{bmatrix}
 \frac{9}{8} & 0 & -\frac{15}{4} \\
  0 & \frac{3}{2} & 0 \\
  -\frac{15}{4} & 0 & \frac{45}{2} \\
 \end{bmatrix}. 
\end{align*}

Again, we compute the asymptotic variance of our minimum distance estimator. Note, however, that now we have both odd and even order polynomials in our basis $P(u)$, therefore we include two higher-order polynomials, that is, we set $Q(u)=(u^{2j}, u^{2j+1})^\Trans$. The asymptotic variance of our minimum distance estimator is
\begin{align*}
\mathrm{Asy}\Var\begin{bmatrix}
\hat{f}_{\mathtt{MD}}(\Evapointx)\\
\hat{f}_{\mathtt{MD}}^{(1)}(\Evapointx)\\
\hat{f}_{\mathtt{MD}}^{(2)}(\Evapointx)
\end{bmatrix} &= f(\Evapointx)\begin{bmatrix}
 \frac{9 (4 j+15)}{16 (2 j+7)} & 0 & -\frac{15 (4 j+17)}{8 (2 j+7)} \\
  0 & \frac{12 j+39}{8 j+20} & 0 \\
  -\frac{15 (4 j+17)}{8 (2 j+7)} & 0 & \frac{45 (4 j+19)}{8 j+28} \\
\end{bmatrix},
\end{align*}
which, again, asymptotes to the variance bound as $j\to \infty$. See also Table \ref{table:density variance} for the efficiency gain of employing the minimum distance technique. 
\qed
\end{example}

\begin{table}[!t]
\renewcommand{\arraystretch}{1.4}
\centering
\caption{Variance Comparison.}\label{table:density variance}

\subfloat[Density ${f}(\Evapointx)$]{\resizebox{0.55\columnwidth}{!}{
\begin{tabular}{lrrrr}
\hline\hline
\multicolumn{1}{l}{}&\multicolumn{1}{c}{$p=1$}&\multicolumn{1}{c}{$p=2$}&\multicolumn{1}{c}{$p=3$}&\multicolumn{1}{c}{$p=4$}\tabularnewline
\hline
{\bfseries Kernel Function}&&&&\tabularnewline
~~Uniform&$0.600$&$0.600$&$1.250$&$1.250$\tabularnewline
~~Triangular&$0.743$&$0.743$&$1.452$&$1.452$\tabularnewline
~~Epanechnikov&$0.714$&$0.714$&$1.407$&$1.407$\tabularnewline
\hline
{\bfseries MD Variance Bound}&$0.500$&$0.500$&$1.125$&$1.125$\tabularnewline
\hline
\end{tabular}
}}
\vskip1em
\subfloat[Density Derivative ${f}^{(1)}(\Evapointx)$]{\resizebox{0.55\columnwidth}{!}{
\begin{tabular}{lrrrr}
\hline\hline
\multicolumn{1}{l}{}&\multicolumn{1}{c}{$p=2$}&\multicolumn{1}{c}{$p=3$}&\multicolumn{1}{c}{$p=4$}&\multicolumn{1}{c}{$p=5$}\tabularnewline
\hline
{\bfseries Kernel Function}&&&&\tabularnewline
~~Uniform&$2.143$&$2.143$&$11.932$&$11.932$\tabularnewline
~~Triangular&$3.498$&$3.498$&$17.353$&$17.353$\tabularnewline
~~Epanechnikov&$3.182$&$3.182$&$15.970$&$15.970$\tabularnewline
\hline
{\bfseries MD Variance Bound}&$1.500$&$1.500$&$ 9.375$&$ 9.375$\tabularnewline
\hline
\end{tabular}
}}
\begin{flushleft}
\footnotesize
\textit{Notes}: Panel (a) compares asymptotic variance of the local polynomial density estimator of \cite*{Cattaneo-Jansson-Ma_2020_JASA} for different polynomial orders ($p=1$, $2$, $3$, and $4$) and different kernel functions (uniform, triangular and Epanechnikov). Also shown are the variance bound of the minimum distance estimator (MD Variance Bound), calculated according to Theorem \ref{thm:variance bound}. Panel(b) provides the same information for the estimated density derivative. All comparisons assume an interior evaluation point $\Evapointx$. 
\end{flushleft}
\end{table}

\begin{example}[Local $p=5$ Minimum Distance Estimation]\label{example:p=5 v=1 interior}
We consider the same setting in Example \ref{example:p=1 v=1 interior} and \ref{example:p=3 v=1 interior}, but with $p=5$: $P(u) = (u, \frac{1}{2}u^2,\cdots, \frac{1}{5!}u^5)^\Trans$. 

The variance bound in Theorem \ref{thm:variance bound} is 
\begin{align*}
 f(\Evapointx)\left(\int_{-1}^1 \dot{P}(u)\dot{P}(u)^\Trans \diff u\right)^{-1}  =  f(\Evapointx)
 \begin{bmatrix}
 \frac{225}{128} & 0 & -\frac{525}{32} & 0 & \frac{2835}{16} \\
  0 & \frac{75}{8} & 0 & -\frac{315}{4} & 0 \\
  -\frac{525}{32} & 0 & \frac{2205}{8} & 0 & -\frac{14175}{4} \\
  0 & -\frac{315}{4} & 0 & \frac{1575}{2} & 0 \\
  \frac{2835}{16} & 0 & -\frac{14175}{4} & 0 & \frac{99225}{2} \\
 \end{bmatrix}. 
\end{align*}

Again, we include two higher order polynomials: $Q(u)=(u^{2j}, u^{2j+1})^\Trans$. The asymptotic variance of our minimum distance estimator is
\begin{align*}
\mathrm{Asy}\Var\begin{bmatrix}
\hat{f}_{\mathtt{MD}}(\Evapointx)\\
\hat{f}_{\mathtt{MD}}^{(1)}(\Evapointx)\\
\hat{f}_{\mathtt{MD}}^{(2)}(\Evapointx)\\
\hat{f}_{\mathtt{MD}}^{(3)}(\Evapointx)\\
\hat{f}_{\mathtt{MD}}^{(4)}(\Evapointx)
\end{bmatrix} &= f(\Evapointx)\begin{bmatrix}
 \frac{225 (4 j+19)}{256 (2 j+9)} & 0 & -\frac{525 (4 j+21)}{64 (2 j+9)} & 0 & \frac{2835 (4 j+23)}{32 (2 j+9)} \\
  0 & \frac{75 (4 j+17)}{16 (2 j+7)} & 0 & -\frac{315 (4 j+19)}{8 (2 j+7)} & 0 \\
  -\frac{525 (4 j+21)}{64 (2 j+9)} & 0 & \frac{2205 (4 j+23)}{16 (2 j+9)} & 0 & -\frac{14175 (4 j+25)}{8 (2 j+9)} \\
  0 & -\frac{315 (4 j+19)}{8 (2 j+7)} & 0 & \frac{1575 (4 j+21)}{8 j+28} & 0 \\
  \frac{2835 (4 j+23)}{32 (2 j+9)} & 0 & -\frac{14175 (4 j+25)}{8 (2 j+9)} & 0 & \frac{99225 (4 j+27)}{8 j+36} \\
\end{bmatrix},
\end{align*}
which converges to the variance bound as $j\to \infty$. See also Table \ref{table:density variance} for the efficiency gain of employing the minimum distance technique. 
\qed
\end{example}

Before closing this section, we make several remarks on the variance bound derived in Theorem \ref{thm:variance bound}, as well as to what extent it is achievable. 

\begin{remark}[Achievability of the Variance Bound]
The previous two examples suggest that the variance bound derived in Theorem \ref{thm:variance bound} can be achieved by employing a minimum distance estimator with two additional regressors, one higher-order even polynomial and one higher-order odd polynomial. With analytic calculation, we verify that this is indeed the case for $p\leq 10$ when a uniform kernel function is used. 
\qed
\end{remark}

\begin{remark}[Optimality of the Variance Bound]
\cite{Granovsky-Muller_1991_ISR} discuss the problem of finding the optimal kernel function for kernel-type estimators. To be precise, consider the following
\begin{align*}
\frac{1}{nh^{\ell+1}}\sum_{i=1}^n\phi_{\ell,k}\left( \frac{x_i-\Evapointx}{h} \right),
\end{align*}
where $\phi_{\ell,k}(u)$ is a function satisfying
\begin{align*}
\int_{-1}^1 u^{j}\phi_{\ell,k}(u)\diff u &= \begin{cases}
0 & 0\leq j< k,\ j\neq \ell\\
\ell! & j=\ell
\end{cases},\qquad \int_{-1}^1 u^{k}\phi_{\ell,k}(u)\diff u\neq 0.
\end{align*}
Then it is easy to see that, with a Taylor expansion argument, 
\begin{align*}
\Expectation\left[\frac{1}{nh^{\ell+1}}\sum_{i=1}^n\phi_{\ell,k}\left( \frac{x_i-\Evapointx}{h} \right)\right] &= \frac{1}{h^{\ell+1}}\int_{-1}^1 \phi_{\ell,k}\left( \frac{u-\Evapointx}{h} \right)f(u)\diff u\\
&= \frac{1}{h^{\ell}}\int_{-1}^1 \phi_{\ell,k}\left( u \right)f(\Evapointx+hu)\diff u\\
&= \frac{1}{h^{\ell}}\int_{-1}^1 \phi_{\ell,k}\left( u \right)\left[\sum_{j=0}^{k-1} \frac{(hu)^j}{j!}f^{(j)}(\Evapointx) + u^kO(h^k) \right]\diff u\\
&= f^{(\ell)}(\Evapointx) + O(h^{k-\ell}). 
\end{align*}
That is, the kernel $\phi_{\ell,k}(u)$ facilitates estimating the $\ell$-th derivative of the density function with a leading bias of order $h^{k-\ell}$. Asymptotic variance of this kernel-type estimator is easily found to be
\begin{align*}
\mathrm{Asy}\Var\left[\frac{1}{nh^{\ell+1}}\sum_{i=1}^n\phi_{\ell,k}\left( \frac{x_i-\Evapointx}{h} \right)\right] &= f(\Evapointx)\int_{-1}^1 \phi_{\ell,k}(u)^2\diff u. 
\end{align*}
\cite{Granovsky-Muller_1991_ISR} provide the exact form of the kernel function $\phi_{\ell,k}(u)$ that minimizes the asymptotic variance subject to the order of the leading bias. 

Take $\ell=0$ and $k=2$, $\phi_{\ell,k}(u)$ takes the following form:
\begin{align*}
\phi_{\ell,k}(u) &= \frac{1}{2}\Indicator(|u|\leq 1),
\end{align*}
which is the uniform kernel and minimizes variance among all second order kernels for density estimation. As illustrated in Example \ref{example:p=1 v=1 interior}, our variance bound matches $f(\Evapointx)\int_{-1}^1 \phi_{\ell,k}(u)^2\diff u$.

Now take $\ell=1$ and $k=3$. This will give an estimator for the density derivative $f^{(1)}(\Evapointx)$ with a leading bias of order $O(h^2)$. The optimal choice of $\phi_{\ell,k}(u)$ is
\begin{align*}
\phi_{\ell,k}(u) &= \frac{3}{2}u\Indicator(|u|\leq 1). 
\end{align*}
to match the order of bias, we consider the minimum distance estimator with $p=3$. Again, the variance bound in Theorem \ref{thm:variance bound} matches $f(\Evapointx)\int_{-1}^1 \phi_{\ell,k}(u)^2\diff u$.

As a final illustration, take $\ell=1$ and $k=5$, which gives an estimator for the density derivative $f^{(1)}(\Evapointx)$ with a leading bias of order $O(h^4)$. The optimal choice of $\phi_{\ell,k}(u)$ is
\begin{align*}
\phi_{\ell,k}(u) &= \left(\frac{75}{8}u - \frac{105}{8}u^3\right)\Indicator(|u|\leq 1). 
\end{align*}
It is easy to see that $f(\Evapointx)\int_{-1}^1 \phi_{\ell,k}(u)^2\diff u = 75f(\Evapointx)/8$. To match the bias order, we take $p=5$ for our minimum distance estimator. The variance bound is $75f(\Evapointx)/8$, which is the same as $f(\Evapointx)\int_{-1}^1 \phi_{\ell,k}(u)^2\diff u$. 

With analytic calculations, we verify that the variance bound stated in Theorem \ref{thm:variance bound} is the same as the minimum variance found in \cite{Granovsky-Muller_1991_ISR}. Together with the previous remark, we reach a much stronger conclusion: including two higher-order polynomials in our minimum distance estimator can help achieve the variance bound in Theorem \ref{thm:variance bound}, which, in turn, is the smallest variance any kernel-type estimator can achieve (given a specific leading bias order). 
\qed
\end{remark}

\begin{remark}[Another Density Estimator Which Achieves the Variance Bound]
The following estimator achieves the bound of Theorem \ref{thm:variance bound}, although it does not belong to the class of estimators we consider in this paper.
\begin{align*}
\hat{\theta}_{\mathtt{ND}} &= \left( \int_{\mathcal{X}}\dot{P}(u-\Evapointx)\dot{P}(u-\Evapointx)^\Trans \frac{1}{h}K\left(\frac{u-\Evapointx}{h}\right) \diff u \right)^{-1}\left(\frac{1}{n}\sum_{i=1}^n \dot{P}(x_i-\Evapointx)\frac{1}{h}K\left(\frac{x_i-\Evapointx}{h}\right)\right),
\end{align*}
where $\dot{P}(u)=(1,u,u^2/2,\cdots,u^{p-1}/(p-1)!)^\Trans$ is the $(p-1)$-th order polynomial basis. The subscript represents ``numerical derivative,'' because the above estimator can be understood as
\begin{align*}
\hat{\theta}_{\mathtt{ND}} &= \left( \int_{\mathcal{X}}\dot{P}(u-\Evapointx)\dot{P}(u-\Evapointx)^\Trans \frac{1}{h}K\left(\frac{u-\Evapointx}{h}\right) \diff u \right)^{-1}\left(\int_{\mathcal{X}} \dot{P}(u-\Evapointx)\frac{1}{h}K\left(\frac{u-\Evapointx}{h}\right)\frac{\diff \hat{F}(u)}{\diff u}\diff u\right)\\
&= \argmin_{\theta}\int_{\mathcal{X}} \left( \frac{\diff \hat{F}(u)}{\diff u} - \dot{P}(u-\Evapointx)^\Trans \theta \right)^2 \frac{1}{h}K\left(\frac{u-\Evapointx}{h}\right)\diff u,
\end{align*}
where the derivative $\diff \hat{F}(u)/\diff u$ is interpreted in the sense of generalized functions. From the above, it is clear that this estimator requires the knowledge of the boundary position (that is, the knowledge of $\mathcal{X}$). 

With straightforward calculations, this estimator has a leading bias
\begin{align*}
\Expectation[\hat{\theta}_{\mathtt{ND}}] &= \left( \int_{\mathcal{X}}\dot{P}(u-\Evapointx)\dot{P}(u-\Evapointx)^\Trans \frac{1}{h}K\left(\frac{u-\Evapointx}{h}\right) \diff u \right)^{-1}\Expectation\left[ \dot{P}(x_i-\Evapointx)\frac{1}{h}K\left(\frac{x_i-\Evapointx}{h}\right) \right]\\
&= \theta + h^{p}\Upsilon_{h}f^{(p)}(\Evapointx)\left( \int_{\frac{\mathcal{X}-\Evapointx}{h}}\dot{P}\left(u\right)\dot{P}\left(u\right)^\Trans K\left(u\right) \diff u \right)^{-1} \int_{\frac{\mathcal{X}-\Evapointx}{h}}\dot{P}\left(u\right)u^pK\left(u\right) \diff u + o(h^{p}\Upsilon_{h}),
\end{align*}
where $\Upsilon_{h}$ is a diagonal matrix containing $1$, $h^{-1}$, $\cdots$, $h^{-(p-1)}$. Its leading variance is also easy to establish:
\begin{align*}
\Var[\hat{\theta}_{\mathtt{ND}}] &= \frac{1}{nh}\Upsilon_{h}f(\Evapointx)\left( \int_{\frac{\mathcal{X}-\Evapointx}{h}}\dot{P}\left(u\right)\dot{P}\left(u\right)^\Trans K\left(u\right) \diff u \right)^{-1} 
\left( \int_{\frac{\mathcal{X}-\Evapointx}{h}}\dot{P}\left(u\right)\dot{P}\left(u\right)^\Trans K\left(u\right)^2 \diff u \right)\\
&\qquad \qquad \qquad\qquad \cdot\left( \int_{\frac{\mathcal{X}-\Evapointx}{h}}\dot{P}\left(u\right)\dot{P}\left(u\right)^\Trans K\left(u\right) \diff u \right)^{-1}\Upsilon_{h}\\ 
&+ o\left(\frac{1}{nh}\Upsilon_{h}^2\right). 
\end{align*}
To reach the efficiency bound in Theorem \ref{thm:variance bound}, it suffices to set $K(\cdot)$ to be the uniform kernel. Section 5.1.1 in \cite{Loader_2006_book_Book}  also discussed this estimator, although it seems its efficiency property has not been realized in the literature. 
\qed
\end{remark}

\section{Uniform Distribution Theory}\label{section:Uniform Distribution Theory}

We establish distribution approximation for $\{\hat\theta_G(\Evapointx),\Evapointx\in\mathcal{I}\}$ and $\{\hat\theta(\Evapointx),\Evapointx\in\mathcal{I}\}$, which can be viewed as processes indexed by the evaluation point $\Evapointx$ in some set $\mathcal{I}\subseteq \mathcal{X}$. Recall the definition of $\Gamma_{h,\Evapointx}$ and $\Sigma_{h,\Evapointx}$ from Section \ref{section:setup}, and we define $\Omega_{h,\Evapointx} = \Gamma_{h,\Evapointx}^{-1}\Sigma_{h,\Evapointx}\Gamma_{h,\Evapointx}^{-1}$. 

We first study the following (infeasible) centered and Studentized process:
\begin{align}\label{eq:centered and studentized process}
\mathfrak{T}_G(\Evapointx) &=  \frac{1}{\sqrt{n}}\sum_{i=1}^n\frac{c_{h,\Evapointx}^\Trans\Upsilon_{h}\Gamma_{h,\Evapointx}^{-1}  \int_{\Intregion} R(u)\Big[\Indicator(x_i\leq \Evapointx + hu) - F(\Evapointx + hu)\Big] K\left(u\right) g(\Evapointx+hu)\diff u}{\sqrt{c_{h,\Evapointx}^\Trans\Upsilon_{h}\Omega_{h,\Evapointx}\Upsilon_{h}c_{h,\Evapointx}}},\ \Evapointx \in \mathcal{I},
\end{align}
where we consider linear combinations through a (known) vector $c_{h,\Evapointx}$, which can depend on the sample size through the bandwidth $h$, and can depend on the evaluation point. Again, we use the subscript $G$ to denote the local $L^2$ approach with $G$ being the design distribution. To economize notation, let
\begin{align*}
\mathcal{K}_{h,\Evapointx}(x) &= \frac{c_{h,\Evapointx}^\Trans\Upsilon_{h}\Gamma_{h,\Evapointx}^{-1}  \int_{\Intregion} R(u)\Big[\Indicator(x\leq \Evapointx + hu) - F(\Evapointx + hu)\Big] K\left(u\right) g(\Evapointx+hu)\diff u}{\sqrt{c_{h,\Evapointx}^\Trans\Upsilon_{h}\Omega_{h,\Evapointx}\Upsilon_{h}c_{h,\Evapointx}}},
\end{align*}
then we can conveniently rewrite \eqref{eq:centered and studentized process} as
\begin{align*}
\mathfrak{T}_G(\Evapointx) &= \frac{1}{\sqrt{n}}\sum_{i=1}^n \mathcal{K}_{h,\Evapointx}(x_i),
\end{align*}
and hence the centered and Studentized process $\mathfrak{T}_G(\cdot)$ takes a kernel form. The difference compared to standard kernel density estimators, however, is that the (equivalent) kernel in our case changes with the evaluation point, which is why our estimator is able to adapt to boundary points automatically. From the pointwise distribution theory developed in Section \ref{section:pointwise distribution theory}, the process $\mathfrak{T}_G(\Evapointx)$ has variance
\begin{align*}
\Var\left[ \mathfrak{T}_G(\Evapointx) \right] = \Expectation\left[ \mathcal{K}_{h,\Evapointx}(x_i)^2 \right] = 1. 
\end{align*}
We can also compute the covariance as
\begin{align*}
\Cov\left[ \mathfrak{T}_G(\Evapointx),\mathfrak{T}_G(\Evapointy) \right] = \Expectation\left[\mathcal{K}_{h,\Evapointx}(x_i)\mathcal{K}_{h,\Evapointy}(x_i)\right] = \frac{c_{h,\Evapointx}^\Trans\Upsilon_{h}\Omega_{h,\Evapointx,\Evapointy}\Upsilon_{h}c_{h,\Evapointy}}{\sqrt{c_{h,\Evapointx}^\Trans\Upsilon_{h}\Omega_{h,\Evapointx}\Upsilon_{h}c_{h,\Evapointx}}\sqrt{c_{h,\Evapointy}^\Trans\Upsilon_{h}\Omega_{h,\Evapointy}\Upsilon_{h}c_{h,\Evapointy}}} + O(h),
\end{align*}
where $\Omega_{h,\Evapointx,\Evapointy} = \Gamma_{h,\Evapointx}^{-1}\Sigma_{h,\Evapointx,\Evapointy}\Gamma_{h,\Evapointy}^{-1}$, and
\begin{align*}
\Sigma_{h,\Evapointx,\Evapointy} &= \int_{\frac{\mathcal{X}-\Evapointy}{h}}\int_{\Intregion} R(u)R(v)^\Trans  \Big[ F((\Evapointx+hu)\wedge (\Evapointy+hv)) - F(\Evapointx+hu)F(\Evapointy+hv) \Big]  \\
&\qquad \qquad \qquad \qquad K(u)K(v)g(\Evapointx+hu)g(\Evapointy+hv)\diff u\diff v.
\end{align*}
Of course one can further expand the above, but this is unnecessary for our purpose. 

For future reference, let 
\begin{align*}
r_1(\epsilon,h)&=\sup_{\Evapointx,\Evapointy\in \mathcal{I},|\Evapointx-\Evapointy|\leq \epsilon}\left|c^\Trans_{h,\Evapointx}\Upsilon_{h} - c^\Trans_{h,\Evapointy}\Upsilon_{h}\right|,\qquad r_2(h) = \sup_{\Evapointx\in\mathcal{I}} \frac{1}{|c_{h,\Evapointx}^\Trans\Upsilon_{h}|}.
\end{align*}

\begin{remark}[On the Order of $r_1(\epsilon,h)$, $r_2(h)$ and $\sup_{\Evapointx\in\mathcal{I}}\varrho(h,\Evapointx)$]\label{remark: order of some errors}
In general, it is not possible to give precise orders of the quantities introduced above. In this remark, we consider the local polynomial estimator of \cite*{Cattaneo-Jansson-Ma_2020_JASA} (see Section \ref{section:setup} for an introduction). The local polynomial estimator employs a polynomial basis, and hence estimates the density function and higher-order derivatives by (it also estimates the distribution function)
\begin{align*}
\hat{F}^{(\ell)}(\Evapointx) &= e_\ell^\Trans \hat{\theta}(\Evapointx),
\end{align*}
where $e_\ell$ is the $(\ell+1)$-th standard basis vector. As a result, $c_{h,\Evapointx} = e_\ell$, which does not depend on the evaluation point. For the scaling matrix $\Upsilon_{h}$, we note that it is diagonal with elements $1, h^{-1},\cdots, h^{-p}$, and hence it does not depend on the evaluation point either. Therefore, we conclude that, for density (and higher-order) derivative estimation using the local polynomial estimator, $r_1(\epsilon,h)$ is identically zero. Similarly, we have that $r_2(h) = h^{\ell}$. Finally, given the discussion in Section \ref{section:setup}, the bias term generally has order $\sup_{\Evapointx\in\mathcal{I}}\varrho(h,\Evapointx) = h^{p+1}$ for the local polynomial density estimator. 

The above discussion restricts to the local polynomial density estimator, but more can be said about $r_2(h)$. We will argue that, in general, one should expect $r_2(h) = O(1)$. Recall that the leading variance of $c_{h,\Evapointx}^\Trans \hat{\theta}(\Evapointx)$ and $c_{h,\Evapointx}^\Trans \hat{\theta}_G(\Evapointx)$ is $\frac{1}{n}c_{h,\Evapointx}^\Trans\Upsilon_{h}\Omega_{h,\Evapointx}\Upsilon_{h}c_{h,\Evapointx}$, and that the maximum eigenvalue of $\Omega_{h,\Evapointx}$ is bounded. Therefore, the variance has order $O(1/(nr_2(h)^2))$. In general, we do not expect the variance to shrink faster than $1/n$, which is why $r_2(h)$ is usually bounded. In fact, for most interesting cases, $c_{h,\Evapointx}^\Trans \hat{\theta}(\Evapointx)$ and $c_{h,\Evapointx}^\Trans \hat{\theta}_G(\Evapointx)$ will be ``nonparametric'' estimators in the sense that they estimate local features of the distribution function. If this is the case, we may even argue that $r_2(h)$ will be vanishing as the bandwidth shrinks. 
\qed
\end{remark}

We also make some additional assumptions. 

\begin{assumption}\label{assumption:uniform}
Let $\mathcal{I}$ be a compact interval.

(i) The density function is twice continuously differentiable and bounded away from zero in $\mathcal{I}$. 

(ii) There exists some $\delta>0$ and compactly supported kernel functions $K^{\dag}(\cdot)$ and $\{K^{\ddag,d}(\cdot) \}_{d\leq \delta}$, such that (ii.1) $\sup_{u\in\mathbb{R}}| K^{\dag}(u) |, \sup_{d\leq \delta,u\in\mathbb{R}} | K^{\ddag,d}(u) |<\infty$; (ii.2) the support of $K^{\ddag,d}(\cdot)$ has Lebesgue measure bounded by $Cd$, where $C$ is independent of $d$; and (ii.3)
for all $u$ and $v$ such that $|u-v|\leq \delta$, 
\begin{align*}
|K(u)-K(v)| \leq |u-v|\cdot K^{\dag}(u)  + K^{\ddag,|u-v|}(u).
\end{align*}

(iii) The basis function $R(\cdot)$ is Lipschitz continuous in $[-1,1]$.

(iv) For all $h$ sufficiently small, the minimum eigenvalues of $\Gamma_{h,\Evapointx}$ and $h^{-1}\Sigma_{h,\Evapointx}$ are bounded away from zero uniformly for $\Evapointx\in\mathcal{I}$. 

(v) $h\to 0$ and $nh/\log n\to\infty$ as $n\to \infty$.

(vi) For some $C_1>0$ and $C_2,\ C_3\geq0$, 
\begin{align*}
r_1(\epsilon,h) = O\left(\epsilon^{C_1}h^{-C_2}\right),\qquad
r_2(h)= O \left(h^{C_3}\right).
\end{align*}
In addition,
\begin{align*}
\frac{\sup_{\Evapointx\in\mathcal{I}}|c_{h,\Evapointx}^\Trans\Upsilon_{h}|}{\inf_{\Evapointx\in\mathcal{I}}|c_{h,\Evapointx}^\Trans\Upsilon_{h}|} = O(1).
\end{align*}
\vskip-2em\qed
\end{assumption}

\begin{assumption}\label{assumption:design uniform}
The design density function $g(\cdot)$ is twice continuously differentiable and is bounded away from zero in $\mathcal{I}$.
\qed
\end{assumption}

For any $h>0$ (and fixed $n$), we can define a centered Gaussian process, $\{\mathfrak{B}_G(\Evapointx):\Evapointx\in\mathcal{I}\}$, which has the same variance-covariance structure as the process $\mathfrak{T}_G(\cdot)$. The following lemma shows that it is possible to construct such a process, and that $\mathfrak{T}_G(\cdot)$ and $\mathfrak{B}_G(\cdot)$ are ``close in distribution.''
 
\begin{thm}[Strong Approximation]\label{thm:strong approximation}
Assume Assumptions \ref{assumption:dgp pointwise}, \ref{assumption:design pointwise}, \ref{assumption:uniform} and \ref{assumption:design uniform} hold. 
Then on a possibly enlarged probability space there exist two processes, $\{\tilde{\mathfrak{T}}_G(\Evapointx):\Evapointx\in\mathcal{I}\}$ and $\{\mathfrak{B}_G(\Evapointx):\Evapointx\in\mathcal{I}\}$, such that (i) $\tilde{\mathfrak{T}}_G(\cdot)$ has the same distribution as $\mathfrak{T}_G(\cdot)$; (ii) $\mathfrak{B}_G(\cdot)$ is a Gaussian process with the same covariance structure as $\mathfrak{T}_G(\cdot)$; and (iii)
\begin{align*}
\Prob\left[ \sup_{\Evapointx\in\mathcal{I}}\left| \tilde{\mathfrak{T}}_G(\Evapointx) - \mathfrak{B}_G(\Evapointx) \right| > \frac{C_4(u+C_5\log n)}{\sqrt{nh}} \right] \leq C_5e^{-C_5u},
\end{align*}
where $C_5$ is some constant that does not depend on $h$ or $n$. 
\qed
\end{thm}

Next we consider the continuity property of the implied (equivalent) kernel of the process $\mathfrak{T}_G(\cdot)$, which will help control the complexity of the Gaussian process $\mathfrak{B}_G(\cdot)$. To be precise, define the pseudo-metric $\sigma_G(\Evapointx,\Evapointy)$ as
\begin{align*}
\sigma_G(\Evapointx,\Evapointy) &= \sqrt{\Var\left[ \mathfrak{T}_G(\Evapointx) - \mathfrak{T}_G(\Evapointy) \right]} = \sqrt{\Expectation\left[ (\mathcal{K}_{h,\Evapointx}(x_i) - \mathcal{K}_{h,\Evapointy}(x_i))^2 \right]},
\end{align*}
we would like to provide an upper bound of $\sigma_G(\Evapointx,\Evapointy)$ in terms of $|\Evapointx-\Evapointy|$ (at least for all $\Evapointx$ and $\Evapointy$ such that $|\Evapointx-\Evapointy|$ is small enough). 

\begin{lem}[VC-type Property]\label{lem:VC-type}
Assume Assumptions \ref{assumption:dgp pointwise}, \ref{assumption:design pointwise}, \ref{assumption:uniform} and \ref{assumption:design uniform} hold. Then for all $\Evapointx,\Evapointy\in\mathcal{I}$ such that $|\Evapointx-\Evapointy|=\epsilon\leq h$,  
\begin{align*}
\sigma_G(\Evapointx,\Evapointy)  = O\left(  \frac{1}{\sqrt{h}}\frac{\epsilon}{h}  + \frac{1}{\sqrt{h}}r_1(\epsilon,h)r_2(h) + \frac{1}{h}r_1(\epsilon,h)^2r_2(h)^2 \right).
\end{align*}
Therefore, 
\begin{align*}
\Expectation\left[ \sup_{\Evapointx\in\mathcal{I}}|\mathfrak{B}_G(\Evapointx)| \right] = O\left(\sqrt{\log n}\right),\qquad \text{and}\qquad \Expectation\left[ \sup_{\Evapointx\in\mathcal{I}}|\mathfrak{T}_G(\Evapointx)| \right] = O\left(\sqrt{\log n} \right).
\end{align*}
\vskip-2em\qed
\end{lem}

\subsection{Local $L^2$ Distribution Estimation}

We first discuss the covariance estimator. For the local $L^2$ distribution estimator, let $\hat{\Omega}_{h,\Evapointx,\Evapointy} = \Gamma_{h,\Evapointx}^{-1}\hat\Sigma_{h,\Evapointx,\Evapointy}\Gamma_{h,\Evapointy}^{-1}$ with $\hat\Sigma_{h,\Evapointx,\Evapointy}$ given by
\begin{align*} \hat\Sigma_{h,\Evapointx,\Evapointy} &= \frac{1}{n}\sum_{i=1}^n\int_{\frac{\mathcal{X}-\Evapointy}{h}}\int_{\Intregion} R(u)R(v)^\Trans\Big[\Indicator(x_i\leq \Evapointx + hu) - \hat{F}(\Evapointx + hu)\Big]\Big[\Indicator(x_i\leq \Evapointy + hv) - \hat{F}(\Evapointy + hv)\Big] \\
&\ \qquad \qquad \qquad K(u)K(v) g(\Evapointx+hu)g(\Evapointy+hv)\diff u\diff v.
\end{align*}
The next lemma characterizes the convergence rate of $\hat{\Omega}_{h,\Evapointx,\Evapointy}$. 

\begin{lem}[Local $L^2$ Distribution Estimation: Covariance Estimation] \label{lemma:standard error, local projection uniform}
Assume Assumptions \ref{assumption:dgp pointwise}, \ref{assumption:design pointwise}, \ref{assumption:uniform} and \ref{assumption:design uniform} hold, and that $nh^2/\log n\to \infty$. Then
\begin{align*}
\sup_{\Evapointx,\Evapointy\in \mathcal{I}}\left|\frac{c^\Trans_{h,\Evapointx}\Upsilon_h(\hat{\Omega}_{h,\Evapointx,\Evapointy}-{\Omega}_{h,\Evapointx,\Evapointy})\Upsilon_h c_{h,\Evapointy} }{\sqrt{c^\Trans_{h,\Evapointx}\Upsilon_h{\Omega}_{h,\Evapointx}\Upsilon_h c_{h,\Evapointx}}\sqrt{c^\Trans_{h,\Evapointy}\Upsilon_h{\Omega}_{h,\Evapointy}\Upsilon_h c_{h,\Evapointy}}}\right| = \Op\left( \sqrt{\frac{\log n}{nh^2}} \right)
\end{align*}
\vskip-2em\qed
\end{lem} 

We now consider the estimator $c^\Trans_{h,\Evapointx}\hat\theta_G(\Evapointx)$. From \eqref{eq:approximation bias, projection} and \eqref{eq:linear variance, projection}, one has
\begin{align}
\nonumber T_G(\Evapointx) &= \frac{\sqrt{n}c_{h,\Evapointx}^\Trans\Big(\hat\theta_{G}(\Evapointx) - \theta(\Evapointx)\Big)}{\sqrt{c_{h,\Evapointx}^\Trans\Upsilon_{h}\hat{\Omega}_{h,\Evapointx}\Upsilon_{h}c_{h,\Evapointx}}}\\
\label{eq:local projection uniform term 1}&= \sqrt{n}\frac{c_{h,\Evapointx}^\Trans\Upsilon_{h}\Gamma_h^{-1}\int_{\Intregion} R(u)\Big[F(\Evapointx + hu) - \theta^\Trans R(u)\Upsilon_h^{-1}\Big] K\left(u\right) g(\Evapointx+hu)\diff u}{\sqrt{c_{h,\Evapointx}^\Trans\Upsilon_{h}\hat{\Omega}_{h,\Evapointx}\Upsilon_{h}c_{h,\Evapointx}}}\\
\label{eq:local projection uniform term 2}&\quad+  \frac{1}{\sqrt{n}}\sum_{i=1}^n\frac{c^\Trans_{h,\Evapointx}\Upsilon_{h}\Gamma_{h,\Evapointx}^{-1}  \int_{\Intregion} R(u)\Big[\Indicator(x_i\leq \Evapointx + hu) - F(\Evapointx + hu)\Big] K\left(u\right) g(\Evapointx+hu)\diff u}{\sqrt{c_{h,\Evapointx}^\Trans\Upsilon_{h}\hat{\Omega}_{h,\Evapointx}\Upsilon_{h}c_{h,\Evapointx}}}.
\end{align}

In the following lemma, we analyze the two terms in the above decomposition. 

\begin{lem}\label{lem:local projection uniform term 1}
Assume Assumptions \ref{assumption:dgp pointwise}, \ref{assumption:design pointwise}, \ref{assumption:uniform} and \ref{assumption:design uniform} hold, and that $nh^2/\log n\to \infty$. Then
\begin{align*}
\sup_{\Evapointx\in\mathcal{I}}\Big|\text{\eqref{eq:local projection uniform term 1}}\Big| &= \Op\left( \sqrt{\frac{n}{h}}\sup_{\Evapointx\in\mathcal{I}}\varrho(h,\Evapointx) \right),\qquad \sup_{\Evapointx\in\mathcal{I}}\Big|\text{\eqref{eq:local projection uniform term 2}} - \mathfrak{T}_G(\Evapointx)\Big| = \Op\left(  \frac{\log n}{\sqrt{nh^2}} \right).
\end{align*}
\vskip-2em\qed
\end{lem}

Now we state the main result on uniform distributional approximation.

\begin{thm}[Local $L^2$ Distribution Estimation: Uniform Distributional Approximation]\label{thm:strong approximation, local projection}
Assume Assumptions \ref{assumption:dgp pointwise}, \ref{assumption:design pointwise}, \ref{assumption:uniform} and \ref{assumption:design uniform} hold, and that $nh^2/\log n\to \infty$. Then on a possibly enlarged probability space there exist two processes, $\{\tilde{\mathfrak{T}}_G(\Evapointx):\Evapointx\in\mathcal{I}\}$ and $\{\mathfrak{B}_G(\Evapointx):\Evapointx\in\mathcal{I}\}$, such that (i) $\tilde{\mathfrak{T}}_G(\cdot)$ has the same distribution as $\mathfrak{T}_G(\cdot)$; (ii) $\mathfrak{B}_G(\cdot)$ is a Gaussian process with the same covariance structure as $\mathfrak{T}_G(\cdot)$; and (iii)
\begin{align*}
\sup_{\Evapointx\in\mathcal{I}}\Big| T_G(\Evapointx) - \mathfrak{T}_G(\Evapointx) \Big| + \sup_{\Evapointx\in\mathcal{I}}\Big| \tilde{\mathfrak{T}}_G(\Evapointx) - \mathfrak{B}_G(\Evapointx) \Big| = \Op\left(\frac{\log n}{\sqrt{nh^2}}+\sqrt{\frac{n}{h}}\sup_{\Evapointx\in\mathcal{I}}\varrho(h,\Evapointx)   \right).
\end{align*}
\vskip-2em\qed
\end{thm}

The following theorem shows that a feasible approximation to the process $\mathfrak{B}_G(\cdot)$ can be achieved by simulating a Gaussian process with covariance estimated from the data. In the following, we use $\Prob^\star$, $\Expectation^\star$ and $\Cov^\star$ to denote the probability, expectation and covariance operator conditioning on the data $X_n=(x_1,x_2,\dots,x_n)'$. 

\begin{thm}[Local $L^2$ Distribution Estimation: Feasible Distributional Approximation]\label{thm:feasible uniform approximation local projection}
Assume Assumptions \ref{assumption:dgp pointwise}, \ref{assumption:design pointwise}, \ref{assumption:uniform} and \ref{assumption:design uniform} hold, and that $nh^2/\log n\to \infty$. Then conditional on the data there exists a centered Gaussian process $\hat{\mathfrak{B}}_G(\cdot)$ with covariance
\begin{align*}
\Cov^\star\left[ \hat{\mathfrak{B}}_G(\Evapointx),\hat{\mathfrak{B}}_G(\Evapointy) \right] &= \frac{c_{h,\Evapointx}^\Trans\Upsilon_{h}\hat{\Omega}_{h,\Evapointx,\Evapointy}\Upsilon_{h}c_{h,\Evapointy}}{\sqrt{c_{h,\Evapointx}^\Trans\Upsilon_{h}\hat{\Omega}_{h,\Evapointx}\Upsilon_{h}c_{h,\Evapointx}}\sqrt{c_{h,\Evapointy}^\Trans\Upsilon_{h}\hat{\Omega}_{h,\Evapointy}\Upsilon_{h}c_{h,\Evapointy}}},
\end{align*}
such that
\begin{align*}
\sup_{u\in \mathbb{R}}\left| \Prob\Big[ \sup_{\Evapointx\in\mathcal{I}}|\mathfrak{B}_G(\Evapointx)| \leq u \Big]
-
\Prob^\star\Big[\sup_{\Evapointx\in\mathcal{I}}|\hat{\mathfrak{B}}_G(\Evapointx)| \leq u \Big] \right| = \Op\left(\left(\frac{\log^{5} n}{nh^2}\right)^\frac{1}{4} \right).
\end{align*}
\vskip-2em\qed
\end{thm}

\begin{remark}[On the Remainders in Theorems \ref{thm:strong approximation, local projection} and \ref{thm:feasible uniform approximation local projection}]\label{remark:bounding errors of strong approximation}
Recall that the local polynomial density estimator employs a polynomial basis, which implies that $\sup_{\Evapointx\in\mathcal{I}}\varrho(h,\Evapointx) = h^{p+1}$, where $p$ is the highest polynomial order. Then the error in Theorem \ref{thm:strong approximation, local projection} reduces to
\begin{align*}
\sqrt{nh^{2p+1}} + \frac{\log n}{\sqrt{nh^2}}.
\end{align*}
Therefore, a sufficient set of conditions for both errors to be negligible is $nh^{2p+1}\to 0$ and $nh^2/\log^5 n\to \infty$. 
\qed
\end{remark}

\subsection{Local Regression Distribution Estimation}

Now we consider the local regression estimator $\{\hat\theta(\Evapointx),\Evapointx\in\mathcal{I}\}$. As before, we first discuss the construction of the covariance $\Omega_{h,\Evapointx,\Evapointy}$. Let $\hat{\Omega}_{h,\Evapointx,\Evapointy} = \hat{\Gamma}_{h,\Evapointx}^{-1}\hat{\Sigma}_{h,\Evapointx,\Evapointy}\hat{\Gamma}_{h,\Evapointy}^{-1}$. Construction of $\hat{\Gamma}_{h,\Evapointx}$ is given in Section \ref{subsection:local regression: pointwise distribution theory}. The following lemma shows that $\hat{\Gamma}_{h,\Evapointx}$ is uniformly consistent. 

\begin{lem}[Uniform Consistency of $\hat{\Gamma}_{h,\Evapointx}$]\label{lem:uniform consistency of Gamma}
Assume Assumptions \ref{assumption:dgp pointwise} and \ref{assumption:uniform} hold. Then
\begin{align*}
\sup_{\Evapointx\in\mathcal{I}}\left|\hat{\Gamma}_{h,\Evapointx}-\Gamma_{h,\Evapointx} \right| = \Op\left(\sqrt{\frac{\log n}{nh}}\right).
\end{align*}
\vskip-2em\qed
\end{lem}

Construction of $\hat{\Sigma}_{h,\Evapointx,\Evapointy}$ also mimics that in Section \ref{subsection:local regression: pointwise distribution theory}. To be precise, we let
\begin{align*}
\hat{\Sigma}_{h,\Evapointx, \Evapointy} &= \frac{1}{n}\sum_{i=1}^n \left[\frac{1}{n}\sum_{j=1}^n \Upsilon_hR(x_j-\Evapointx)\Big[ \Indicator(x_i\leq x_j) -\hat{F}(x_j)  \Big] \frac{1}{h}K\left(\frac{x_j-\Evapointx}{h}\right)\right]\\
&\qquad \qquad \qquad \left[\frac{1}{n}\sum_{j=1}^n \Upsilon_hR(x_j-\Evapointy)\Big[ \Indicator(x_i\leq x_j) -\hat{F}(x_j)  \Big] \frac{1}{h}K\left(\frac{x_j-\Evapointy}{h}\right)\right]^\Trans. 
\end{align*}
where $\hat{F}(\cdot)$ remains to be the empirical distribution function. The following result justifies consistency of $\hat{\Omega}_{h,\Evapointx,\Evapointy}$. 

\begin{lem}[Local Regression Distribution Estimation: Covariance Estimation] \label{lemma:standard error, local regression uniform}
Assume Assumptions \ref{assumption:dgp pointwise} and \ref{assumption:uniform} hold, and that $nh^2/\log n\to \infty$. Then
\begin{align*}
\sup_{\Evapointx,\Evapointy\in \mathcal{I}}\left|\frac{c^\Trans_{h,\Evapointx}\Upsilon_h(\hat{\Omega}_{h,\Evapointx,\Evapointy}-{\Omega}_{h,\Evapointx,\Evapointy})\Upsilon_h c_{h,\Evapointy} }{\sqrt{c^\Trans_{h,\Evapointx}\Upsilon_h{\Omega}_{h,\Evapointx}\Upsilon_h c_{h,\Evapointx}}\sqrt{c^\Trans_{h,\Evapointy}\Upsilon_h{\Omega}_{h,\Evapointy}\Upsilon_h c_{h,\Evapointy}}}\right| = \Op\left( \sqrt{\frac{\log n}{nh^2}} \right)
\end{align*}
\vskip-2em\qed
\end{lem} 

The following is an expansion of $T(\cdot)$. 
\begin{align}
\nonumber T(\Evapointx) &= \frac{\sqrt{n}c_{h,\Evapointx}^\Trans\Big(\hat\theta(\Evapointx) - \theta(\Evapointx)\Big)}{\sqrt{c_{h,\Evapointx}^\Trans\Upsilon_{h}\hat{\Omega}_{h,\Evapointx}\Upsilon_{h}c_{h,\Evapointx}}}\\
\label{eq:local regression uniform term 1}&= \frac{1}{n\sqrt{n}}\sum_{i=1}^n \frac{c_{h,\Evapointx}^\Trans\Upsilon_{h}\hat{\Gamma}_{h,\Evapointx}^{-1}\Upsilon_{h}R(x_i-\Evapointx)[1-F(x_i)]\frac{1}{h}K(\frac{x_i-\Evapointx}{h})}{\sqrt{c_{h,\Evapointx}^\Trans\Upsilon_{h}\hat{\Omega}_{h,\Evapointx}\Upsilon_{h}c_{h,\Evapointx}}}\\
\label{eq:local regression uniform term 2}&+ \frac{1}{\sqrt{n}}\sum_{i=1}^n \frac{c_{h,\Evapointx}^\Trans\Upsilon_{h}\hat{\Gamma}_{h,\Evapointx}^{-1}\Upsilon_{h}R(x_i-\Evapointx)[F(x_i)-\theta(\Evapointx)^\Trans R(x_i-\Evapointx)]\frac{1}{h}K(\frac{x_i-\Evapointx}{h})}{\sqrt{c_{h,\Evapointx}^\Trans\Upsilon_{h}\hat{\Omega}_{h,\Evapointx}\Upsilon_{h}c_{h,\Evapointx}}}\\
\nonumber &+ \frac{1}{n\sqrt{n}}\sum_{i,j=1,i\neq j}^n \frac{1}{\sqrt{c_{h,\Evapointx}^\Trans\Upsilon_{h}\hat{\Omega}_{h,\Evapointx}\Upsilon_{h}c_{h,\Evapointx}}} \Bigg\{c_{h,\Evapointx}^\Trans\Upsilon_{h}\hat{\Gamma}_{h,\Evapointx}^{-1}\Upsilon_hR(x_j-\Evapointx)\Big[ \Indicator(x_i\leq x_j) -F(x_j)  \Big] \frac{1}{h}K\left(\frac{x_j-\Evapointx}{h}\right) \\
\label{eq:local regression uniform term 3}&\qquad \qquad \qquad\qquad- \int_{\Intregion} c_{h,\Evapointx}^\Trans\Upsilon_{h}\hat{\Gamma}_{h,\Evapointx}^{-1}R(u)\Big[\Indicator(x_i\leq \Evapointx + hu) - F(\Evapointx + hu)\Big] K\left(u\right) f(\Evapointx + hu)\diff u\Bigg\}\\
\label{eq:local regression uniform term 4}&+ \frac{n-1}{n\sqrt{n}}\sum_{i=1}^n\frac{c^\Trans_{h,\Evapointx}\Upsilon_{h}\hat{\Gamma}_{h,\Evapointx}^{-1}  \int_{\Intregion} R(u)\Big[\Indicator(x_i\leq \Evapointx + hu) - F(\Evapointx + hu)\Big] K\left(u\right) f(\Evapointx+hu)\diff u}{\sqrt{c_{h,\Evapointx}^\Trans\Upsilon_{h}\hat{\Omega}_{h,\Evapointx}\Upsilon_{h}c_{h,\Evapointx}}}.
\end{align} 

\begin{lem}\label{lem:local regression uniform term 1}
Assume Assumptions \ref{assumption:dgp pointwise} and \ref{assumption:uniform} hold, and that $nh^2/\log n\to \infty$. Then
\begin{align*}
\sup_{\Evapointx\in\mathcal{I}}\Big|\text{\eqref{eq:local regression uniform term 1}}\Big| &= \Op\left( \frac{1}{\sqrt{nh}}\right),\quad
\sup_{\Evapointx\in\mathcal{I}}\Big|\text{\eqref{eq:local regression uniform term 2}}\Big| = \Op\Big( \sqrt{\frac{n}{h}}\sup_{\Evapointx\in\mathcal{I}}\varrho(h,\Evapointx)\Big),\quad
\sup_{\Evapointx\in\mathcal{I}}\Big|\text{\eqref{eq:local regression uniform term 3}}\Big| = \Op\left( \frac{\log n}{\sqrt{nh^2}}\right).
\end{align*}
\vskip-2em\qed
\end{lem}

\begin{lem}\label{lem:local regression uniform term 2}
Assume Assumptions \ref{assumption:dgp pointwise} and \ref{assumption:uniform} hold, and that $nh^2/\log n\to \infty$. Then
\begin{align*}
\sup_{\Evapointx\in\mathcal{I}}\Big|\text{\eqref{eq:local regression uniform term 4}} - \mathfrak{T}_F(\Evapointx)\Big| &= \Op\left(  \frac{\log n}{\sqrt{nh^2}}\right).
\end{align*}
\vskip-2em\qed
\end{lem}

Finally we have the following result on uniform distributional approximation for the local regression distribution estimator, as well as a feasible approximation by simulating from a Gaussian process with estimated covariance.

\begin{thm}[Local Regression Distribution Estimation: Uniform Distributional Approximation]\label{thm:strong approximation, local regression}
Assume Assumptions \ref{assumption:dgp pointwise} and \ref{assumption:uniform} hold, and that $nh^2/\log n\to \infty$. Then on a possibly enlarged probability space there exist two processes, $\{\tilde{\mathfrak{T}}_F(\Evapointx):\Evapointx\in\mathcal{I}\}$ and $\{\mathfrak{B}_F(\Evapointx):\Evapointx\in\mathcal{I}\}$, such that (i) $\tilde{\mathfrak{T}}_F(\cdot)$ has the same distribution as $\mathfrak{T}_F(\cdot)$; (ii) $\mathfrak{B}_F(\cdot)$ is a Gaussian process with the same covariance structure as $\mathfrak{T}_F(\cdot)$; and (iii)
\begin{align*}
\sup_{\Evapointx\in\mathcal{I}}\Big| T(\Evapointx) - \mathfrak{T}_F(\Evapointx) \Big| + \sup_{\Evapointx\in\mathcal{I}}\Big| \tilde{\mathfrak{T}}_F(\Evapointx) - \mathfrak{B}_F(\Evapointx) \Big| = \Op\left(  \frac{\log n}{\sqrt{nh^2}}+\sqrt{\frac{n}{h}}\sup_{\Evapointx\in\mathcal{I}}\varrho(h,\Evapointx)   \right).
\end{align*}
\vskip-2em\qed
\end{thm}

\begin{thm}[Local Regression Distribution Estimation: Feasible Distributional Approximation]\label{thm:feasible uniform approximation local regression}
Assume Assumptions \ref{assumption:dgp pointwise} and \ref{assumption:uniform} hold, and that $nh^2/\log n\to \infty$. Then conditional on the data there exists a centered Gaussian process $\hat{\mathfrak{B}}_F(\cdot)$ with covariance
\begin{align*}
\Cov^\star\left[ \hat{\mathfrak{B}}_F(\Evapointx),\hat{\mathfrak{B}}_F(\Evapointy) \right] &= \frac{c_{h,\Evapointx}^\Trans\Upsilon_{h}\hat{\Omega}_{h,\Evapointx,\Evapointy}\Upsilon_{h}c_{h,\Evapointy}}{\sqrt{c_{h,\Evapointx}^\Trans\Upsilon_{h}\hat{\Omega}_{h,\Evapointx}\Upsilon_{h}c_{h,\Evapointx}}\sqrt{c_{h,\Evapointy}^\Trans\Upsilon_{h}\hat{\Omega}_{h,\Evapointy}\Upsilon_{h}c_{h,\Evapointy}}},
\end{align*}
such that
\begin{align*}
\sup_{u\in \mathbb{R}}\left| \Prob\Big[ \sup_{\Evapointx\in\mathcal{I}}|\mathfrak{B}_F(\Evapointx)| \leq u \Big]
-
\Prob^\star\Big[\sup_{\Evapointx\in\mathcal{I}}|\hat{\mathfrak{B}}_F(\Evapointx)| \leq u \Big] \right| = \Op\left(\left(\frac{\log^{5} n}{nh^2}\right)^{\frac{1}{4}} \right).
\end{align*}
\vskip-2em\qed
\end{thm}

\clearpage
{\footnotesize

\section{Proofs}

\subsection{Proof of Theorem \ref{thm:local projection: asymptotic normality}}

\subsubsection*{Part (i)}
The bias term can be bounded by
\begin{align*}
\left|\int_{\Intregion} R(u)\Big[F(\Evapointx + hu) - \theta^\Trans R(u)\Upsilon_h^{-1}\Big] K\left(u\right) \diff u\right| &\leq \sup_{u\in[-1,1]}\Big|F(\Evapointx + hu) - \theta^\Trans R(u)\Upsilon_h^{-1}\Big|\int_{\Intregion} |R(u)| K\left(u\right) \diff u\\
&= \varrho(h)\int_{\Intregion} |R(u)| K\left(u\right) \diff u.
\end{align*}

\subsubsection*{Part (ii)}
The variance can be found as
\begin{align*}
&\ \Var\left[ \frac{1}{\sqrt{n}}\sum_{i=1}^n  \int_{\Intregion} R(u)\Big[\Indicator(x_i\leq \Evapointx + hu) - F(\Evapointx + hu)\Big] K\left(u\right) g(\Evapointx+hu)\diff u \right]\\
=&\ \iint_{\Intregion} R(u)R(v)^\Trans K(u)K(v) \Big[ F(\Evapointx+h(u\wedge v)) - F(\Evapointx+hu)F(\Evapointx+hv) \Big]  g(\Evapointx+hu)g(\Evapointx+hv)\diff u\diff v.
\end{align*}

To establish asymptotic normality, we verify the Lyapunov condition with a fourth moment calculation. Take $c$ to be a nonzero vector of conformable dimension, and we employ the Cramer-Wold device:
\begin{align*}
\frac{1}{n}\left(c^\Trans \Sigma_h c\right)^{-2}\Expectation\left[  \left(\int_{\Intregion} c^\Trans R(u)\Big[\Indicator(x_i\leq \Evapointx + hu) - F(\Evapointx + hu)\Big] K\left(u\right) g(\Evapointx+hu)\diff u\right)^4\right].
\end{align*}
If $c^\Trans \Sigma_h c$ is bounded away from zero as the bandwidth decreases, the above will have order $n^{-1}$, as $K(\cdot)$ is bounded and compactly supported and $R(\cdot)$ is locally bounded. Therefore, the Lyapunov condition holds in this case. The more challenging case is when $c^\Trans \Sigma_h c$ is of order $h$. In this case, it implies 
\begin{align*}
F(\Evapointx)(1-F(\Evapointx)) \left|\iint_{\Intregion} c^\Trans R(u)  K(u)g_u  \diff u\right|^2 = O(h).
\end{align*}
Now consider the fourth moment. The leading term is
\begin{align*}
F(\Evapointx)(1-F(\Evapointx))(3F(\Evapointx)^2-3F(\Evapointx)+1)\left|\int_{\Intregion} c^\Trans R(u) K(u)  g(\Evapointx+hu)\diff u\right|^4 = O(h),
\end{align*}
meaning that for the Lyapunov condition to hold, we need the requirement that $nh\to \infty$. 

\subsubsection*{Part (iii)}
This follows immediately from Part (i) and (ii).

\subsection{Proof of Theorem \ref{thm:local projection: standard error pointwise}}

To study the property of $\hat\Sigma_{h}$, we make the following decomposition:
\begin{align*}
\tag{I}\hat\Sigma_{h} &= \frac{1}{n}\sum_{i=1}^n\iint_{\Intregion} R(u)R(v)^\Trans\Big[\Indicator(x_i\leq \Evapointx + hu) - {F}(\Evapointx + hu)\Big]\Big[\Indicator(x_i\leq \Evapointx + hv) - {F}(\Evapointx + hv)\Big] K(u)K(v) g(\Evapointx+hu)g(\Evapointx+hv)\diff u\diff v\\
\tag{II}&- \iint_{\Intregion} R(u)R(v)^\Trans\Big[\hat{F}( \Evapointx + hu) - {F}(\Evapointx + hu)\Big]\Big[\hat{F}(\Evapointx + hv) - {F}(\Evapointx + hv)\Big] K(u)K(v) g(\Evapointx+hu)g(\Evapointx+hv)\diff u\diff v.
\end{align*}

First, it is obvious that term (II) is of order $\Op(1/n)$. Term (I) requires more delicate analysis. Let $c$ be a vector of unit length and suitable dimension, and define
\begin{align*}
c_i &= \iint_{\Intregion} c^\Trans R(u)R(v)^\Trans c\Big[\Indicator(x_i\leq \Evapointx + hu) - {F}(\Evapointx + hu)\Big]\Big[\Indicator(x_i\leq \Evapointx + hv) - {F}(\Evapointx + hv)\Big] K(u)K(v) g(\Evapointx+hu)g(\Evapointx+hv)\diff u\diff v. 
\end{align*}
Then 
\begin{align*}
c^\Trans \text{(I)} c &= \Expectation[c^\Trans \text{(I)} c] + \Op\left(\sqrt{\Var[c^\Trans \text{(I)} c]}\right) = \Expectation[c_i] + \Op\left(\sqrt{\frac{1}{n}\Big(\Expectation[c_i^2] - (\Expectation[c_i])^2 \Big)}\right),
\end{align*}
which implies that 
\begin{align*}
\frac{c^\Trans \text{(I)} c}{\Expectation[c^\Trans \text{(I)} c]} - 1 &=  \Op\left(\sqrt{\frac{1}{n}\left(\frac{\Expectation[c_i^2]}{(\Expectation[c_i])^2} - 1 \right)}\right).
\end{align*}
With the same argument used in the proof of Theorem \ref{thm:local projection: asymptotic normality}, one can show that
\begin{align*}
\frac{\Expectation[c_i^2]}{(\Expectation[c_i])^2} = O\left(\frac{1}{h}\right),
\end{align*}
which implies
\begin{align*}
\frac{c^\Trans \text{(I)} c}{c^\Trans\Sigma_{h} c} - 1 &= \Op\left(  \sqrt{\frac{1}{nh}}\right).
\end{align*}

\subsection{Proof of Theorem \ref{thm:local regression: asymptotic normality}}

\subsubsection*{Part (i)}

For the ``denominator,'' its variance is bounded by
\begin{align*}
&\ \left|\Var\left[\frac{1}{n}\sum_{i=1}^n\Upsilon_hR(x_i-\Evapointx)R(x_i-\Evapointx)^\Trans \Upsilon_h \frac{1}{h}K\left(\frac{x_i-\Evapointx}{h}\right)\right]\right|
\leq \frac{1}{n}\Expectation\left[ \left|\Upsilon_hR(x_i-\Evapointx)R(x_i-\Evapointx)^\Trans \Upsilon_h\right|^2 \frac{1}{h^2}K\left(\frac{x_i-\Evapointx}{h}\right)^2 \right]\\
=&\ \frac{1}{n}\int_{\mathcal{X}} \left|\Upsilon_hR(u-\Evapointx)R(u-\Evapointx)^\Trans \Upsilon_h\right|^2 \frac{1}{h^2}K\left(\frac{u-\Evapointx}{h}\right)^2 f(u)\diff u
= \frac{1}{nh}\int_{\Intregion} \left|R(u)R(u)^\Trans \right|^2 K\left(u\right)^2 f(\Evapointx+hu)\diff u\\
=&\ O\left(\frac{1}{nh}\right).
\end{align*}
Therefore, under the assumption that $h\to 0$ and $nh\to \infty$, we have
\begin{align*}
\left|\frac{1}{n}\sum_{i=1}^n\Upsilon_hR(x_i-\Evapointx)R(x_i-\Evapointx)^\Trans \Upsilon_h \frac{1}{h}K\left(\frac{x_i-\Evapointx}{h}\right)\ -\ \Gamma_h\right| = \Op\left(\sqrt{\frac{1}{nh}} \right),
\end{align*}
which further implies that
\begin{align*}
\hat\theta - \theta &= \Upsilon_h\Gamma_h^{-1}\left( \frac{1}{n}\sum_{i=1}^n \Upsilon_hR(x_i-\Evapointx)\Big[ \hat{F}(x_i) -R(x_i-\Evapointx)^\Trans\theta_0  \Big] \frac{1}{h}K\left(\frac{x_i-\Evapointx}{h}\right)\right)\big(1 + \op(1)\big).
\end{align*}

\subsubsection*{Part (ii)}

The order of the leave-in bias is clearly $1/n$. For the approximation bias \eqref{eq:approximation bias}, we obtained its mean in the proof of Theorem \ref{thm:local projection: asymptotic normality} by setting $G=F$, which has an order of $\varrho(h)$. The approximation bias has a variance of order
\begin{align*}
&\ \left|\Var\left[\frac{1}{n}\sum_{j=1}^n \Upsilon_hR(x_j-\Evapointx)\Big[ F(x_j) -R(x_j-\Evapointx)^\Trans\theta_0  \Big] \frac{1}{h}K\left(\frac{x_j-\Evapointx}{h}\right)\right]\right|\\
\leq&\ \frac{1}{n}\Expectation\left[ \left|\Upsilon_hR(x_j-\Evapointx)\Big[ F(x_j) -R(x_j-\Evapointx)^\Trans\theta_0  \Big] \frac{1}{h}K\left(\frac{x_j-\Evapointx}{h}\right)\right|^2\right]\\
=&\ \frac{1}{n}\int_{\mathcal{X}}\left|\Upsilon_hR(u-\Evapointx)\Big[ F(u) -R(u-\Evapointx)^\Trans\theta_0  \Big]\right|^2 \frac{1}{h^2}K\left(\frac{u-\Evapointx}{h}\right)^2f(u)\diff u\\
=&\ \frac{1}{nh}\int_{\Intregion}\left|R(u)\Big[ F(\Evapointx + hu) -R(u-\Evapointx)^\Trans\theta_0  \Big]\right|^2 K(u)^2 f(\Evapointx + hu)\diff u\\
\leq&\ \frac{1}{nh}\varrho(h)^2\int_{\Intregion}|R(u)|^2 K(u)^2 f(\Evapointx + hu)\diff u
=\ O\left(\frac{\rho(h)^2}{nh}\right).
\end{align*}
Therefore, 
\begin{align*}
\text{\eqref{eq:approximation bias}} &= \Op\left( \varrho(h) +\varrho(h)\sqrt{\frac{1}{nh}}   \right) = \Op(\varrho(h)),
\end{align*}
provided that $nh\to \infty$. 

\subsubsection*{Part (iii)}

We compute the variance of the U-statistic \eqref{eq:quadratic variance}. For simplicity, define
\begin{align*}
u_{ij} &= \Upsilon_hR(x_j-\Evapointx)\Big[ \Indicator(x_i\leq x_j) -F(x_j)  \Big] \frac{1}{h}K\left(\frac{x_j-\Evapointx}{h}\right) - \int_{\Intregion} R(u)\Big[\Indicator(x_i\leq \Evapointx + hu) - F(\Evapointx + hu)\Big] K\left(u\right) f(\Evapointx + hu)\diff u,
\end{align*}
which satisfies $\Expectation[u_{ij}]=\Expectation[u_{ij}|x_i]=\Expectation[u_{ij}|x_j]=0$. Therefore
\begin{align*}
\Var\left[ \text{\eqref{eq:quadratic variance}}  \right] =& \frac{1}{n^4}\sum_{i,j=1,i\neq j}^n\sum_{i',j'=1,i\neq j}^n \Expectation\left[u_{ij}u_{i'j'}^\Trans\right]
= \frac{1}{n^4}\sum_{i,j=1,i\neq j}^n \Expectation\left[u_{ij}u_{ij}^\Trans\right] + \Expectation\left[u_{ij}u_{ji}^\Trans\right], 
\end{align*}
meaning that
\begin{align*}
\text{\eqref{eq:quadratic variance}} &= \Op\left( \sqrt{\frac{1}{n^2h}} \right).
\end{align*}

\subsubsection*{Part (iv)}

This follows immediately from Part (i)--(iii) and Theorem \ref{thm:local projection: asymptotic normality}. 

\subsection{Proof of Theorem \ref{thm:local regression: standard error pointwise}}

We first decompose $\hat{\Sigma}_{h}$ into two terms, 
\begin{align*}
\text{(I)} &= \frac{1}{n^3}\sum_{i,j,k=1}^n \Upsilon_{h}R_jR_k^\Trans \Upsilon_{h} W_jW_k \Big(\Indicator(x_i\leq x_j) - F(x_j)\Big) \Big(\Indicator(x_i\leq x_k) - F(x_k)\Big)\\
\text{(II)} &= -\frac{1}{n^2}\sum_{j,k=1}^n \Upsilon_{h}R_jR_k^\Trans \Upsilon_{h} W_jW_k \Big(\hat{F}(x_j)-{F}(x_j)\Big) \Big(\hat{F}(x_k) - {F}(x_k)\Big),
\end{align*}
where we use $R_i = R(x_i-\Evapointx)$ and $W_i = K((x_i-\Evapointx)/h)/h$ to conserve space. 

(II) satisfies
\begin{align*}
\left|\text{(II)}\right| \leq \sup_{x}|\hat{F}(x) - F(x)|^2\frac{1}{n^2}\sum_{j,k=1}^n \left| \Upsilon_{h}R_jR_k^\Trans \Upsilon_{h} W_jW_k\right|.
\end{align*}
It is obvious that 
\begin{align*}
\sup_{x}|\hat{F}(x) - F(x)|^2 = \Op\left( \frac{1}{n}\right). 
\end{align*}
As for the second part, we have
\begin{align*}
\frac{1}{n^2}\sum_{j,k=1}^n\Expectation\left[ \left| \Upsilon_{h}R_jR_k^\Trans \Upsilon_{h} W_jW_k\right| \right]
=&  \frac{n-1}{n}\Expectation\Big[ \left| \Upsilon_{h}R_jR_k^\Trans \Upsilon_{h} W_jW_k\right|\ \Big| j\neq k \Big] + \frac{1}{n}\Expectation\Big[ \left| \Upsilon_{h}R_kR_k^\Trans \Upsilon_{h} W_kW_k\right| \Big]\\
=& \Op\left( 1 + \frac{1}{nh}\right) = \Op\left(1\right), 
\end{align*}
which holds as long as $nh\to \infty$. Then it further implies that
\begin{align*}
\text{(II)} = \Op\left(\frac{1}{n}\right).
\end{align*}

To analyze (I), we further expand this term into ``diagonal'' and ``non-diagonal'' sums:
\begin{align*}
\tag{I.1}\text{(I)} &= \frac{1}{n^3}\sum_{\substack{i,j,k=1 \\ \text{distinct}}}^n \Upsilon_{h}R_jR_k^\Trans \Upsilon_{h} W_jW_k \Big(\Indicator(x_i\leq x_j) - F(x_j)\Big) \Big(\Indicator(x_i\leq x_k) - F(x_k)\Big)\\
\tag{I.2}&\quad + \frac{1}{n^3}\sum_{\substack{i,k=1 \\ \text{distinct}}}^n \Upsilon_{h}R_iR_k^\Trans \Upsilon_{h} W_iW_k \Big(\Indicator(x_i\leq x_i) - F(x_i)\Big) \Big(\Indicator(x_i\leq x_k) - F(x_k)\Big)\\
\tag{I.3}&\quad + \frac{1}{n^3}\sum_{\substack{i,j=1 \\ \text{distinct}}}^n \Upsilon_{h}R_jR_i^\Trans \Upsilon_{h} W_jW_i \Big(\Indicator(x_i\leq x_j) - F(x_j)\Big) \Big(\Indicator(x_i\leq x_i) - F(x_i)\Big)\\
\tag{I.4}&\quad + \frac{1}{n^3}\sum_{\substack{i,j=1 \\ \text{distinct}}}^n \Upsilon_{h}R_jR_j^\Trans \Upsilon_{h} W_jW_j \Big(\Indicator(x_i\leq x_j) - F(x_j)\Big) \Big(\Indicator(x_i\leq x_j) - F(x_j)\Big)\\
\tag{I.5}&\quad + \frac{1}{n^3}\sum_{i} \Upsilon_{h}R_iR_i^\Trans \Upsilon_{h} W_iW_i \Big(\Indicator(x_i\leq x_i) - F(x_i)\Big) \Big(\Indicator(x_i\leq x_i) - F(x_i)\Big).
\end{align*}
By calculating the expectation of the absolute value of the summands above, it is straightforward to show
\begin{align*}
\text{(I.2)} = \Op\left(\frac{1}{n}\right),\quad 
\text{(I.3)} = \Op\left(\frac{1}{n}\right),\quad 
\text{(I.4)} = \Op\left(\frac{1}{nh}\right),\quad 
\text{(I.5)} = \Op\left(\frac{1}{n^2h}\right).
\end{align*}
Therefore, we have
\begin{align*}
\hat{\Sigma}_{h} &= \text{(I.1)} + \Op\left(\frac{1}{nh}\right)
= \frac{1}{n^3}\sum_{\substack{i,j,k=1 \\ \text{distinct}}}^n \Upsilon_{h}R_jR_k^\Trans \Upsilon_{h} W_jW_k \Big(\Indicator(x_i\leq x_j) - F(x_j)\Big) \Big(\Indicator(x_i\leq x_k) - F(x_k)\Big) + \Op\left(\frac{1}{nh}\right).
\end{align*}

To proceed, define 
\begin{align*}
u_{ij} &= \Upsilon_{h}R_j W_j \Big(\Indicator(x_i\leq x_j) - F(x_j)\Big)\quad \text{and}\quad \bar{u}_i = \Expectation[u_{ij}|x_i;i\neq j]. 
\end{align*}
Then we can further decompose (I.1) into
\begin{align*}
\text{(I.1)} &= \frac{1}{n^3}\sum_{\substack{i,j,k=1 \\ \text{distinct}}}^n u_{ij}u_{ik}^\Trans
= \frac{1}{n^3}\sum_{\substack{i,j,k=1 \\ \text{distinct}}}^n \Expectation[u_{ij}u_{ik}^\Trans|x_i] 
\quad + \frac{1}{n^3}\sum_{\substack{i,j,k=1 \\ \text{distinct}}}^n \Big(u_{ij}u_{ik}^\Trans - \Expectation[u_{ij}u_{ik}^\Trans|x_i]\Big)\\
&= \underbrace{\frac{(n-1)(n-2)}{n^3}\sum_{i=1}^n \bar{u}_i\bar{u}_i^\Trans}_{\textstyle \text{(I.1.1)}}
\ +\ \underbrace{\frac{1}{n^3}\sum_{\substack{i,j,k=1 \\ \text{distinct}}}^n \Big(u_{ij}u_{ik}^\Trans - \bar{u}_i\bar{u}_i^\Trans\Big)}_{\textstyle \text{(I.1.2)}}.
\end{align*}
We have already analyzed (I.1.1) in Theorem \ref{thm:local projection: standard error pointwise}, which suggests
\begin{align*}
\text{(I.1.1)}&= \Sigma_{h} + \Op\left( \frac{1}{\sqrt{n}} \right). 
\end{align*}
Now we study (I.1.2), which satisfies
\begin{align*}
\text{(I.1.2)} &= \underbrace{\frac{n-2}{n^3}\sum_{\substack{i,j=1 \\ \text{distinct}}}^n \Big(u_{ij} - \bar{u}_i\Big)\bar{u}_i^\Trans}_{\textstyle \text{(I.1.2.1)}}
\ +\ \underbrace{\frac{n-2}{n^3}\sum_{\substack{i,j=1 \\ \text{distinct}}}^n \bar{u}_i\Big(u_{ij} - \bar{u}_i\Big)^\Trans}_{\textstyle \text{(I.1.2.2)}}
\ +\ \underbrace{\frac{1}{n^3}\sum_{\substack{i,j,k=1 \\ \text{distinct}}}^n \Big(u_{ij} - \bar{u}_i\Big)\Big(u_{ik} - \bar{u}_i\Big)^\Trans}_{\textstyle \text{(I.1.2.3)}}.
\end{align*}
With variance calculation, it is easy to see that 
\begin{align*}
\text{(I.1.2.3)} &= \Op\left( \frac{1}{nh} \right). 
\end{align*}
Therefore we have
\begin{align*}
\frac{c^\Trans (\hat\Sigma_{h} - \Sigma_{h}) c}{c^\Trans \Sigma_{h} c} &= \Op\left( \frac{1}{\sqrt{nh^2}} \right) + 2\frac{c^\Trans \text{(I.1.2.1)} c}{c^\Trans \Sigma_{h} c},
\end{align*}
since (I.1.2.1) and (I.1.2.2) are transpose of each other. To close the proof, we calculate the variance of the last term in the above. 
\begin{align*}
\Var\left[ \frac{c^\Trans \text{(I.1.2.1)} c}{c^\Trans \Sigma_{h} c} \right] &= \frac{1}{(c^\Trans \Sigma_{h} c)^2}\frac{(n-2)^2}{n^6}\Expectation\left[\sum_{\substack{i,j=1 \\ \text{distinct}}}^n\sum_{\substack{i',j'=1 \\ \text{distinct}}}^n c^\Trans\Big(u_{ij} - \bar{u}_i\Big)\bar{u}_i^\Trans c c^\Trans\Big(u_{i'j'} - \bar{u}_{i'}\Big)\bar{u}_{i'}^\Trans c \right]\\
&= \frac{1}{(c^\Trans \Sigma_{h} c)^2}\frac{(n-2)^2}{n^6}\Expectation\left[\sum_{\substack{i,j,i'=1 \\ \text{distinct}}}^n c^\Trans u_{ij}\bar{u}_i^\Trans c c^\Trans u_{i'j}\bar{u}_{i'}^\Trans c \right] + \text{higher order terms}.
\end{align*}
The expectation is further given by (note that $i$, $j$ and $i'$ are assumed to be distinct indices)
\begin{align*}
&\ \Expectation\left[ c^\Trans u_{ij}\bar{u}_i^\Trans c c^\Trans u_{i'j}\bar{u}_{i'}^\Trans c \right]\\
=&\ \Expectation\iint_{\Intregion}  W_j^2\left[c^\Trans \Upsilon_h R_jR(u) c c^\Trans \Upsilon_h  R_jR(v) c\right]  K(u)  K(v) \\
&\qquad \left[ F(x_j\wedge (\Evapointx + hu)) - F(x_j)F(\Evapointx + hu) \right]\left[ F(x_j\wedge (\Evapointx + hv)) - F(x_j)F(\Evapointx + hv) \right]f(\Evapointx + hu)f(\Evapointx + hv)\diff u \diff v\\
=&\ \frac{1}{h}\iiint_{\Intregion}  \left[c^\Trans R(w)R(u) c c^\Trans R(w)R(v) c\right]  K(u)  K(v) K(w)^2 \\
&\qquad \left[ F(\Evapointx + h(w\wedge u)) - F(\Evapointx + hw)F(\Evapointx + hu) \right]\left[ F(\Evapointx + h(w\wedge v)) - F(\Evapointx + hw)F(\Evapointx + hv) \right]f(\Evapointx + hw)f(\Evapointx + hu)f(\Evapointx + hv)\diff w\diff u \diff v\\
=&\ \frac{1}{h}F(\Evapointx)^2(1-F(\Evapointx))^2\iiint_{\Intregion}  \left[c^\Trans R(w)R(u) c c^\Trans R(w)R(v) c\right]  K(u)  K(v) K(w)^2  f_wf_uf_v\diff w\diff u \diff v + \text{higher-order terms}.
\end{align*}
If $c^\Trans \Sigma_{h} c = O(1)$, then the above will have order $h$, which means 
\begin{align*}
\Var\left[ \frac{c^\Trans \text{(I.1.2.1)} c}{c^\Trans \Sigma_{h} c} \right] &= O\left(\frac{1}{nh}\right). 
\end{align*}
If $c^\Trans \Sigma_{h} c = O(h)$, however, $\Expectation\left[ c^\Trans u_{ij}\bar{u}_i^\Trans c c^\Trans u_{i'j}\bar{u}_{i'}^\Trans c \right]$ will be $O(1)$, which will imply that
\begin{align*}
\Var\left[ \frac{c^\Trans \text{(I.1.2.1)} c}{c^\Trans \Sigma_{h} c} \right] &= O\left(\frac{1}{nh^2}\right). 
\end{align*}
As a result, we have
\begin{align*}
\frac{c^\Trans (\hat\Sigma_{h} - \Sigma_{h}) c}{c^\Trans \Sigma_{h} c} &= \Op\left( \frac{1}{\sqrt{nh^2}} \right).
\end{align*}

Now consider
\begin{align*}
\frac{c^\Trans \hat\Gamma_{h}^{-1} \hat\Sigma_{h} \hat\Gamma_{h}^{-1} c}{c^\Trans \Gamma_{h}^{-1} \Sigma_{h} \Gamma_{h}^{-1} c} - 1 &= \frac{c^\Trans \hat\Gamma_{h}^{-1} (\hat\Sigma_{h} - \Sigma_h) \hat\Gamma_{h}^{-1} c}{c^\Trans \Gamma_{h}^{-1} \Sigma_{h} \Gamma_{h}^{-1} c} 
+ \frac{c^\Trans (\hat\Gamma_{h}^{-1} -\Gamma_{h}^{-1})  \Sigma_h \hat\Gamma_{h}^{-1} c}{c^\Trans \Gamma_{h}^{-1} \Sigma_{h} \Gamma_{h}^{-1} c} 
+ \frac{c^\Trans (\hat\Gamma_{h}^{-1} -\Gamma_{h}^{-1})  \Sigma_h \Gamma_{h}^{-1} c}{c^\Trans \Gamma_{h}^{-1} \Sigma_{h} \Gamma_{h}^{-1} c}\\
&= \frac{c^\Trans \hat\Gamma_{h}^{-1} (\hat\Sigma_{h} - \Sigma_h) \hat\Gamma_{h}^{-1} c}{c^\Trans \Gamma_{h}^{-1} \Sigma_{h} \Gamma_{h}^{-1} c} 
+ 2\frac{c^\Trans (\hat\Gamma_{h}^{-1} -\Gamma_{h}^{-1})  \Sigma_h \Gamma_{h}^{-1} c}{c^\Trans \Gamma_{h}^{-1} \Sigma_{h} \Gamma_{h}^{-1} c}
+ \frac{c^\Trans (\hat\Gamma_{h}^{-1} -\Gamma_{h}^{-1})  \Sigma_h (\hat\Gamma_{h}^{-1} -\Gamma_{h}^{-1}) c}{c^\Trans \Gamma_{h}^{-1} \Sigma_{h} \Gamma_{h}^{-1} c} 
.
\end{align*}

From the analysis of $\hat\Sigma_h$, we have
\begin{align*}
\frac{c^\Trans \hat\Gamma_{h}^{-1} (\hat\Sigma_{h} - \Sigma_h) \hat\Gamma_{h}^{-1} c}{c^\Trans \Gamma_{h}^{-1} \Sigma_{h} \Gamma_{h}^{-1} c}  = \Op\left(  \frac{1}{\sqrt{nh^2}} \right). 
\end{align*}
For the second term, we have
\begin{align*}
\left|\frac{c^\Trans (\hat\Gamma_{h}^{-1} -\Gamma_{h}^{-1})  \Sigma_h \hat\Gamma_{h}^{-1} c}{c^\Trans \Gamma_{h}^{-1} \Sigma_{h} \Gamma_{h}^{-1} c}\right| 
&\leq \frac{|c^\Trans (\hat\Gamma_{h}^{-1} -\Gamma_{h}^{-1})  \Sigma_h^{1/2}|\cdot|c^\Trans \Gamma_{h}^{-1}  \Sigma_h^{1/2}|}{|c^\Trans \Gamma_{h}^{-1}\Sigma_h^{1/2} |^2}\\
&= \frac{|c^\Trans (\hat\Gamma_{h}^{-1} -\Gamma_{h}^{-1})  \Sigma_h^{1/2}|}{|c^\Trans \Gamma_{h}^{-1}\Sigma_h^{1/2} |} = \Op\left( \sqrt{\frac{1}{nh^2}} \right).
\end{align*}
The third term has order
\begin{align*}
\frac{c^\Trans (\hat\Gamma_{h}^{-1} -\Gamma_{h}^{-1})  \Sigma_h (\hat\Gamma_{h}^{-1} -\Gamma_{h}^{-1}) c}{c^\Trans \Gamma_{h}^{-1} \Sigma_{h} \Gamma_{h}^{-1} c} = \Op\left(  \frac{1}{nh^2} \right). 
\end{align*}

\subsection{Proof of Corollary \ref{coro:asy normal loc pol projection estimator}}

This follows directly from Theorem \ref{thm:local projection: asymptotic normality}. 

\subsection{Proof of Corollary \ref{coro:asy normal ortho loc pol projection estimator}}

To understand \eqref{eq:local orthogonal polynomial projection estimator}, note that
\begin{align*}
\hat\theta_{F}^\perp &= \left( \int_{\mathcal{X}} \Lambda_h^\Trans R(u-\Evapointx)R(u-\Evapointx)^\Trans\Lambda_h \frac{1}{h}K\left(\frac{u-\Evapointx}{h}\right) \diff F(u) \right)^{-1}\left( \int_{\mathcal{X}} \Lambda_h^\Trans R(u-\Evapointx)\hat{F}(u) \frac{1}{h}K\left(\frac{u-\Evapointx}{h}\right) \diff F(u) \right)\\
&= \Lambda_h^{-1}\left( \int_{\mathcal{X}} R(u-\Evapointx)R(u-\Evapointx)^\Trans \frac{1}{h}K\left(\frac{u-\Evapointx}{h}\right) \diff F(u) \right)^{-1}\left( \int_{\mathcal{X}} R(u-\Evapointx)\hat{F}(u) \frac{1}{h}K\left(\frac{u-\Evapointx}{h}\right) \diff F(u) \right),
\end{align*}
which means $\hat\theta_{F}^\perp = \Lambda_h^{-1}\hat\theta_{F}$. Then we have (up to an approximation bias term)
\begin{align*}
&\ \hat\theta_{F}^\perp - \Lambda_h^{-1}\theta_0 = \Lambda_h^{-1}(\hat\theta_{F} - \theta_0) \\
=&\ \Lambda_h^{-1}\left( \int_{\mathcal{X}} R(u-\Evapointx)R(u-\Evapointx)^\Trans \frac{1}{h}K\left(\frac{u-\Evapointx}{h}\right) \diff F(u) \right)^{-1}\left( \int_{\mathcal{X}} R(u-\Evapointx)\Big(\hat{F}(u) -  F(u)\Big) \frac{1}{h}K\left(\frac{u-\Evapointx}{h}\right) \diff F(u) \right)\\
=&\ \Lambda_h^{-1}\Upsilon_h\left( \int_{\Intregion} R(u)R(u)^\Trans K\left(u\right) f(\Evapointx+hu)\diff u \right)^{-1}\left( \int_{\Intregion} R(u)\Big(\hat{F}(\Evapointx+hu) -   F(\Evapointx+hu) \Big) K\left(u\right) f(\Evapointx+hu)\diff u \right)\\
=&\ \Lambda_h^{-1}\Upsilon_h\Lambda_h\left( \int_{\Intregion} R^\perp(u)R^\perp(u)^{ \Trans} K\left(u\right) f(\Evapointx+hu)\diff u \right)^{-1}\left( \int_{\Intregion} R^\perp(u)\Big(\hat{F}(\Evapointx+hu) -   F(\Evapointx+hu) \Big) K\left(u\right) f(\Evapointx+hu)\diff u \right).
\end{align*}

We first discuss the transformed parameter vector $\Lambda_h^{-1}\theta_0$. By construction, the matrix $\Lambda_h$ takes the following form:
\begin{align*}
\Lambda_h &= \begin{bmatrix}
1 & c_{1,2} & c_{1,3} & \cdots  &c_{1,p+2} \\
0 & 1       & 0       & \cdots  &c_{2,p+2} \\
0 & 0       & 1       & \cdots  &c_{2,p+2} \\
\vdots &  \vdots      & \vdots       & \ddots  &\vdots \\
0 & 0       & 0       & \cdots  & 1 
\end{bmatrix}
\end{align*}
where $c_{i,j}$ are some constants (possibly depending on $h$). Therefore, the above matrix differs from the identity matrix only in its first row and in the last column. This observation also holds for $\Lambda_h^{-1}$. Since the last component of $\theta_0$ is zero (because the extra regressor $Q_h(\cdot)$ is redundant), we conclude that, except for the first element, $\Lambda_h\theta$ and $\theta$ are identical. More specifically, let $I_{-1}$ be the identity matrix excluding the first row:
\begin{align*}
I_{-1} &= \begin{bmatrix}
0 & 1 & 0 & 0 & \cdots & 0\\
0 & 0 & 1 & 0 & \cdots & 0\\
0 & 0 & 0 & 1 & \cdots & 0\\
\vdots & \vdots & \vdots & \vdots & \ddots & \vdots\\
0 & 0 & 0 & 0 & \cdots & 1\\
\end{bmatrix},
\end{align*}
which is used to extract all elements of a vector except for the first one, then by Theorem \ref{thm:local projection: asymptotic normality}, 
\begin{align*}
\sqrt{n}\left(I_{-1}(\Lambda_h^{-1}\Upsilon_h\Lambda_h)(\Gamma_h^\perp)^{-1}\Sigma_h^\perp(\Gamma_h^\perp)^{-1}(\Lambda_h^{-1}\Upsilon_h\Lambda_h)^\Trans I_{-1}^\Trans\right)^{-1/2} \begin{bmatrix}
\hat\theta^\perp_{P,F} - \theta_{P} \\
\hat\theta^\perp_{Q,F}
\end{bmatrix} \rightsquigarrow \mathcal{N}(0, I),
\end{align*}
where $\theta^\perp_{P,F}$ contains the second to the $p+1$-th element of $\theta^\perp_{F}$, and $\theta^\perp_{Q,F}$ is the last element. 

Now we discuss the covariance matrix in the above display. Due to orthogonalization, $\Gamma_h^\perp$ is block diagonal. To be precise, 
\begin{align*}
\Gamma_h^\perp = f(\Evapointx)
\begin{bmatrix}
\Gamma_{1,h}^\perp & 0 & 0\\
 & \Gamma_{P,h}^\perp & 0\\
0 & 0 & \Gamma_{Q,h}^\perp
\end{bmatrix},\ \Gamma_{1,h}^\perp = \int_{\Intregion} K(u)\diff u,\ \Gamma_{P,h}^\perp = \int_{\Intregion} P^\perp(u)P^\perp(u)^\Trans K(u)\diff u,\ \Gamma_{Q,h}^\perp = \int_{\Intregion} Q^\perp(u)^2 K(u)\diff u.
\end{align*}
Finally, using the structure of $\Lambda_h$ and $\Upsilon_h$, we have
\begin{align*}
I_{-1}(\Lambda_h^{-1}\Upsilon_h\Lambda_h)(\Gamma_h^\perp)^{-1} &= I_{-1}\Upsilon_h(\Gamma_h^\perp)^{-1}.
\end{align*}

The form of $\Sigma_h^\perp$ is quite involved, but with some algebra, and using the fact that the basis $R(\cdot)$ (or $R^\perp(\cdot)$) includes a constant and polynomials, one can show the following:
\begin{align*}
(\Lambda_h^{-1}\Upsilon_h\Lambda_h)(\Gamma_h^\perp)^{-1}\Sigma_h^\perp(\Gamma_h^\perp)^{-1}(\Lambda_h^{-1}\Upsilon_h\Lambda_h)^\Trans = hf(\Evapointx)\Upsilon_{-1, h}(\Gamma_{-1,h}^\perp)^{-1}\Sigma_{-1,h}^\perp(\Gamma_{-1,h}^\perp)^{-1}\Upsilon_{-1, h},
\end{align*} 
where $\Upsilon_{-1, h}$, $\Gamma_{-1,h}^\perp$ and $\Sigma_{-1,h}^\perp$ are obtained by excluding the first row and the first column of $\Upsilon_{h}$, $\Gamma_{h}^\perp$ and $\Sigma_{h}^\perp$, respectively:
\begin{align*}
\Upsilon_{-1, h} &= \begin{bmatrix}
h^{-1} & 0 & 0 & \cdots & 0\\
0 & h^{-2} & 0 & \cdots & 0\\
0 & 0 & h^{-3} & \cdots & 0\\
\vdots & \vdots & \vdots & \ddots & \vdots\\
0 & 0 & 0 & \cdots & \upsilon_h\\
\end{bmatrix},\ \Gamma_{-1,h}^\perp = f(\Evapointx)\begin{bmatrix}
\Gamma_{P,h}^\perp & 0\\
0 & \Gamma_{Q,h}^\perp
\end{bmatrix},\ \Sigma_{-1,h}^\perp = f(\Evapointx)^3\begin{bmatrix}
\Sigma_{PP,h}^\perp & \Sigma_{PQ,h}^\perp \\
\Sigma_{QP,h}^\perp & \Sigma_{QQ,h}^\perp
\end{bmatrix},
\end{align*}
and
\begin{align*}
\Sigma_{PP,h}^\perp &= \iint_{\Intregion} K(u)K(v)P^\perp(u)P^\perp(v)^\Trans (u\wedge v) \diff u\diff v,\qquad \Sigma_{QQ,h}^\perp &= \iint_{\Intregion} K(u)K(v)Q^\perp(u)Q^\perp(v) (u\wedge v) \diff u\diff v\\ 
\Sigma_{PQ,h}^\perp &= (\Sigma_{QP,h}^\perp)^\Trans = \iint_{\Intregion} K(u)K(v)P^\perp(u)Q^\perp(v) (u\wedge v) \diff u\diff v.
\end{align*}

\subsection{Proof of Lemma \ref{lem:survival transformation}	}

\subsubsection*{Part (i)}
To start, 
\begin{align*}
\int_{\Intregion \cap [-1,1]} \mathcal{H}(g_1)(u)\mathcal{H}(g_2)(u) \diff u &= \int_{\Intregion \cap [-1,1]} \left(\int_{\Intregion} \Indicator(v_1\geq u)K(v_1)g(v_1)\diff v_1\right)\left(\int_{\Intregion} \Indicator(v_2\geq u)K(v_2)g(v_2)\diff v_2\right) \diff u\\
&= \iint_{\Intregion}K(v_1)K(v_2)g(v_1)g(v_2)\left(\int_{\Intregion \cap [-1,1]}  \Indicator(v_1\geq u) \Indicator(v_2\geq u) \diff u\right)\diff v_1\diff v_2\\
&= \iint_{\Intregion}K(v_1)K(v_2)g(v_1)g(v_2)
\left[(v_1\wedge v_2)\wedge \left( \frac{\overline{x}-\Evapointx}{h}\wedge 1 \right) - \left( \frac{\underline{x}-\Evapointx}{h}\vee (-1) \right)\right]
\diff v_1\diff v_2\\
&= \iint_{\Intregion}K(v_1)K(v_2)g(v_1)g(v_2)(v_1\wedge v_2)\diff v_1\diff v_2,
\end{align*}
where to show the last equality, we used the fact that $v_1\leq \frac{\overline{x}-\Evapointx}{h}\wedge 1$ and $v_2\leq \frac{\overline{x}-\Evapointx}{h}\wedge 1$ for the outer double integral. 
\subsubsection*{Part (ii)}
For this part, 
\begin{align*}
\int_{\Intregion \cap [-1,1]} \mathcal{H}(g_1)(u)\dot{g}_2(u) \diff u &= \int_{\Intregion \cap [-1,1]} \left(\int_{\Intregion} \Indicator(v\geq u)K(v)g_1(v)\diff v\right)\dot{g}_2(u) \diff u\\
&=  \int_{\Intregion}K(v)g_1(v) \left(\int_{\Intregion \cap [-1,1]}\Indicator(v\geq u) \dot{g}_2(u) \diff u \right)\diff v\\
&=  \int_{\Intregion}K(v)g_1(v)  \left[g_2\left(v\wedge  \frac{\overline{x}-\Evapointx}{h}\wedge 1 \right) - g_2\left( \frac{\underline{x}-\Evapointx}{h}\vee (-1) \right) \right] \diff v\\
&=  \int_{\Intregion}K(v)g_1(v)  g_2(v) \diff v.
\end{align*}
Again, to show the last equality, we used the fact that $v\leq \frac{\overline{x}-\Evapointx}{h}\wedge 1$ for the outer integral.

\subsection{Proof of Theorem \ref{thm:variance bound}}

To find a bound of the maximization problem, we note that for any $c\in\mathbb{R}^{p-1}$, one has 
\begin{align*}
\int_{\Intregion\cap[-1,1]} \mathcal{H}(p_\ell)(u)\mathcal{H}(q)(u)\diff u = \int_{\Intregion\cap[-1,1]} \Big[\mathcal{H}(p_\ell)(u)+c^\Trans \dot{P}(u)\Big]\mathcal{H}(q)(u)\diff u, 
\end{align*} 
due to the constraint. Therefore, an upper bound of the objective function is (due to the Cauchy-Schwartz inequality)
\begin{align*}
&\ \inf_{c} \int_{\Intregion\cap[-1,1]} \Big[\mathcal{H}(p_\ell)(u)+c^\Trans \dot{P}(u)\Big]^2\diff u\\
=&\ \inf_{c} \int_{\Intregion\cap[-1,1]} \Big[\mathcal{H}(p_\ell)(u)^2 + 2c^\Trans \dot{P}(u)\mathcal{H}(p_\ell)(u) + c^\Trans \dot{P}(u)\dot{P}(u)^\Trans c\Big]\diff u\\
=&\ \int_{\Intregion\cap[-1,1]}\mathcal{H}(p_\ell)(u)^2\diff u + \inf_{c} \int_{\Intregion\cap[-1,1]} \Big[2c^\Trans \dot{P}(u)\mathcal{H}(p_\ell)(u) + c^\Trans \dot{P}(u)\dot{P}(u)^\Trans c\Big]\diff u\\
=&\ \int_{\Intregion\cap[-1,1]}\mathcal{H}(p_\ell)(u)^2\diff u + \inf_{c}\left[ 2c^\Trans \left(\int_{\Intregion} K(u)P(u)p_{\ell}(u)\diff u\right) + c^\Trans \left(\int_{\Intregion\cap[-1,1]} \dot{P}(u)\dot{P}(u)^\Trans \diff u\right)c \right],
\end{align*}
which is minimized by setting
\begin{align*}
c &= -\left(\int_{\Intregion\cap[-1,1]} \dot{P}(u)\dot{P}(u)^\Trans \diff u\right)^{-1} \left(\int_{\Intregion} K(u)P(u)p_{\ell}(u)\diff u\right).
\end{align*}
As a result, an upper bound of \eqref{eq:maximization} is
\begin{align*}
&\ \int_{\Intregion\cap[-1,1]}\mathcal{H}(p_\ell)(u)^2\diff u -  \left(\int_{\Intregion} K(u)P(u)p_{\ell}(u)\diff u\right)^\Trans \left(\int_{\Intregion\cap[-1,1]} \dot{P}(u)\dot{P}(u)^\Trans \diff u\right)^{-1} \left(\int_{\Intregion} K(u)P(u)p_{\ell}(u)\diff u\right).
\end{align*}
We may further simplify the above. First, 
\begin{align*}
\int_{\Intregion\cap[-1,1]}\mathcal{H}(p_\ell)(u)^2\diff u &= e_\ell^\Trans(\Gamma_{P,h}^\perp)^{-1} \Sigma_{PP,h}^\perp (\Gamma_{P,h}^\perp)^{-1}e_\ell.
\end{align*}
Second, note that 
\begin{align*}
\int_{\Intregion} K(u)P(u)p_{\ell}(u)\diff u &= \left(\int_{\Intregion} K(u)P(u)P^\perp(u)^\Trans \diff u\right) (\Gamma_{P,h}^\perp)^{-1}e_\ell
= \left(\int_{\Intregion} K(u)P^\perp(u)P^\perp(u)^\Trans \diff u\right) (\Gamma_{P,h}^\perp)^{-1}e_\ell
= e_\ell.
\end{align*}
As a result, an upper bound of \eqref{eq:maximization} is 
\begin{align*}
&\ e_\ell^\Trans(\Gamma_{P,h}^\perp)^{-1} \Sigma_{PP,h}^\perp (\Gamma_{P,h}^\perp)^{-1}e_\ell - e_\ell^\Trans \left(\int_{\Intregion\cap[-1,1]} \dot{P}(u)\dot{P}(u)^\Trans \diff u\right)^{-1} e_{\ell}\\
&= e_\ell^\Trans\left[ (\Gamma_{P,h}^\perp)^{-1} \Sigma_{PP,h}^\perp (\Gamma_{P,h}^\perp)^{-1} - \left(\int_{\Intregion\cap[-1,1]} \dot{P}(u)\dot{P}(u)^\Trans \diff u\right)^{-1} \right]e_\ell.
\end{align*}

\subsection{Additional Preliminary Lemmas}

\begin{lem}\label{lem:uniform convergence of iid sum}
Assume $\{u_{i,h}(a):\ a\in A\subset\mathbb{R}^d\}$ are independent across $i$, and $\Expectation[u_{i,h}(a)]=0$ for all $a\in A$ and all $h > 0$. In addition, assume for each $\epsilon>0$ there exists $\{u_{i,h,\epsilon}(a): a\in A\}$, such that
\begin{align*}
|a-b|\leq \epsilon\quad \Rightarrow\quad |u_{i,h}(a)-u_{i,h}(b)|\leq u_{i,h,\epsilon}(a).
\end{align*} 
Define
\begin{alignat*}{2}
C_1 &= \sup_{a\in A}\max_{1\leq i\leq n}\Var[u_{i,h}(a)],\qquad C_2 &&= \sup_{a\in A}\max_{1\leq i\leq n}|u_{i,h}(a)|\\
C_{1,\epsilon} &= \sup_{a\in A}\max_{1\leq i\leq n}\Var[u_{i,h,\epsilon}(a)],\quad C_{2,\epsilon} &&= \sup_{a\in A}\max_{1\leq i\leq n}|u_{i,h,\epsilon}(a) - \Expectation[u_{i,h,\epsilon}(a)]|,\quad C_{3,\epsilon}=\sup_{a\in A}\max_{1\leq i\leq n}\Expectation[|u_{i,h,\epsilon}(a)|].
\end{alignat*}
Then
\begin{align*}
\sup_{a\in A}\left|\frac{1}{n}\sum_{i=1}^n u_{i,h}(a)\right| = \Op\left( \gamma + \gamma_{\epsilon} + C_{3,\epsilon}\right),
\end{align*}
where $\gamma$ and $\gamma_{\epsilon}$ are any sequences satisfying
\begin{align*}
&\frac{\gamma^2n}{(C_{1} + \frac{1}{3} \gamma C_{2})\log N(\epsilon, A, |\cdot|)}\quad \text{and} \quad
\frac{\gamma_{\epsilon}^2n}{(C_{1,\epsilon} + \frac{1}{3} \gamma_\epsilon C_{2,\epsilon})\log N(\epsilon, A, |\cdot|)}\quad \text{are bounded from below},
\end{align*}
and $N(\epsilon, A, |\cdot|)$ is the covering number of $A$. 
\qed
\end{lem}

\begin{remark}
Provided that $u_{i,h}(\cdot)$ is reasonably smooth, one can always choose $\epsilon$ (as a function of $n$ and $h$) small enough, and the leading order will be given by $\gamma$ (and hence is determined by $C_1$ and $C_2$). \qed
\end{remark}

\noindent\textbf{Proof.} Let $A_\epsilon$ be an $\epsilon$-covering of $A$, then 
\begin{align*}
&\ \sup_{a\in A} \left|\frac{1}{n}\sum_{i=1}^n u_{i,h}(a) \right| \leq \sup_{a\in A_\epsilon} \left|\frac{1}{n}\sum_{i=1}^n u_{i,h}(a) \right| 
+\sup_{a\in A_\epsilon,b\in A,|a-b|\leq \epsilon}\left|\frac{1}{n}\sum_{i=1}^n  u_{i,h}(a)-u_{i,h}(b)\right|. 
\end{align*}
Next we apply the union bound and Bernstein's inequality:
\begin{align*}
\Prob\left[ \sup_{a\in A_\epsilon} \left|\frac{1}{n}\sum_{i=1}^n u_{i,h}(a) \right| \geq \gamma u\right] &\leq N(\epsilon, A, |\cdot|) \sup_{a\in A}\Prob\left[ \left|\frac{1}{n}\sum_{i=1}^n u_{i,h}(a) \right| \geq \gamma u\right]\\
&\leq 2N(\epsilon, A, |\cdot|)\exp\left\{ -\frac{1}{2}\frac{\gamma^2nu^2}{ C_{1} + \frac{1}{3} \gamma C_{2} u } \right\}\\
&= 2\exp\left\{ -\frac{1}{2}\frac{\gamma^2nu^2}{ C_{1} + \frac{1}{3} \gamma C_{2} u } + \log N(\epsilon, A, |\cdot|) \right\}.
\end{align*}
Now take $u$ sufficiently large, then the above is further bounded by:
\begin{align*}
\Prob\left[ \sup_{a\in A_\epsilon} \left|\frac{1}{n}\sum_{i=1}^n u_{i,h}(a) \right| \geq \gamma u\right] &\leq 2\exp\left\{ -\log N(\epsilon, A, |\cdot|)\left[\frac{1}{2}\frac{1}{\log N(\epsilon, A, |\cdot|)}\frac{\gamma^2n}{ C_{1} + \frac{1}{3} \gamma C_{2}  }u - 1 \right] \right\} ,
\end{align*}
which tends to zero if $\log N(\epsilon, A, |\cdot|)\to\infty$ and 
\begin{align*}
\frac{\gamma^2n}{(C_{1} + \frac{1}{3} \gamma C_{2})\log N(\epsilon, A, |\cdot|)}\ \text{is bounded from below},
\end{align*}
in which case we have
\begin{align*}
\sup_{a\in A_\epsilon} \left|\frac{1}{n}\sum_{i=1}^n u_{i,h}(a) \right| = \Op\left( \gamma \right).
\end{align*}
We can apply the same technique to the other term, and obtain
\begin{align*}
\sup_{a\in A_\epsilon,b\in A,|a-b|\leq \epsilon}\left|\frac{1}{n}\sum_{i=1}^n  u_{i,h}(a)-u_{i,h}(b)\right| = \Op\left( \gamma_{\epsilon} \right),
\end{align*}
where $\gamma_{\epsilon}$ is any sequence satisfying
\begin{align*}
\frac{\gamma_{\epsilon}^2n}{(C_{1,\epsilon} + \frac{1}{3} \gamma_\epsilon C_{2,\epsilon})\log N(\epsilon, A, |\cdot|)}\ \text{is bounded from below}.
\end{align*}
\qed

\begin{lem}[Corollary 5.1 in \citealt{chernozhukov2019central}]\label{lem:gaussian comparison}
Let $\bz_1,\bz_2\in\mathbb{R}^{\ell_n}$ be two mean-zero Gaussian random vectors with covariance matrices $\bOmega_{1}$ and $\bOmega_{2}$, respectively. Further assume that the diagonal elements in $\bOmega_{1}$ are all one. Then
\begin{align*}
\sup_{\substack{A\subseteq \mathbb{R}^{\ell_n} \\ A\ \text{rectangular}}} \left| \Prob\left[ \bz_1 \in A \right]
-
\Prob\left[ \bz_2 \in A \right] \right| \leq C\sqrt{\Vert \bOmega_{1} - \bOmega_{2} \Vert_{\infty}}\log \ell_n,
\end{align*}
where $\Vert \cdot\Vert_\infty$ denotes the supremum norm, and $C$ is an absolute constant.
\qed
\end{lem}

\begin{lem}[Equation (3.5) in \citealt*{Gine-Latala-Zinn_2000_Ustat}]\label{lem:ustatistic concentration inequality}
For a degenerate and decoupled second order U-statistic, $\sum_{i,j=1,i\neq j}^n h_{ij}(x_i,\tilde{x}_j)$, the following holds:
\begin{align*}
\Prob\left[ \left|\sum_{i,j,i\neq j}^n u_{ij}(x_i,\tilde{x}_j)\right| > t \right] \leq C\exp\left\{ -\frac{1}{C}\min\left[ \frac{t}{D},\ \left(\frac{t}{B}\right)^{\frac{2}{3}},\ \left( \frac{t}{A} \right)^{\frac{1}{2}} \right] \right\},
\end{align*}
where $C$ is some universal constant, and $A$, $B$ and $D$ are any constants satisfying
\begin{align*}
A   &\geq \max_{1\leq i,j\leq n}\sup_{u,v}| u_{ij}(u,v) |\\
B^2 &\geq \max_{1\leq i,j\leq n}\left[  \sup_{v}\left|\sum_{i=1}^n \Expectation u_{ij}(x_i,v)^2\right| ,\  \sup_{u}\left|\sum_{j=1}^n \Expectation u_{ij}(u,\tilde{x}_j)^2\right|  \right]\\
D^2 &\geq \sum_{i,j=1,i\neq j}^n \Expectation u_{ij}(x_i,\tilde{x}_j)^2.
\end{align*}
where $\{{x}_i, 1\leq i\leq n\}$ are independent random variables, and $\{\tilde{x}_i, 1\leq i\leq n\}$ is an independent copy of $\{{x}_i, 1\leq i\leq n\}$. \qed
\end{lem}

\begin{remark}
To apply the above lemma, an additional decoupling step is usually needed. Fortunately, the decoupling step only introduces an extra constant, but will not affect the order of the tail probability bound. Formally, 
\begin{lem}[\citealt{DeLaPena-MontgomerySmith_1995_AoP}]
Consider the setting of Lemma \ref{lem:ustatistic concentration inequality}. Then 
\begin{align*}
\Prob\left[ \left|\sum_{i,j,i\neq j}^n u_{ij}(x_i,{x}_j)\right| > t \right] \leq C\cdot \Prob\left[ C\left|\sum_{i,j,i\neq j}^n u_{ij}(x_i,\tilde{x}_j)\right| > t \right],
\end{align*}
where $C$ is a universal constant.\qed
\end{lem}
As a result, we will apply Lemma \ref{lem:ustatistic concentration inequality} without explicitly mentioning the decoupling step or the extra constant it introduces.\qed
\end{remark}

\subsection{Proof of Theorem \ref{thm:strong approximation}}

To bound the distance between the two processes, $\tilde{\mathfrak{T}}_G(\cdot)$ and $\mathfrak{B}_G(\cdot)$, we employ the proof strategy of \cite*{Gine-Koltchinskii-Sakhanenko_2004_PTRF}. Recall that $F$ denotes the distribution of $x_i$, and we define
\begin{align*}
\mathcal{K}_{h,\Evapointx}\circ F^{-1}(x) = \mathcal{K}_{h,\Evapointx}(F^{-1}(x)). 
\end{align*}
Take $v<v'$ in $[0,1]$, we have
\begin{align*}
&\ \left|\mathcal{K}_{h,\Evapointx}\circ F^{-1}(v) - \mathcal{K}_{h,\Evapointx}\circ F^{-1}(v')\right|\\
&= \left|\frac{  \int_{\Intregion} c_{h,\Evapointx}^\Trans\Upsilon_{h}\Gamma_{h,\Evapointx}^{-1}R(u)\Big[\Indicator(F^{-1}(v)\leq \Evapointx + hu) - \Indicator(F^{-1}(v')\leq \Evapointx + hu)\Big] K\left(u\right) g(\Evapointx+hu)\diff u}{\sqrt{c_{h,\Evapointx}^\Trans\Upsilon_{h}\Omega_{h,\Evapointx}\Upsilon_{h}c_{h,\Evapointx}}}\right|\\
&\leq \frac{  \int_{\Intregion} \left|c_{h,\Evapointx}^\Trans\Upsilon_{h}\Gamma_{h,\Evapointx}^{-1}R(u)\right|\Big[\Indicator(F^{-1}(v)\leq \Evapointx + hu) - \Indicator(F^{-1}(v')\leq \Evapointx + hu)\Big] K\left(u\right) g(\Evapointx+hu)\diff u}{\sqrt{c_{h,\Evapointx}^\Trans\Upsilon_{h}\Omega_{h,\Evapointx}\Upsilon_{h}c_{h,\Evapointx}}}.
\end{align*}
Therefore, the function $\mathcal{K}_{h,\Evapointx}\circ F^{-1}(\cdot)$ has a total variation bounded by
\begin{align*}
&\ \frac{  \int_{\Intregion} \left|c_{h,\Evapointx}^\Trans\Upsilon_{h}\Gamma_{h,\Evapointx}^{-1}R(u)\right|\Big[\Indicator(F^{-1}(0)\leq \Evapointx + hu) - \Indicator(F^{-1}(1)\leq \Evapointx + hu)\Big] K\left(u\right) g(\Evapointx+hu)\diff u}{\sqrt{c_{h,\Evapointx}^\Trans\Upsilon_{h}\Omega_{h,\Evapointx}\Upsilon_{h}c_{h,\Evapointx}}}\\
&= \frac{  \int_{-1}^1 \left|c_{h,\Evapointx}^\Trans\Upsilon_{h}\Gamma_{h,\Evapointx}^{-1}R(u)\right| K\left(u\right) g(\Evapointx+hu)\diff u}{\sqrt{c_{h,\Evapointx}^\Trans\Upsilon_{h}\Omega_{h,\Evapointx}\Upsilon_{h}c_{h,\Evapointx}}}\leq C_4\frac{1}{\sqrt{h}}.
\end{align*}

It is well-known that functions of bounded variation can be approximated (pointwise) by convex combination of indicator functions of half intervals. To be more precise, 
\begin{align*}
\Big\{\mathcal{K}_{h,\Evapointx}\circ F^{-1}(\cdot):\ \Evapointx\in\mathcal{I}\Big\}\subset C_4\frac{1}{\sqrt{h}}\overline{\mathrm{conv}}\Big\{ \pm\Indicator(\cdot\leq t), \pm\Indicator(\cdot\geq  t)\Big\}.
\end{align*}
Following (2.3) and (2.4) of \cite*{Gine-Koltchinskii-Sakhanenko_2004_PTRF}, we have
\begin{align*}
\Prob\left[ \sup_{\Evapointx\in\mathcal{I}}\left| \tilde{\mathfrak{T}}_G(\Evapointx) - \mathfrak{B}_G(\Evapointx) \right| > \frac{C_4(u+C_5\log n)}{\sqrt{nh}} \right] \leq C_5e^{-C_5u},
\end{align*}
where $C_5$ is some universal constant.

\subsection{Proof of Lemma \ref{lem:VC-type}}

Take $|\Evapointx-\Evapointy|\leq \epsilon$ to be some small number, then
\begin{align*}
&\ \mathcal{K}_{h,\Evapointx}(x) - \mathcal{K}_{h,\Evapointy}(x) 
= \frac{c_{h,\Evapointx}^\Trans\Upsilon_{h}\Gamma_{h,\Evapointx}^{-1}  \int_{\Intregion} R(u)\Big[\Indicator(x\leq \Evapointx + hu) - F(\Evapointx + hu)\Big] K\left(u\right) g(\Evapointx+hu)\diff u}{\sqrt{c_{h,\Evapointx}^\Trans\Upsilon_{h}\Omega_{h,\Evapointx}\Upsilon_{h}c_{h,\Evapointx}}}\\
&\qquad\qquad\qquad\qquad\qquad\qquad - \frac{c_{h,\Evapointy}^\Trans\Upsilon_{h}\Gamma_{h,\Evapointy}^{-1}  \int_{\frac{\mathcal{X}-\Evapointy}{h}} R(u)\Big[\Indicator(x\leq \Evapointy + hu) - F(\Evapointy + hu)\Big] K\left(u\right) g(\Evapointy+hu)\diff u}{\sqrt{c_{h,\Evapointy}^\Trans\Upsilon_{h}\Omega_{h,\Evapointy}\Upsilon_{h}c_{h,\Evapointy}}}\\
=&\ \left(\frac{c_{h,\Evapointx}^\Trans\Upsilon_{h}\Gamma_{h,\Evapointx}^{-1}}{\sqrt{c_{h,\Evapointx}^\Trans\Upsilon_{h}\Omega_{h,\Evapointx}\Upsilon_{h}c_{h,\Evapointx}}} - \frac{c_{h,\Evapointy}^\Trans\Upsilon_{h}\Gamma_{h,\Evapointy}^{-1}}{\sqrt{c_{h,\Evapointy}^\Trans\Upsilon_{h}\Omega_{h,\Evapointy}\Upsilon_{h}c_{h,\Evapointy}}}\right)\left(\int_{\Intregion} R(u)\Big[\Indicator(x\leq \Evapointx + hu) - F(\Evapointx + hu)\Big] K\left(u\right) g(\Evapointx+hu)\diff u\right)\\
+&\left( \frac{c_{h,\Evapointy}^\Trans\Upsilon_{h}\Gamma_{h,\Evapointy}^{-1}}{\sqrt{c_{h,\Evapointy}^\Trans\Upsilon_{h}\Omega_{h,\Evapointy}\Upsilon_{h}c_{h,\Evapointy}}}\right) 
\left(\frac{1}{h}\int_{\mathcal{X}} \Big[R\left(\frac{u-\Evapointx}{h}\right)K\left(\frac{u-\Evapointx}{h}\right) - R\left(\frac{u-\Evapointy}{h}\right)K\left(\frac{u-\Evapointy}{h}\right)\Big]\Big[\Indicator(x\leq u) - F(u)\Big]  g(u)\diff u\right)\\
\tag{I}=&\ \frac{1}{\sqrt{c_{h,\Evapointy}^\Trans\Upsilon_{h}\Omega_{h,\Evapointy}\Upsilon_{h}c_{h,\Evapointy}}}\left(c_{h,\Evapointx}^\Trans\Upsilon_{h} - c_{h,\Evapointy}^\Trans\Upsilon_{h}\right)\Gamma_{h,\Evapointx}^{-1}\left(\int_{\Intregion} R(u)\Big[\Indicator(x\leq \Evapointx + hu) - F(\Evapointx + hu)\Big] K\left(u\right) g(\Evapointx+hu)\diff u\right)\\
\tag{II}+&\ \frac{1}{\sqrt{c_{h,\Evapointy}^\Trans\Upsilon_{h}\Omega_{h,\Evapointy}\Upsilon_{h}c_{h,\Evapointy}}}c_{h,\Evapointy}^\Trans\Upsilon_{h}\left(\Gamma_{h,\Evapointx}^{-1} - \Gamma_{h,\Evapointy}^{-1}\right)\left(\int_{\Intregion} R(u)\Big[\Indicator(x\leq \Evapointx + hu) - F(\Evapointx + hu)\Big] K\left(u\right) g(\Evapointx+hu)\diff u\right)\\
\tag{III}+& \left(\frac{1}{\sqrt{c_{h,\Evapointx}^\Trans\Upsilon_{h}\Omega_{h,\Evapointx}\Upsilon_{h}c_{h,\Evapointx}}} - \frac{1}{\sqrt{c_{h,\Evapointy}^\Trans\Upsilon_{h}\Omega_{h,\Evapointy}\Upsilon_{h}c_{h,\Evapointy}}}\right)
c_{h,\Evapointx}^\Trans\Upsilon_{h}\Gamma_{h,\Evapointx}^{-1}\left(\int_{\Intregion} R(u)\Big[\Indicator(x\leq \Evapointx + hu) - F(\Evapointx + hu)\Big] K\left(u\right) g(\Evapointx+hu)\diff u\right)\\
\tag{IV}+&\left( \frac{c_{h,\Evapointy}^\Trans\Upsilon_{h}\Gamma_{h,\Evapointy}^{-1}}{\sqrt{c_{h,\Evapointy}^\Trans\Upsilon_{h}\Omega_{h,\Evapointy}\Upsilon_{h}c_{h,\Evapointy}}}\right) 
\left(\frac{1}{h}\int_{\mathcal{X}} \Big[R\left(\frac{u-\Evapointx}{h}\right)K\left(\frac{u-\Evapointx}{h}\right) - R\left(\frac{u-\Evapointy}{h}\right)K\left(\frac{u-\Evapointy}{h}\right)\Big]\Big[\Indicator(x\leq u) - F(u)\Big]  g(u)\diff u\right).
\end{align*}

For term (I), its variance (replace the placeholder $x$ by $x_i$) is 
\begin{align*}
\Var[\text{(I)}] &= \frac{1}{c_{h,\Evapointy}^\Trans\Upsilon_{h}\Omega_{h,\Evapointy}\Upsilon_{h}c_{h,\Evapointy}}\left(c_{h,\Evapointx}^\Trans\Upsilon_{h} - c_{h,\Evapointy}^\Trans\Upsilon_{h}\right)\Omega_{h,\Evapointx}\left(c_{h,\Evapointx}^\Trans\Upsilon_{h} - c_{h,\Evapointy}^\Trans\Upsilon_{h}\right)^\Trans  = O\left(\frac{1}{h}r_1(\epsilon,h)^2r_2(h)^2\right).
\end{align*}

Term (II) has variance
\begin{align*}
\Var[\text{(II)}] &= \frac{1}{c_{h,\Evapointy}^\Trans\Upsilon_{h}\Omega_{h,\Evapointy}\Upsilon_{h}c_{h,\Evapointy}}c_{h,\Evapointy}^\Trans\Upsilon_{h}\left(\Gamma_{h,\Evapointx}^{-1} - \Gamma_{h,\Evapointy}^{-1}\right)\Sigma_{h,\Evapointx}\left(\Gamma_{h,\Evapointx}^{-1} - \Gamma_{h,\Evapointy}^{-1}\right)^\Trans \left(c_{h,\Evapointy}^\Trans\Upsilon_{h}\right)^\Trans
= O \left(\frac{1}{h}\left(\frac{\epsilon}{h}\wedge 1\right)^2\right),
\end{align*}
where the order $\frac{\epsilon}{h}\wedge 1$ comes from the difference $\Gamma_{h,\Evapointx}^{-1} - \Gamma_{h,\Evapointy}^{-1}$. 

Next for term (III), we have
\begin{align*}
\Var[\text{(III)}] &= \left(\frac{1}{\sqrt{c_{h,\Evapointx}^\Trans\Upsilon_{h}\Omega_{h,\Evapointx}\Upsilon_{h}c_{h,\Evapointx}}} - \frac{1}{\sqrt{c_{h,\Evapointy}^\Trans\Upsilon_{h}\Omega_{h,\Evapointy}\Upsilon_{h}c_{h,\Evapointy}}}\right)^2c_{h,\Evapointx}^\Trans\Upsilon_{h}\Omega_{h,\Evapointx}\Upsilon_{h}c_{h,\Evapointx}\\
&= \left(1 - \sqrt{1+\frac{c_{h,\Evapointx}^\Trans\Upsilon_{h}\Omega_{h,\Evapointx}\Upsilon_{h}c_{h,\Evapointx}-c_{h,\Evapointy}^\Trans\Upsilon_{h}\Omega_{h,\Evapointy}\Upsilon_{h}c_{h,\Evapointy}}{c_{h,\Evapointy}^\Trans\Upsilon_{h}\Omega_{h,\Evapointy}\Upsilon_{h}c_{h,\Evapointy}}}\right)^2\\
&\asymp \left(\frac{c_{h,\Evapointx}^\Trans\Upsilon_{h}\Omega_{h,\Evapointx}\Upsilon_{h}c_{h,\Evapointx}-c_{h,\Evapointy}^\Trans\Upsilon_{h}\Omega_{h,\Evapointy}\Upsilon_{h}c_{h,\Evapointy}}{c_{h,\Evapointy}^\Trans\Upsilon_{h}\Omega_{h,\Evapointy}\Upsilon_{h}c_{h,\Evapointy}}\right)^2\\
&= \left(\frac{c_{h,\Evapointx}^\Trans\Upsilon_{h}(\Omega_{h,\Evapointx}-\Omega_{h,\Evapointy})\Upsilon_{h}c_{h,\Evapointx}}{c_{h,\Evapointy}^\Trans\Upsilon_{h}\Omega_{h,\Evapointy}\Upsilon_{h}c_{h,\Evapointy}}
+\frac{(c_{h,\Evapointx}^\Trans\Upsilon_{h}-c_{h,\Evapointy}^\Trans\Upsilon_{h})\Omega_{h,\Evapointy}\Upsilon_{h}c_{h,\Evapointx}+(c_{h,\Evapointx}^\Trans\Upsilon_{h}-c_{h,\Evapointy}^\Trans\Upsilon_{h})\Omega_{h,\Evapointy}\Upsilon_{h}c_{h,\Evapointy}}{c_{h,\Evapointy}^\Trans\Upsilon_{h}\Omega_{h,\Evapointy}\Upsilon_{h}c_{h,\Evapointy}}\right)^2.
\end{align*}
The first term has bound
\begin{align*}
\frac{c_{h,\Evapointx}^\Trans\Upsilon_{h}(\Omega_{h,\Evapointx}-\Omega_{h,\Evapointy})\Upsilon_{h}c_{h,\Evapointx}}{c_{h,\Evapointy}^\Trans\Upsilon_{h}\Omega_{h,\Evapointy}\Upsilon_{h}c_{h,\Evapointy}} = O \left(\frac{\epsilon}{h}\right).
\end{align*}
The third term has bound
\begin{align*}
\frac{(c_{h,\Evapointx}^\Trans\Upsilon_{h}-c_{h,\Evapointy}^\Trans\Upsilon_{h})\Omega_{h,\Evapointy}\Upsilon_{h}c_{h,\Evapointy}}{c_{h,\Evapointy}^\Trans\Upsilon_{h}\Omega_{h,\Evapointy}\Upsilon_{h}c_{h,\Evapointy}} 
&\precsim \frac{|(c_{h,\Evapointx}^\Trans\Upsilon_{h}-c_{h,\Evapointy}^\Trans\Upsilon_{h})\Omega_{h,\Evapointy}^{1/2}|}{\sqrt{c_{h,\Evapointy}^\Trans\Upsilon_{h}\Omega_{h,\Evapointy}\Upsilon_{h}c_{h,\Evapointy}}} 
= O \left(\frac{1}{\sqrt{h}}r_1(\epsilon,h)r_2(h)\right).
\end{align*}
Finally, the second term can be bounded as
\begin{align*}
\frac{(c_{h,\Evapointx}^\Trans\Upsilon_{h}-c_{h,\Evapointy}^\Trans\Upsilon_{h})\Omega_{h,\Evapointy}\Upsilon_{h}c_{h,\Evapointx}}{c_{h,\Evapointy}^\Trans\Upsilon_{h}\Omega_{h,\Evapointy}\Upsilon_{h}c_{h,\Evapointy}} &= \frac{(c_{h,\Evapointx}^\Trans\Upsilon_{h}-c_{h,\Evapointy}^\Trans\Upsilon_{h})\Omega_{h,\Evapointy}\Upsilon_{h}c_{h,\Evapointy}+(c_{h,\Evapointx}^\Trans\Upsilon_{h}-c_{h,\Evapointy}^\Trans\Upsilon_{h})\Omega_{h,\Evapointy}(c_{h,\Evapointx}^\Trans\Upsilon_{h}-c_{h,\Evapointy}^\Trans\Upsilon_{h})^\Trans}{c_{h,\Evapointy}^\Trans\Upsilon_{h}\Omega_{h,\Evapointy}\Upsilon_{h}c_{h,\Evapointy}}\\
&= O \left(\frac{1}{\sqrt{h}}r_1(\epsilon,h)r_2(h) + \frac{1}{h}r_1(\epsilon,h)^2r_2(h)^2\right).
\end{align*}
Overall, we have that
\begin{align*}
\Var[\text{(III)}]  = O \left(\frac{\epsilon^2}{h^2} + \frac{1}{h}r_1(\epsilon,h)^2r_2(h)^2 + \frac{1}{h^2}r_1(\epsilon,h)^4r_2(h)^4\right).
\end{align*}

Given our assumptions on the basis function and on the kernel function, it is obvious that term (IV) has variance
\begin{align*}
\Var[\text{(IV)}] &= O \left(\frac{1}{h}\left(\frac{\epsilon}{h}\wedge 1\right)^2\right).
\end{align*}

The bound on $\Expectation[\sup_{\Evapointx\in\mathcal{I}}|\mathfrak{B}_G(\Evapointx)|]$ can be found by standard entropy calculation, and the bound on $\Expectation[\sup_{\Evapointx\in\mathcal{I}}|\mathfrak{T}_G(\Evapointx)|]$ is obtained by the following fact
\begin{align*}
\Expectation\left[\sup_{\Evapointx\in\mathcal{I}}|\mathfrak{T}_G(\Evapointx)|\right] \leq \Expectation\left[\sup_{\Evapointx\in\mathcal{I}}|\mathfrak{B}_G(\Evapointx)|\right] + \Expectation\left[\sup_{\Evapointx\in\mathcal{I}}|\tilde{\mathfrak{T}}_G(\Evapointx)-B_G(\Evapointx)|\right],
\end{align*}
and that
\begin{align*}
\Expectation\left[\sup_{\Evapointx\in\mathcal{I}}|\tilde{\mathfrak{T}}_G(\Evapointx)-\mathfrak{B}_G(\Evapointx)|\right] &= \int_0^\infty \Prob\left[\sup_{\Evapointx\in\mathcal{I}}|\tilde{\mathfrak{T}}_G(\Evapointx)-\mathfrak{B}_G(\Evapointx)|>u\right]\diff u = O\left(\frac{\log n}{\sqrt{nh}}\right) = o(\sqrt{\log n}),
\end{align*}
which follows from Theorem \ref{thm:strong approximation} and our assumption that $\log n/(nh)\to 0$.

\subsection{Proof of Lemma \ref{lemma:standard error, local projection uniform}}

We adopt the following decomposition (the integration is always on $\frac{\mathcal{X}-\Evapointy}{h}\times\Intregion$, unless otherwise specified):
\begin{align*}
\tag{I}&\frac{1}{n}\sum_{i=1}^n\iint R(u)R(v)^\Trans\Big[\Indicator(x_i\leq \Evapointx + hu) - {F}(\Evapointx + hu)\Big]\Big[\Indicator(x_i\leq \Evapointy + hv) - {F}(\Evapointy + hv)\Big] K(u)K(v) g(\Evapointx+hu)g(\Evapointy+hv)\diff u\diff v\\
\tag{II}&- \iint R(u)R(v)^\Trans\Big[\hat{F}(\Evapointx + hu) - {F}(\Evapointx + hu)\Big]\Big[\hat{F}(\Evapointy + hv) - {F}(\Evapointy + hv)\Big] K(u)K(v) g(\Evapointx+hu)g(\Evapointy+hv)\diff u\diff v.
\end{align*}

By the uniform convergence of the empirical distribution function, we have that
\begin{align*}
\sup_{\Evapointx,\Evapointy\in \mathcal{I}}|\text{(II)}| =\Op\left(\frac{1}{n}\right).
\end{align*}

From the definition of $\Sigma_{h,\Evapointx,\Evapointy}$, we know that
\begin{align*}
\Expectation[\text{(I)}] &= \Sigma_{h,\Evapointx,\Evapointy}.
\end{align*}
As (I) is a sum of bounded terms, we can apply Lemma \ref{lem:uniform convergence of iid sum} and easily show that
\begin{align*}
\sup_{\Evapointx,\Evapointy\in \mathcal{I}}\left|\text{(I)} - \Sigma_{h,\Evapointx,\Evapointy}\right| + \Op\left( \sqrt{\frac{\log n}{n}}\right).
\end{align*}

\subsection{Proof of Lemma \ref{lem:local projection uniform term 1}}

We rewrite \eqref{eq:local projection uniform term 1} as
\begin{align*}
|\text{\eqref{eq:local projection uniform term 1}}| 
&= \left|\sqrt{n}\sqrt{\frac{{c_{h,\Evapointx}^\Trans\Upsilon_{h}{\Omega}_{h,\Evapointx}\Upsilon_{h}c_{h,\Evapointx}}}{{c_{h,\Evapointx}^\Trans\Upsilon_{h}\hat{\Omega}_{h,\Evapointx}\Upsilon_{h}c_{h,\Evapointx}}}}\frac{c_{h,\Evapointx}^\Trans\Upsilon_{h}{\Gamma}_h^{-1}\int_{\Intregion} R(u)\Big[F(\Evapointx + hu) - \theta^\Trans R(u)\Upsilon_h^{-1}\Big] K\left(u\right) g(\Evapointx+hu)\diff u}{\sqrt{c_{h,\Evapointx}^\Trans\Upsilon_{h}{\Omega}_{h,\Evapointx}\Upsilon_{h}c_{h,\Evapointx}}}\right|\\
&\leq \sqrt{\frac{n}{h}}\left[\sup_{\Evapointx\in\mathcal{I}}\sqrt{\frac{{c_{h,\Evapointx}^\Trans\Upsilon_{h}{\Omega}_{h,\Evapointx}\Upsilon_{h}c_{h,\Evapointx}}}{{c_{h,\Evapointx}^\Trans\Upsilon_{h}\hat{\Omega}_{h,\Evapointx}\Upsilon_{h}c_{h,\Evapointx}}}}\right]\left[\sup_{\Evapointx\in\mathcal{I}} \left|\int_{\Intregion} R(u)\Big[F(\Evapointx + hu) - \theta^\Trans R(u)\Upsilon_h^{-1}\Big] K\left(u\right) g(\Evapointx+hu)\diff u\right| \right]\\
&= \Op\left( \sqrt{\frac{n}{h}}\sup_{\Evapointx\in\mathcal{I}}\varrho(h,\Evapointx) \right),
\end{align*}
where the final bound holds uniformly for $\Evapointx\in\mathcal{I}$. 

Next, we expand term \eqref{eq:local projection uniform term 2} as
\begin{align*}
\text{\eqref{eq:local projection uniform term 2}} 
&= \frac{1}{\sqrt{n}}\sum_{i=1}^n\frac{c^\Trans_{h,\Evapointx}\Upsilon_{h}{\Gamma}_{h,\Evapointx}^{-1}  \int_{\Intregion} R(u)\Big[\Indicator(x_i\leq \Evapointx + hu) - F(\Evapointx + hu)\Big] K\left(u\right) g(\Evapointx+hu)\diff u}{\sqrt{c_{h,\Evapointx}^\Trans\Upsilon_{h}{\Omega}_{h,\Evapointx}\Upsilon_{h}c_{h,\Evapointx}}}\\
&\qquad+ \frac{1}{\sqrt{n}}\sum_{i=1}^n\left[ 1- \sqrt{\frac{c_{h,\Evapointx}^\Trans\Upsilon_{h}{\Omega}_{h,\Evapointx}\Upsilon_{h}c_{h,\Evapointx}}{c_{h,\Evapointx}^\Trans\Upsilon_{h}\hat{\Omega}_{h,\Evapointx}\Upsilon_{h}c_{h,\Evapointx}}} \right]\frac{c^\Trans_{h,\Evapointx}\Upsilon_{h}{\Gamma}_{h,\Evapointx}^{-1}  \int_{\Intregion} R(u)\Big[\Indicator(x_i\leq \Evapointx + hu) - F(\Evapointx + hu)\Big] K\left(u\right) g(\Evapointx+hu)\diff u}{\sqrt{c_{h,\Evapointx}^\Trans\Upsilon_{h}{\Omega}_{h,\Evapointx}\Upsilon_{h}c_{h,\Evapointx}}}\\
&= \mathfrak{T}_G(\Evapointx) + \underbrace{\left[ 1- \sqrt{\frac{c_{h,\Evapointx}^\Trans\Upsilon_{h}{\Omega}_{h,\Evapointx}\Upsilon_{h}c_{h,\Evapointx}}{c_{h,\Evapointx}^\Trans\Upsilon_{h}\hat{\Omega}_{h,\Evapointx}\Upsilon_{h}c_{h,\Evapointx}}}\ \right]\mathfrak{T}_G(\Evapointx)}_{\textstyle \text{(I)}}. 
\end{align*}

Term (I) can be easily bounded by
\begin{align*}
\sup_{\Evapointx\in\mathcal{I}}|\text{(I)}| =\Op\left( \left(\sqrt{\frac{\log n}{nh^2}}\right)\Expectation\left[\sup_{\Evapointx\in\mathcal{I}}|\mathfrak{T}_G(\Evapointx)|\right] \right) = \Op\left( \frac{\log n}{\sqrt{nh^2}} \right).
\end{align*}

\subsection{Proof of Theorem \ref{thm:strong approximation, local projection}}

The claim follows from Theorem \ref{thm:strong approximation} and previous lemmas.

\subsection{Proof of Theorem \ref{thm:feasible uniform approximation local projection}}

Let $\mathcal{I}_\epsilon$ be an $\epsilon$-covering (with respect to the Euclidean metric) of $\mathcal{I}$, and assume $\epsilon\leq h$. Then the process ${\mathfrak{B}}_G(\cdot)$ can be decomposed into:
\begin{align*}
{\mathfrak{B}}_G(\Evapointx) &= {\mathfrak{B}}_G(\Pi_{\mathcal{I}_\epsilon}(\Evapointx)) + {\mathfrak{B}}_G(\Evapointx) - {\mathfrak{B}}_G(\Pi_{\mathcal{I}_\epsilon}(\Evapointx)), 
\end{align*}
where $\Pi_{\mathcal{I}_\epsilon}:\mathcal{I}\to \mathcal{I}_\epsilon$ is a mapping satisfying:
\begin{align*}
\Pi_{\mathcal{I}_\epsilon}(\Evapointx) &= \argmin_{\Evapointy\in \mathcal{I}_\epsilon}|\Evapointy-\Evapointx|. 
\end{align*}

We first study the properties of ${\mathfrak{B}}_G(\Evapointx) - {\mathfrak{B}}_G(\Pi_{\mathcal{I}_\epsilon}(\Evapointx))$. With standard entropy calculation, one has:
\begin{align*}
\Expectation\left[\sup_{\Evapointx\in\mathcal{I}} |{\mathfrak{B}}_G(\Evapointx) - {\mathfrak{B}}_G(\Pi_{\mathcal{I}_\epsilon}(\Evapointx))|\right] &\leq \Expectation\left[ \sup_{\Evapointx,\Evapointy\in \mathcal{I},|\Evapointx-\Evapointy|\leq \epsilon} |{\mathfrak{B}}_G(\Evapointx)-{\mathfrak{B}}_G(\Evapointy)| \right]
\leq \Expectation\left[ \sup_{\Evapointx,\Evapointy\in \mathcal{I},\sigma(\Evapointx,\Evapointy)\leq \delta(\epsilon)} |{\mathfrak{B}}_G(\Evapointx)-{\mathfrak{B}}_G(\Evapointy)| \right]\\
&\precsim \int_0^{\delta(\epsilon)} \sqrt{\log  N(\lambda, \mathcal{I}, \sigma_G)}\diff \lambda,
\end{align*}
where 
\begin{align*}
\delta(\epsilon) &= C\left(\frac{1}{\sqrt{h}}\frac{\epsilon}{h}  + \frac{1}{\sqrt{h}}r_1(\epsilon,h)r_2(h) + \frac{1}{h}r_1(\epsilon,h)^2r_2(h)^2\right),
\end{align*}
for some $C>0$ that does not depend on $\epsilon$ and $h$, and $N(\lambda, \mathcal{I}, \sigma_G)$ is the covering number of $\mathcal{I}$ measured by the pseudo metric $\sigma_G(\cdot,\cdot)$, which satisfies
\begin{align*}
 N(\lambda, \mathcal{I}, \sigma_G) \precsim \frac{1}{\delta^{-1}(\lambda)}. 
\end{align*}
Therefore, we have
\begin{align*}
\tag{I}\Expectation\left[\sup_{\Evapointx\in\mathcal{I}} |{\mathfrak{B}}_G(\Evapointx) - {\mathfrak{B}}_G(\Pi_{\mathcal{I}_\epsilon}(\Evapointx))|\right] &\precsim \left(\frac{1}{\sqrt{h}}\frac{\epsilon}{h}  + \frac{1}{\sqrt{h}}r_1(\epsilon,h)r_2(h) + \frac{1}{h}r_1(\epsilon,h)^2r_2(h)^2\right)\sqrt{\log n}. 
\end{align*}
A similar bound holds for the process $\hat{\mathfrak{B}}_G(\cdot)$ due to the uniform consistency of the covariance estimator. 

Now consider the discretized version of ${\mathfrak{B}}_G(\cdot)$ and $\hat{\mathfrak{B}}_G(\cdot)$. By applying Lemmas \ref{lemma:standard error, local projection uniform} and \ref{lem:gaussian comparison}, we directly obtain the following bound:
\begin{align*}
\tag{II}\sup_{ A\ \text{rectangular}} \left| \Prob\Big[ \Big\{\mathfrak{B}_G(\Pi_{\mathcal{I}_\epsilon}(\Evapointx)),\Evapointx\in\mathcal{I}\Big\} \in A \Big]
-
\Prob^\star\Big[ \Big\{\hat{\mathfrak{B}}_G(\Pi_{\mathcal{I}_\epsilon}(\Evapointx)),\Evapointx\in\mathcal{I}\Big\} \in A \Big] \right| = \Op\left( \left(\frac{\log n}{nh^2}\right)^{\frac{1}{4}}\log \frac{1}{\epsilon} \right).
\end{align*}
As $\epsilon$ appears in (I) polynomially but only logarithmically in (II), it is possible to choose $\epsilon$ sufficiently small so that the discretization error becomes negligible. Therefore, 
\begin{align*}
\sup_{u\in \mathbb{R}}\left| \Prob\Big[ \sup_{\Evapointx\in\mathcal{I}}|\mathfrak{B}_G(\Evapointx)| \leq u \Big]
-
\Prob^\star\Big[\sup_{\Evapointx\in\mathcal{I}}|\hat{\mathfrak{B}}_G(\Evapointx)| \leq u \Big] \right| = \Op\left(\frac{\log^{\frac{5}{4}} n}{(nh^2)^{\frac{1}{4}}} \right).
\end{align*}

\subsection{Proof of Lemma \ref{lem:uniform consistency of Gamma}}

We apply Lemma \ref{lem:uniform convergence of iid sum}. For simplicity, assume $R(\cdot)$ is scalar, and let
\begin{align*}
u_{i,h}(\Evapointx) =  R\left( \frac{x_i-\Evapointx}{h} \right)^2\frac{1}{h}K\left(\frac{x_i-\Evapointx}{h}\right) - \Gamma_{h,\Evapointx}.
\end{align*}
Then it is easy to see that
\begin{align*}
\sup_{\Evapointx\in\mathcal{I}}\max_{1\leq i\leq n}\Var[u_{i,h}(\Evapointx)] = O(h^{-1}), \qquad \sup_{\Evapointx\in\mathcal{I}}\max_{1\leq i\leq n}|u_{i,h}(\Evapointx)| = O(h^{-1}).
\end{align*}
Let $|\Evapointx - \Evapointy|\leq \epsilon\leq h$, we also have
\begin{align*}
\left|u_{i,h}(\Evapointx) - u_{i,h}(\Evapointy)\right| &\leq \left|R\left( \frac{x_i-\Evapointx}{h} \right)^2\frac{1}{h}K\left(\frac{x_i-\Evapointx}{h}\right) - R\left( \frac{x_i-\Evapointy}{h} \right)^2\frac{1}{h}K\left(\frac{x_i-\Evapointy}{h}\right)\right| + \left|\Gamma_{h,\Evapointx} - \Gamma_{h,\Evapointy} \right|\\
&\leq \left|R\left( \frac{x_i-\Evapointx}{h} \right)^2-R\left( \frac{x_i-\Evapointy}{h} \right)^2\right|\frac{1}{h}K\left(\frac{x_i-\Evapointx}{h}\right)  + 
R\left( \frac{x_i-\Evapointy}{h} \right)^2\frac{1}{h}\left|K\left(\frac{x_i-\Evapointx}{h}\right) - K\left(\frac{x_i-\Evapointy}{h}\right)\right|  + \left|\Gamma_{h,\Evapointx} - \Gamma_{h,\Evapointy} \right|\\
&\leq M\left[\frac{\epsilon}{h}\frac{1}{h}K\left(\frac{x_i-\Evapointx}{h}\right) + \frac{\epsilon}{h}\frac{1}{h}K^{\dag}\left(\frac{x_i-\Evapointx}{h}\right) + \frac{1}{h}K^{\ddag}\left(\frac{x_i-\Evapointx}{h}\right) +  \frac{\epsilon}{h}\right].
\end{align*}
where $M$ is some constant that does not depend on $n$, $h$ or $\epsilon$. Then it is easy to see that
\begin{align*}
\sup_{\Evapointx\in\mathcal{I}}\max_{1\leq i\leq n}\Var[u_{i,h,\epsilon}(\Evapointx)] = O\left(\frac{\epsilon}{h^2}\right), \qquad \sup_{\Evapointx\in\mathcal{I}}\max_{1\leq i\leq n}|u_{i,h,\epsilon}(\Evapointx) - \Expectation[u_{i,h,\epsilon}(\Evapointx)]| = O(h^{-1}),\quad \sup_{\Evapointx\in\mathcal{I}}\max_{1\leq i\leq n}\Expectation[|u_{i,h,\epsilon}(\Evapointx)|] = O\left(\frac{\epsilon}{h}\right). 
\end{align*}
Now take $\epsilon = \sqrt{h\log n/n}$, then $\log N(\epsilon, \mathcal{I}, |\cdot|)=O(\log n)$. Lemma \ref{lem:uniform convergence of iid sum} implies that
\begin{align*}
\sup_{\Evapointx\in \mathcal{I}}\left|\frac{1}{n}\sum_{i=1}^n R\left( \frac{x_i-\Evapointx}{h} \right)^2\frac{1}{h}K\left(\frac{x_i-\Evapointx}{h}\right) - \Gamma_{h,\Evapointx}\right| = \Op\left( \sqrt{\frac{\log n}{nh}} \right).
\end{align*}

\subsection{Proof of Lemma \ref{lemma:standard error, local regression uniform}}

Let $R_i(\Evapointx) = R(x_i-\Evapointx)$ and $W_i(\Evapointx) = K((x_i-\Evapointx)/h)/h$, then we split $\hat{\Sigma}_{h,\Evapointx,\Evapointy}$ into two terms, 
\begin{align*}
\text{(I)} &= \frac{1}{n^3}\sum_{i,j,k} \Upsilon_{h}R_j(\Evapointx)R_k(\Evapointy)^\Trans \Upsilon_{h} W_j(\Evapointx)W_k(\Evapointy) \Big(\Indicator(x_i\leq x_j) - F(x_j)\Big) \Big(\Indicator(x_i\leq x_k) - F(x_k)\Big)\\
\text{(II)} &= -\frac{1}{n^2}\sum_{j,k} \Upsilon_{h}R_j(\Evapointx)R_k(\Evapointy)^\Trans \Upsilon_{h} W_j(\Evapointx)W_k(\Evapointy) \Big(\hat{F}(x_j)-{F}(x_j)\Big) \Big(\hat{F}(x_k) - {F}(x_k)\Big).
\end{align*}

(II) satisfies
\begin{align*}
\sup_{\Evapointx,\Evapointy\in \mathcal{I}}\left|\text{(II)}\right| \leq \sup_{x}|\hat{F}(x) - F(x)|^2\left(\sup_{\Evapointx\in \mathcal{I}}\frac{1}{n}\sum_{j} \left| \Upsilon_{h}R_j(\Evapointx)W_j(\Evapointx)\right|\right)^2.
\end{align*}
It is obvious that 
\begin{align*}
\sup_{x}|\hat{F}(x) - F(x)|^2 = \Op\left( \frac{1}{n}\right). 
\end{align*}
As for the second part, one can employ the same technique used to prove Lemma \ref{lem:uniform consistency of Gamma} and show that 
\begin{align*}
\sup_{\Evapointx\in \mathcal{I}}\frac{1}{n}\sum_{j} \left| \Upsilon_{h}R_j(\Evapointx)W_j(\Evapointx)\right| = \Op(1),
\end{align*}
implying that
\begin{align*}
\sup_{\Evapointx,\Evapointy\in \mathcal{I}}\left|\text{(II)}\right| = \Op\left(\frac{1}{n}\right).
\end{align*} 

For (I), we first define
\begin{align*}
u_{ij}(\Evapointx) &= \Upsilon_{h}R_j(\Evapointx) W_j(\Evapointx) \Big(\Indicator(x_i\leq x_j) - F(x_j)\Big),
\end{align*}
and
\begin{align*}
\bar{u}_i(\Evapointx) &= \Expectation[u_{ij}(\Evapointx)|x_i;i\neq j],\qquad \hat{u}_i(\Evapointx) = \frac{1}{n}\sum_j u_{ij}(\Evapointx).
\end{align*}
Then
\begin{align*}
\text{(I)} &= \frac{1}{n}\sum_i\left( \frac{1}{n}\sum_j u_{ij}(\Evapointx) \right)\left( \frac{1}{n}\sum_j u_{ij}(\Evapointy) \right)^\Trans = \frac{1}{n}\sum_i \hat{u}_i(\Evapointx)\hat{u}_i(\Evapointy)^\Trans\\
&= \frac{1}{n}\sum_i \bar{u}_i(\Evapointx)\bar{u}_i(\Evapointy)^\Trans + \frac{1}{n}\sum_i\left(\hat{u}_i(\Evapointx) - \bar{u}_i(\Evapointx)\right)\hat{u}_i(\Evapointy)^\Trans + \frac{1}{n}\sum_i\bar{u}_i(\Evapointx)\left(\hat{u}_i(\Evapointy) - \bar{u}_i(\Evapointy)\right)^\Trans\\
&= \underbrace{\frac{1}{n}\sum_i \bar{u}_i(\Evapointx)\bar{u}_i(\Evapointy)^\Trans}_{\text{(I.1)}} + \underbrace{\frac{1}{n}\sum_i\left(\hat{u}_i(\Evapointx) - \bar{u}_i(\Evapointx)\right)\bar{u}_i(\Evapointy)^\Trans}_{\text{(I.2)}} + \underbrace{\frac{1}{n}\sum_i\bar{u}_i(\Evapointx)\left(\hat{u}_i(\Evapointy) - \bar{u}_i(\Evapointy)\right)^\Trans}_{\text{(I.3)}}\\
&\qquad\qquad\qquad + \underbrace{\frac{1}{n}\sum_i\left(\hat{u}_i(\Evapointx) - \bar{u}_i(\Evapointx)\right)\left(\hat{u}_i(\Evapointy) - \bar{u}_i(\Evapointy)\right)^\Trans}_{\text{(I.4)}}.
\end{align*}

Term (I.1) has been analyzed in Lemma \ref{lemma:standard error, local projection uniform}, which satisfies
\begin{align*}
\sup_{\Evapointx,\Evapointy\in \mathcal{I}}\left|\text{(I.1)} - \Sigma_{h,\Evapointx,\Evapointy}\right| = \Op\left( \sqrt{\frac{\log n}{n}} \right). 
\end{align*}

Term (I.2) has expansion:
\begin{align*}
\text{(I.2)} &= \frac{1}{n^2}\sum_{i,j}\left(u_{ij}(\Evapointx) - \bar{u}_i(\Evapointx)\right)\bar{u}_i(\Evapointy)^\Trans
= \underbrace{\frac{1}{n^2}\sum_{\substack{i,j\\ \text{distinct}}}\left(u_{ij}(\Evapointx) - \bar{u}_i(\Evapointx)\right)\bar{u}_i(\Evapointy)^\Trans}_{\text{(I.2.1)}} 
+ \underbrace{\frac{1}{n^2}\sum_{i}\left(u_{ii}(\Evapointx) - \bar{u}_i(\Evapointx)\right)\bar{u}_i(\Evapointy)^\Trans}_{\text{(I.2.2)}}.
\end{align*}
By the same technique of Lemma \ref{lem:uniform consistency of Gamma}, one can show that
\begin{align*}
\sup_{\Evapointx,\Evapointy\in \mathcal{I}}|\text{(I.2.2)}| = \Op\left(\frac{1}{n}\right). 
\end{align*}
We need a further decomposition to make (I.2.1) a degenerate U-statistic:
\begin{align*}
\text{(I.2.1)} &= \underbrace{\frac{n-1}{n^2}\sum_{j}\Expectation\left[\left.\left(u_{ij}(\Evapointx) - \bar{u}_i(\Evapointx)\right)\bar{u}_i(\Evapointy)^\Trans\right|x_j\right]}_{\text{(I.2.1.1)}}\\
&\qquad + \underbrace{\frac{1}{n^2}\sum_{\substack{i,j\\ \text{distinct}}}\Bigg\{\left(u_{ij}(\Evapointx) - \bar{u}_i(\Evapointx)\right)\bar{u}_i(\Evapointy)^\Trans - \Expectation\left[\left.\left(u_{ij}(\Evapointx) - \bar{u}_i(\Evapointx)\right)\bar{u}_i(\Evapointy)^\Trans\right|x_j\right]\Bigg\}}_{\text{(I.2.1.2)}}.
\end{align*}

(I.2.1) has zero mean. By discretizing $\mathcal{I}$ and apply Bernstein's inequality, one can show that the (I.2.1.1) has order $\Op\left(\sqrt{\log n/n}\right)$.

For (I.2.1.2), we first discretize $\mathcal{I}$ and then apply a Bernstein-type inequality (Lemma \ref{lem:ustatistic concentration inequality}) for degenerate U-statistics, which gives an order
\begin{align*}
\sup_{\Evapointx,\Evapointy\in \mathcal{I}}\left|\text{(I.2.1.2)} \right| = \Op\left( \frac{\log n}{\sqrt{n^2h}} \right).
\end{align*}

Overall, we have
\begin{align*}
\sup_{\Evapointx,\Evapointy\in \mathcal{I}}|\text{(I.2)}| = \Op\left( \frac{1}{n} +  \sqrt{\frac{\log n}{n}} + \frac{\log n}{\sqrt{n^2h}} \right) = \Op\left(  \sqrt{\frac{\log n}{n}}  \right),
\end{align*}
and the same bound applies to (I.3).

For (I.4), one can show that
\begin{align*}
\sup_{\Evapointx\in \mathcal{I}}\sup_{x\in\mathcal{X}} \left| \frac{1}{n}\sum_j \Upsilon_{h}R_j(\Evapointx) W_j(\Evapointx) \Big(\Indicator(x\leq x_j) - F(x_j)\Big) - \Expectation\left[\Upsilon_{h}R_j(\Evapointx) W_j(\Evapointx) \Big(\Indicator(x\leq x_j) - F(x_j)\Big)\right] \right| = \Op\left( \sqrt{\frac{\log n}{nh}} \right),
\end{align*}
which means
\begin{align*}
\sup_{\Evapointx,\Evapointy\in \mathcal{I}} |\text{(I.4)}| = \Op\left( \frac{\log n}{nh} \right) = \Op\left( \sqrt{\frac{\log n}{n}} \right),
\end{align*}
under our assumption that $\log n/(nh^2)\to 0$. 

As a result, we have
\begin{align*}
\sup_{\Evapointx,\Evapointy\in \mathcal{I}} \left| \hat\Sigma_{h,\Evapointx, \Evapointy} - \Sigma_{h,\Evapointx, \Evapointy} \right| = \Op\left( \sqrt{\frac{\log n}{n}} \right). 
\end{align*}

Now take $c$ to be a generic vector. Then we have
\begin{align*}
\frac{c^\Trans_{h,\Evapointx}\Upsilon_h(\hat{\Omega}_{h,\Evapointx,\Evapointy}-{\Omega}_{h,\Evapointx,\Evapointy})\Upsilon_h c_{h,\Evapointy} }{\sqrt{c^\Trans_{h,\Evapointx}\Upsilon_h{\Omega}_{h,\Evapointx}\Upsilon_h c_{h,\Evapointx}}\sqrt{c^\Trans_{h,\Evapointy}\Upsilon_h{\Omega}_{h,\Evapointy}\Upsilon_h c_{h,\Evapointy}}}
&= \frac{c^\Trans_{h,\Evapointx}\Upsilon_h \hat\Gamma_{h,\Evapointx}^{-1} (\hat\Sigma_{h,\Evapointx,\Evapointy} - \Sigma_{h,\Evapointx,\Evapointy}) \hat\Gamma_{h,\Evapointy}^{-1} \Upsilon_h c_{h,\Evapointy}}{\sqrt{c^\Trans_{h,\Evapointx}\Upsilon_h{\Omega}_{h,\Evapointx}\Upsilon_h c_{h,\Evapointx}}\sqrt{c^\Trans_{h,\Evapointy}\Upsilon_h{\Omega}_{h,\Evapointy}\Upsilon_h c_{h,\Evapointy}}} \\
&\quad+ \frac{c^\Trans_{h,\Evapointx}\Upsilon_h (\hat\Gamma_{h,\Evapointx}^{-1} -\Gamma_{h,\Evapointx}^{-1})  \Sigma_{h,\Evapointx,\Evapointy} \hat\Gamma_{h,\Evapointy}^{-1} \Upsilon_h c_{h,\Evapointy}}{\sqrt{c^\Trans_{h,\Evapointx}\Upsilon_h{\Omega}_{h,\Evapointx}\Upsilon_h c_{h,\Evapointx}}\sqrt{c^\Trans_{h,\Evapointy}\Upsilon_h{\Omega}_{h,\Evapointy}\Upsilon_h c_{h,\Evapointy}}} \\
&\quad+ \frac{c^\Trans_{h,\Evapointx}\Upsilon_h \Gamma_{h,\Evapointx}^{-1} \Sigma_{h,\Evapointx,\Evapointy}(\hat\Gamma_{h,\Evapointy}^{-1} -\Gamma_{h,\Evapointy}^{-1})   \Upsilon_h c_{h,\Evapointy}}{\sqrt{c^\Trans_{h,\Evapointx}\Upsilon_h{\Omega}_{h,\Evapointx}\Upsilon_h c_{h,\Evapointx}}\sqrt{c^\Trans_{h,\Evapointy}\Upsilon_h{\Omega}_{h,\Evapointy}\Upsilon_h c_{h,\Evapointy}}}
.
\end{align*}
From the analysis of $\hat\Sigma_{h,\Evapointx,\Evapointy}$, we have
\begin{align*}
\sup_{\Evapointx,\Evapointy\in \mathcal{I}}\left|\frac{c^\Trans_{h,\Evapointx}\Upsilon_h \hat\Gamma_{h,\Evapointx}^{-1} (\hat\Sigma_{h,\Evapointx,\Evapointy} - \Sigma_{h,\Evapointx,\Evapointy}) \hat\Gamma_{h,\Evapointy}^{-1} \Upsilon_h c_{h,\Evapointy}}{\sqrt{c^\Trans_{h,\Evapointx}\Upsilon_h{\Omega}_{h,\Evapointx}\Upsilon_h c_{h,\Evapointx}}\sqrt{c^\Trans_{h,\Evapointy}\Upsilon_h{\Omega}_{h,\Evapointy}\Upsilon_h c_{h,\Evapointy}}}\right|  = \Op\left( \sqrt{\frac{\log n}{nh^2}} \right). 
\end{align*}
For the second term, we have
\begin{align*}
\left|\frac{c^\Trans_{h,\Evapointx}\Upsilon_h (\hat\Gamma_{h,\Evapointx}^{-1} -\Gamma_{h,\Evapointx}^{-1})  \Sigma_{h,\Evapointx,\Evapointy} \hat\Gamma_{h,\Evapointy}^{-1} \Upsilon_h c_{h,\Evapointy}}{\sqrt{c^\Trans_{h,\Evapointx}\Upsilon_h{\Omega}_{h,\Evapointx}\Upsilon_h c_{h,\Evapointx}}\sqrt{c^\Trans_{h,\Evapointy}\Upsilon_h{\Omega}_{h,\Evapointy}\Upsilon_h c_{h,\Evapointy}}}\right| 
&\leq \frac{|c^\Trans_{h,\Evapointx}\Upsilon_h (\hat\Gamma_{h,\Evapointx}^{-1} -\Gamma_{h,\Evapointx}^{-1})  \Sigma_{h,\Evapointx}^{1/2}|\cdot|c^\Trans_{h,\Evapointy}\Upsilon_h \hat{\Gamma}_{h,\Evapointy}^{-1}  \Sigma_{h,\Evapointy}^{1/2}|}{\sqrt{c^\Trans_{h,\Evapointx}\Upsilon_h{\Omega}_{h,\Evapointx}\Upsilon_h c_{h,\Evapointx}}\sqrt{c^\Trans_{h,\Evapointy}\Upsilon_h{\Omega}_{h,\Evapointy}\Upsilon_h c_{h,\Evapointy}}}\\
&= \Op\left(  \sqrt{\frac{\log n}{nh^2}} \right).
\end{align*}
The same bound holds for the third term.

\subsection{Proof of Lemma \ref{lem:local regression uniform term 1}}

We decompose \eqref{eq:local regression uniform term 1} as
\begin{align*}
\sup_{\Evapointx\in\mathcal{I}}|\text{\eqref{eq:local regression uniform term 1}}| &\leq 
\frac{1}{\sqrt{n}}
\underbrace{\left[\sup_{\Evapointx\in\mathcal{I}}\left|\frac{c_{h,\Evapointx}^\Trans\Upsilon_{h}\hat{\Gamma}_{h,\Evapointx}^{-1}}{\sqrt{c_{h,\Evapointx}^\Trans\Upsilon_{h}\hat{\Omega}_{h,\Evapointx}\Upsilon_{h}c_{h,\Evapointx}}}\right|\right]}_{\text{(I)}}
\underbrace{\left[\sup_{\Evapointx\in\mathcal{I}}\left|\frac{1}{n}\sum_{i=1}^n {R((x_i-\Evapointx)/h)[1-F(x_i)]\frac{1}{h}K(\frac{x_i-\Evapointx}{h})}  \right|\right]}_{\text{(II)}}.
\end{align*}
As both $\hat\Gamma_{h,\Evapointx}$ and $c_{h,\Evapointx}^\Trans\Upsilon_{h}\hat{\Omega}_{h,\Evapointx}\Upsilon_{h}c_{h,\Evapointx}$ are uniformly consistent, term (I) has order
\begin{align*}
\text{(I)} = \Op\left(\sqrt{\frac{1}{h}}\right).
\end{align*}

For (II), we can employ the same technique used to prove Lemma \ref{lem:uniform consistency of Gamma} and show that
\begin{align*}
\text{(II)} &= \Op\left( 1 + \sqrt{\frac{\log n}{nh}} \right) = \Op(1),
\end{align*}
where the leading order in the above represents the mean of ${R((x_i-\Evapointx)/h)[1-F(x_i)]\frac{1}{h}K(\frac{x_i-\Evapointx}{h})}$. 

Next, term \eqref{eq:local regression uniform term 2} is bounded by
\begin{align*}
\sup_{\Evapointx\in\mathcal{I}}|\eqref{eq:local regression uniform term 2}| \leq 
\sqrt{n}
\underbrace{\left[ \sup_{\Evapointx\in\mathcal{I}}\left| \frac{c_{h,\Evapointx}^\Trans\Upsilon_{h}\hat{\Gamma}_{h,\Evapointx}^{-1}}{\sqrt{c_{h,\Evapointx}^\Trans\Upsilon_{h}\hat{\Omega}_{h,\Evapointx}\Upsilon_{h}c_{h,\Evapointx}}} \right| \right]}_{\text{(I)}}
\underbrace{\left[\sup_{\Evapointx\in\mathcal{I}}\left|\frac{1}{n}\sum_{i=1}^n {R((x_i-\Evapointx)/h)[F(x_i)-\theta(\Evapointx)^\Trans R(x_i-\Evapointx)]\frac{1}{h}K(\frac{x_i-\Evapointx}{h})}\right|\right]}_{\text{(II)}}.
\end{align*}
Employing the same argument used to prove Lemma \ref{lem:local regression uniform term 1}, we have
\begin{align*}
\text{(I)} = \Op\left(\sqrt{\frac{1}{h}}\right). 
\end{align*}

To bound term (II), recall that $K(\cdot)$ is supported on $[-1,1]$, meaning that
\begin{align*}
&\ \sup_{\Evapointx\in\mathcal{I}}\left|\frac{1}{n}\sum_{i=1}^n {R((x_i-\Evapointx)/h)[F(x_i)-\theta(\Evapointx)^\Trans R(x_i-\Evapointx)]\frac{1}{h}K(\frac{x_i-\Evapointx}{h})}\right|\\
&= \sup_{\Evapointx\in\mathcal{I}}\left|\frac{1}{n}\sum_{i=1}^n {R((x_i-\Evapointx)/h)[F(x_i)-\theta(\Evapointx)^\Trans R(x_i-\Evapointx)]\Indicator(|x_i-\Evapointx|\leq h)\frac{1}{h}K(\frac{x_i-\Evapointx}{h})}\right|\\
&\leq \underbrace{\left[\sup_{\Evapointx\in\mathcal{I}}\frac{1}{n}\sum_{i=1}^n \left|{R((x_i-\Evapointx)/h)\frac{1}{h}K(\frac{x_i-\Evapointx}{h})}\right|\right]}_{\text{(II.1)}} \underbrace{\left[\sup_{\Evapointx\in\mathcal{I}}\sup_{u\in [\Evapointx-h,\Evapointx+h]}\left|\Big[F(u)-\theta(\Evapointx)^\Trans R(u-\Evapointx)\Big]\right|\right]}_{\text{(II.2)}}.
\end{align*}

Term (II.2) has the bound $\sup_{\Evapointx\in\mathcal{I}}\varrho(h,\Evapointx)$. Term (II.1) can be bounded by mean and variance calculations and adopting the proof of Lemma \ref{lem:uniform consistency of Gamma}, which leads to 
\begin{align*}
\text{(II.1)} = \Op\left(1 + \sqrt{\frac{\log n}{nh}}\right) = \Op(1). 
\end{align*}

To show the last conclusion, define the following:
\begin{align*}
u_{ij}(\Evapointx) &= \Upsilon_hR(x_j-\Evapointx)\Big[ \Indicator(x_i\leq x_j) -F(x_j)  \Big] \frac{1}{h}K\left(\frac{x_j-\Evapointx}{h}\right) - \int_{\Intregion} R(u)\Big[\Indicator(x_i\leq \Evapointx + hu) - F(\Evapointx + hu)\Big] K\left(u\right) f(\Evapointx + hu)\diff u,
\end{align*}
then $n^{-2}\sum_{i,j=1, i\neq j}^n u_{ij}(\Evapointx)$ is a degenerate U-statistic. We rewrite \eqref{eq:local regression uniform term 3} as
\begin{align*}
\sup_{\Evapointx\in\mathcal{I}}|\text{\eqref{eq:local regression uniform term 3}}| \leq
\sqrt{n}
\underbrace{\left[ \sup_{\Evapointx\in\mathcal{I}}\left| \frac{c_{h,\Evapointx}^\Trans\Upsilon_{h}\hat{\Gamma}_{h,\Evapointx}^{-1}}{\sqrt{c_{h,\Evapointx}^\Trans\Upsilon_{h}\hat{\Omega}_{h,\Evapointx}\Upsilon_{h}c_{h,\Evapointx}}} \right| \right]}_{\text{(I)}}
\underbrace{\left[ \sup_{\Evapointx\in\mathcal{I}}\left| \frac{1}{n^2}\sum_{i,j=1,i\neq j}^n u_{ij} \right| \right]}_{\text{(II)}}. 
\end{align*}
As before, we have
\begin{align*}
\text{(I)} = \Op\left(\sqrt{\frac{1}{h}}\right). 
\end{align*}

Now we consider (II). Let $\mathcal{I}_\epsilon$ be an $\frac{\epsilon}{2}$-covering of $\mathcal{I}$, we have
\begin{align*}
\sup_{\Evapointx\in\mathcal{I}}\left|\frac{1}{n^2}\sum_{i,j=1, i\neq j}^n u_{ij}(\Evapointx)\right| &\leq \underbrace{\max_{\Evapointx\in\mathcal{I}_\epsilon}\left|\frac{1}{n^2}\sum_{i,j=1, i\neq j}^n u_{ij}(\Evapointx)\right|}_{\text{(II.1)}}+ \underbrace{\max_{\Evapointx\in\mathcal{I}_\epsilon,\Evapointy\in\mathcal{I},|\Evapointx-\Evapointy|\leq \epsilon}\left|\frac{1}{n^2}\sum_{i,j=1, i\neq j}^n \Big(u_{ij}(\Evapointx)-u_{ij}(\Evapointy)\Big)\right|}_{\text{(II.2)}}.
\end{align*}

We rely on the concentration inequality in Lemma \ref{lem:ustatistic concentration inequality} for degenerate second order U-statistics. By our assumptions, $A$ can be chosen to be $C_1h^{-1}$ where $C_1$ is some constant that is independent of $\Evapointx$. Similarly, $B$ can be chosen to be $C_2\sqrt{n}h^{-1}$ for some constant $C_2$ which is independent of $\Evapointx$, and $D$ can be chosen as $C_3nh^{-1/2}$ for some $C_3$ independent of $\Evapointx$. Therefore, by setting $\eta=K\log n/\sqrt{n^2h}$ for some large constant $K$, we have
\begin{align*}
\Prob\left[ \text{(II.1)}\geq \eta \right] &\leq C\frac{1}{\epsilon}\max_{\Evapointx\in\mathcal{I}_\epsilon}\Prob\left[ \left|\sum_{i,j=1,i\neq j}^n u_{ij}(\Evapointx)\right|\geq n^2\eta \right]\\
&\leq C\frac{1}{\epsilon} \exp\left\{ -\frac{1}{C}\min\left[ \frac{n^2h^{1/2}\eta}{nc_3},\ \left(\frac{n^2h\eta}{n^{1/2}c_2}\right)^{\frac{2}{3}},\ \left( \frac{n^2h\eta}{c_1} \right)^{\frac{1}{2}} \right] \right\}\\
&= C\frac{1}{\epsilon} \exp\left\{ -\frac{1}{C}\min\left[ \frac{K \log n}{c_3},\ \left(\frac{K\sqrt{nh}\log n}{c_2}\right)^{\frac{2}{3}},\ \left( \frac{K\sqrt{n^2h}\log n}{c_1} \right)^{\frac{1}{2}} \right] \right\}.
\end{align*}
As $\epsilon$ is at most polynomial in $n$, the above tends to zero for all $K$ large enough, which implies
\begin{align*}
\text{(II.1)} = \Op\left( \frac{\log n}{\sqrt{n^2h}} \right).
\end{align*}
With tedious but still straightforward calculations, it can be shown that
\begin{align*}
\text{(II.2)} &= \Op\left( \frac{\epsilon}{h} + \frac{\log n}{\sqrt{n^2h}} + \frac{\epsilon}{h}\frac{\log n}{\sqrt{n^2h}} \right),
\end{align*}
and to match the rates, let $\epsilon = h\log n/\sqrt{n^2h}$. 

\subsection{Proof of Lemma \ref{lem:local regression uniform term 2}}

The proof resembles that of of Lemma \ref{lem:local projection uniform term 1}. 

\subsection{Proof of Theorem \ref{thm:strong approximation, local regression}}

The proof resembles that of Theorem \ref{thm:strong approximation, local projection}.

\subsection{Proof of Theorem \ref{thm:feasible uniform approximation local regression}}

The proof resembles that of Theorem \ref{thm:feasible uniform approximation local projection}.

}


\singlespacing
\bibliographystyle{econometrica}
\bibliography{Cattaneo-Jansson-Ma_2021_JoE}

\clearpage

\clearpage